\documentclass[aps,prd,amsmath,floats,floatfix,twocolumn,nofootinbib,superscriptaddress,showpacs]{revtex4-1}
\usepackage{graphicx}
\usepackage{multirow}
\usepackage{color}
\usepackage{latexsym}
\usepackage{amsfonts}
\usepackage{amsmath}
\usepackage[colorlinks]{hyperref}
\hypersetup{
  colorlinks=true,        
  linkcolor=blue,         
  citecolor=cyan,         
}
\graphicspath{{figures/}} 

\newcommand{\be}{\begin{eqnarray}}
\newcommand{\ee}{\end{eqnarray}}

\begin{document}

	\title{The Image of Scalar Hairy Black Holes with Asymmetric Potential}

\author{Carlos~A.~Benavides-Gallego}
\email{cabenavidesg@sjtu.edu.cn}
\affiliation{School of Aeronautics and Astronautics, Shanghai Jiao Tong University, Shanghai 200240, PRC.}
\affiliation{School of Physics and Astronomy, Shanghai Jiao Tong University, 800 Dongchuan Road, Minhang, Shanghai 200240, PRC.}
\affiliation{Shanghai Frontiers Science Center of Gravitational Wave Detection, 800 Dongchuan Road, Minhang, Shanghai 200240, People's Republic of China}
    
\author{Eduard Larra\~{n}aga}
\email[]{ealarranga@unal.edu.co}
\affiliation{ Universidad Nacional de Colombia, Sede Bogotá. Facultad de Ciencias. Observatorio Astronómico Nacional. Ciudad Universitaria. Bogotá, 111321, Colombia. }

\date{\today}

\begin{abstract}
Black hole accretion disks are a fascinating topic in astrophysics, as they play a crucial role in several high-energy situations. This paper investigates the optical appearance of scalar hairy black holes (SHBHs) with asymmetric potential, a numerical solution obtained in Phys. Rev. D 73, 084002 (2006) and discussed in Phys.Rev.D 108 (2023) 4, 044020. Since the solution is spherically symmetric and surrounded by a thin accretion disk, we base our analysis on the work of J.~P. Lumininet (1979). We discuss the behavior of the effective potential for massive and massless particles, the innermost stable circular orbits (ISCO), and the photon sphere radius for different SHBHs. The study includes the plots of isoradial curves and spectral shifts arising from gravitational and Doppler shifts by considering direct and secondary images. Based on the work of Page and Thorne (1974), we also investigate the intrinsic intensity of radiation emitted by the disk at a given radius, which allows the calculation of the distribution of observed bolometric flux. We use the angular size of the shadow reported by the EHT for Sagittarius A* and M87* to constrain the SHBHs parameters. 
\end{abstract}
	
	\maketitle
	
\section{Introduction}
    
    The black holes predicted by general relativity (GR) are characterized uniquely by the mass, the angular momentum, and the electric charge; the well-known non-hair theorem/conjecture~\cite{Israel:1967wq, Israel:1967za, Carter:1971zc, Hawking:1972qk, Robinson:1975bv}. The conjecture considers two assumptions: firstly, the existence of an event horizon that encloses the singularity~\cite{Penrose:1969pc}, and secondly, there are no closed time-like loops in the exterior domain of the black hole~\cite{Johannsen:2010ru}. Nevertheless, when it comes to astrophysical black holes, the presence of the plasma around them quickly discharges the BHs, and the electric charge is usually neglected~\cite{Zajacek:2019kla}. Therefore, scientists believe that astrophysical black holes are best modeled by the Kerr space-time, which only considers the mass and the angular momentum, the Kerr hypothesis~\cite{Bambi:2011mj}. 

    On the other hand, some research works have demonstrated the possibility of eluding the non-hair theorem/conjecture under special conditions~\cite{Bizon:1990sr, Bartnik:1988am}. In~\cite{Bizon:1990sr}, for example, P.~Bizon numerically investigated the static and spherically symmetric Einstein-Yang-Mills equations with $SU(2)$ gauge group showing that these equations have asymptotically flat solutions with regular event horizons that approximates the Schwarzschild solution. A similar analysis was performed previously by R.~Bartnik and J.~McKinnon in~\cite{Bartnik:1988am}, where the authors also show the existence of non-singular and asymptotically flat numerical solutions that, in a near-field region, behave approximately to the Reissner-Nordstrom with Dirac monopole curvature source and, asymptotically, behave as Schwarzschild space-time. According to P.~Bizon, ``\textit{the existence of these solutions is incompatible with the idea of the no-hair conjecture since the Yang-Mills hair is not associated with any global charge which would forbid it to be radiated away to infinity.}''~\cite{Bizon:1990sr}. There are more examples of numerical black hole solutions violating the no-hair theorem/conjecture in~\cite{Kuenzle:1990is, Greene:1992fw, Lavrelashvili:1992ia, Corichi:2005pa, Chew:2022enh, Chew:2023olq}. Nevertheless, some of these solutions are not stable~\cite{Straumann:1989tf, Straumann:1990as, Bizon:1991nt, Zhou:1991nu}. In this sense, although the non-hair theorem/conjecture only involves the existence of solutions and not their stability, this is still a crucial question to consider when obtaining these kinds of solutions~\cite{Greene:1992fw}.  
    
    Another way to construct numerical solutions not compatible with the no-hair theorem/conjecture is to minimally couple the Einstein gravity with a scalar field in which the scalar potential is not strictly positive so that the weak energy condition is violated~\cite{Herdeiro:2015waa}. One example of such construction was proposed in~\cite{Corichi:2005pa}. There, the authors solved numerically the field equations for static solutions of a self-gravitating scalar field with an asymmetric scalar potential containing a global minimum, a local minimum, and a local maximum. In this way, A.~Corichi et al. construct the asymptotically flat black hole solution by considering the local minimum of the scalar potential to be zero and then analyzing the empirical mass formulas, finding that the ADM mass of the space-time is similar to other ``hairy'' theories for small black holes, but its behavior changes drastically in the case of larger black holes. Similar solutions following this procedure were obtained in \cite{Nucamendi:1995ex, Karakasis:2021rpn, Anabalon:2012ih, Nikonov:2008zz, Winstanley:2005fu, Martinez:2004nb, Bronnikov:2001ah, Dennhardt:1996cz, Bechmann:1995sa}.    
    
    From the observational point of view, we have seen strong evidence of the existence of black holes in the universe from different perspectives. On the one hand, we have the observation of the dynamics of stars near the center of our galaxy (Sgr A*) showing that the central object is undoubtedly a supermassive black hole~\cite{Ghez:1998ph, Ghez:2000ay, Ghez:2003rt, Ghez:2008ms}. On the other hand, we have the data from the gravitational wave observatories LIGO-VIRGO~\cite{LIGOScientific:2016aoc} and the observations from the Event Horizon Telescope (EHT) of the supermassive black holes M87*~\cite{EventHorizonTelescope:2019dse} and Sagittarius A*~\cite{EventHorizonTelescope:2022wkp}. In this sense, the possibility of testing the non-hair theorem/conjecture has increased in the last decades. In this sense, there has been relevant research work on potential tests of the no-hair theorem using gravitational waves~\cite{Ryan:1995wh, Ryan:1997kh, Ryan:1997hg, Datta:2019euh} or the electromagnetic spectrum~\cite{Johannsen:2010ru, Glampedakis:2023eek}. 

    In this work, we use the numerical black hole solution obtained by A.~Corichi et al. in \cite{Corichi:2005pa} to investigate the image cast by this scalar hairy black hole (SHBH) using a ray tracing code. We organize the paper as follows: in Sec.~\ref{SecII}, we discuss the most important aspects and considerations for obtaining the numerical black hole solution. Then, in Sec.~\ref{secIII}, we introduce the Hamiltonian formalism and obtain the equations of motion needed in the ray tracing code. In this section, we also define the effective potential for both photons and massive particles, from which we obtain the photon sphere radius ($r_\text{ps}$) and the innermost stable circular orbit (ISCO). In Sec.~\ref{secIV}, we focus on the images of SHBHs. We divide the section into two subsections, where we investigate the behavior of the angular diameter of the shadow, the specific intensity, the accretion disk, the photon rings, the redshift factor, and the energy flux, comparing our results with the Schwarzschild black hole. When analyzing the shadow and the photon rings, we consider a static, spherically symmetric, optically, and geometrically thin accretion flow, following Refs.~\cite{Zeng:2020dco, Jaroszynski:1997bw, Bambi:2013nla}. On the other hand, when considering the redshift, the energy flux, and the image of the SHBH, we assume a thin accretion disk on the equatorial plane using the same physical setup as Luminent~\cite{Luminet1979}. Finally, in Sec.~\ref{secV}, we discuss our results. In the manuscript, we use dimensionless coordinates, $x^\alpha$, with $\alpha$ running from 0 to 3 and $x^0$, $x^1$, $x^2$, and $x^3$, corresponding to $t$, $r$, $\theta$, and $\varphi$, respectively.
    
    

 \section{Hairy Black Holes \label{SecII}}		
    The action for Einstein-Klein-Gordon (EKG) system with an asymmetric potential $V(\phi)$ of a scalar field $\phi$ is given by~\cite{Corichi:2005pa, Chew:2022enh}
    \begin{equation} 
    \label{EHaction}
    S=  \int d^4 x \sqrt{-g}  \left[  \frac{R}{16 \pi G} - \frac{1}{2} \nabla_\mu \phi \nabla^\mu \phi - V(\phi) \right]  \,,
    \end{equation}
    where
    \begin{equation}
    \label{Sec_II_eq2}
    \begin{aligned}
    V(\phi)=& \frac{V_0\left( \phi - a \right)^2}{12}  \big[ 3 \left( \phi-a\right)^2 - 4 (\phi-a) (\phi_0 + \phi_1) \\
    &+ 6 \phi_0 \phi_1  \big] \,,
    \end{aligned}
    \end{equation}
    with $a$, $V_0$, $\phi_0$ and $\phi_1$ are constant parameters. At $\phi=a$, this potential has a local minimum equal to zero. Note that the asymptotic value of the scalar field $\phi$ at the infinity is given by $\phi=a$. $V(\phi)$ also possesses a local maximum at $\phi=a+\phi_0$ and a global minimum at $\phi=a+\phi_1$. Therefore, $0<2\phi_0<\phi_1$. The variation of the action \eqref{EHaction} with respect to the metric and scalar field yields the Einstein equation and the Klein-Gordon equation,
    \begin{equation}
    \label{einstein_KG}
    \begin{aligned} 
    \kappa T_{\mu \nu}&=R_{\mu \nu} - \frac{1}{2} g_{\mu \nu} R\,, \\
    \frac{d V}{d \phi}&=\nabla_\mu \nabla^\mu \phi\,,
    \end{aligned}
    \end{equation}
    where $\kappa=8\pi G$ and the stress-energy tensor, $T_{\mu\nu}$, is given by~\cite{Chew:2022enh} 
    \begin{equation}
     T_{\mu \nu} =-g_{\mu\nu}\left(\frac{1}{2}\nabla_\alpha\phi\nabla^\alpha\phi+V(\phi)\right)+\nabla_\mu\phi\nabla_\nu\phi\,.
    \end{equation}
    \begin{figure*}[t]
    \centering
    \mbox{
    (a)
    \includegraphics[scale=0.55]{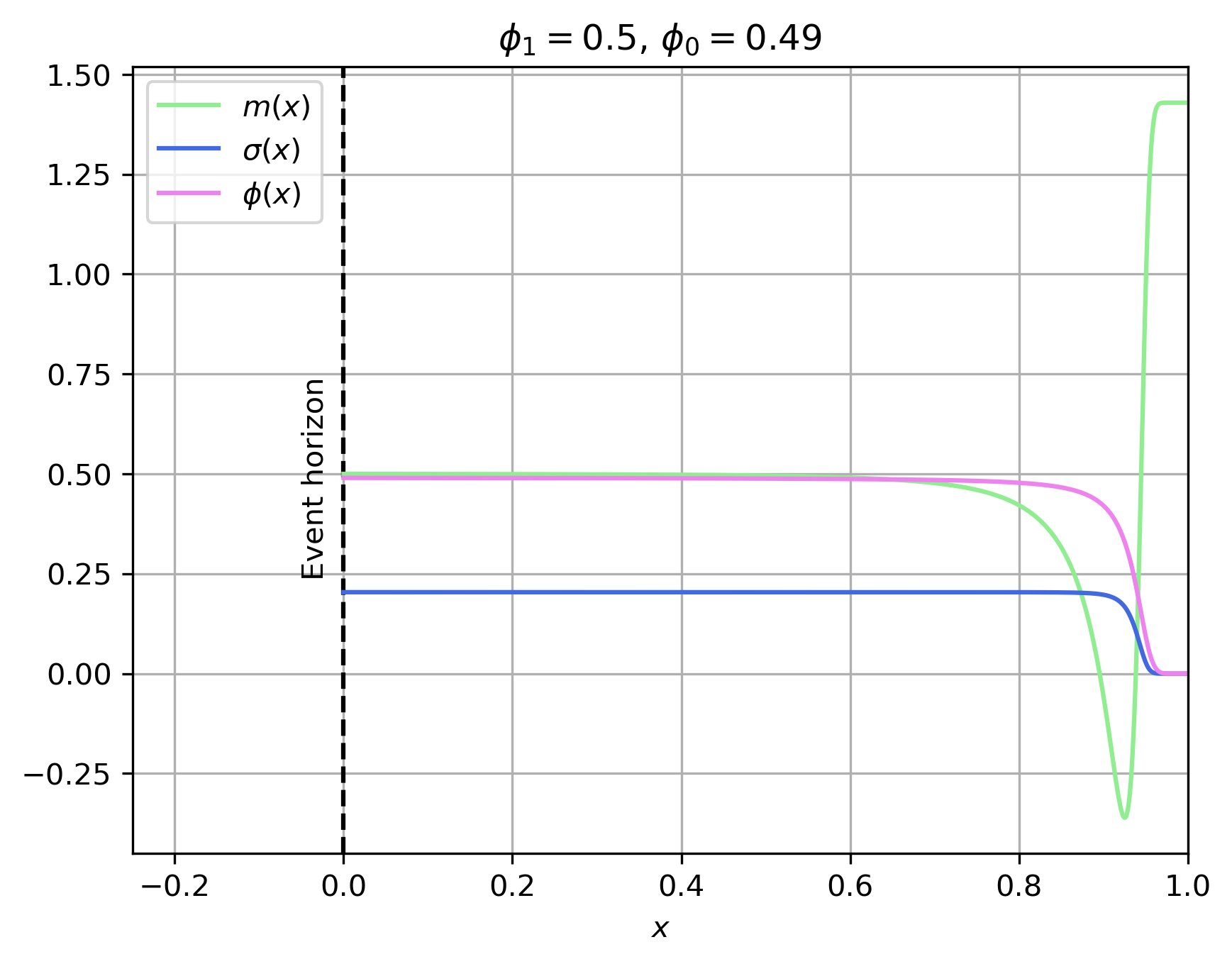}
    (b)
    \includegraphics[scale=0.55]{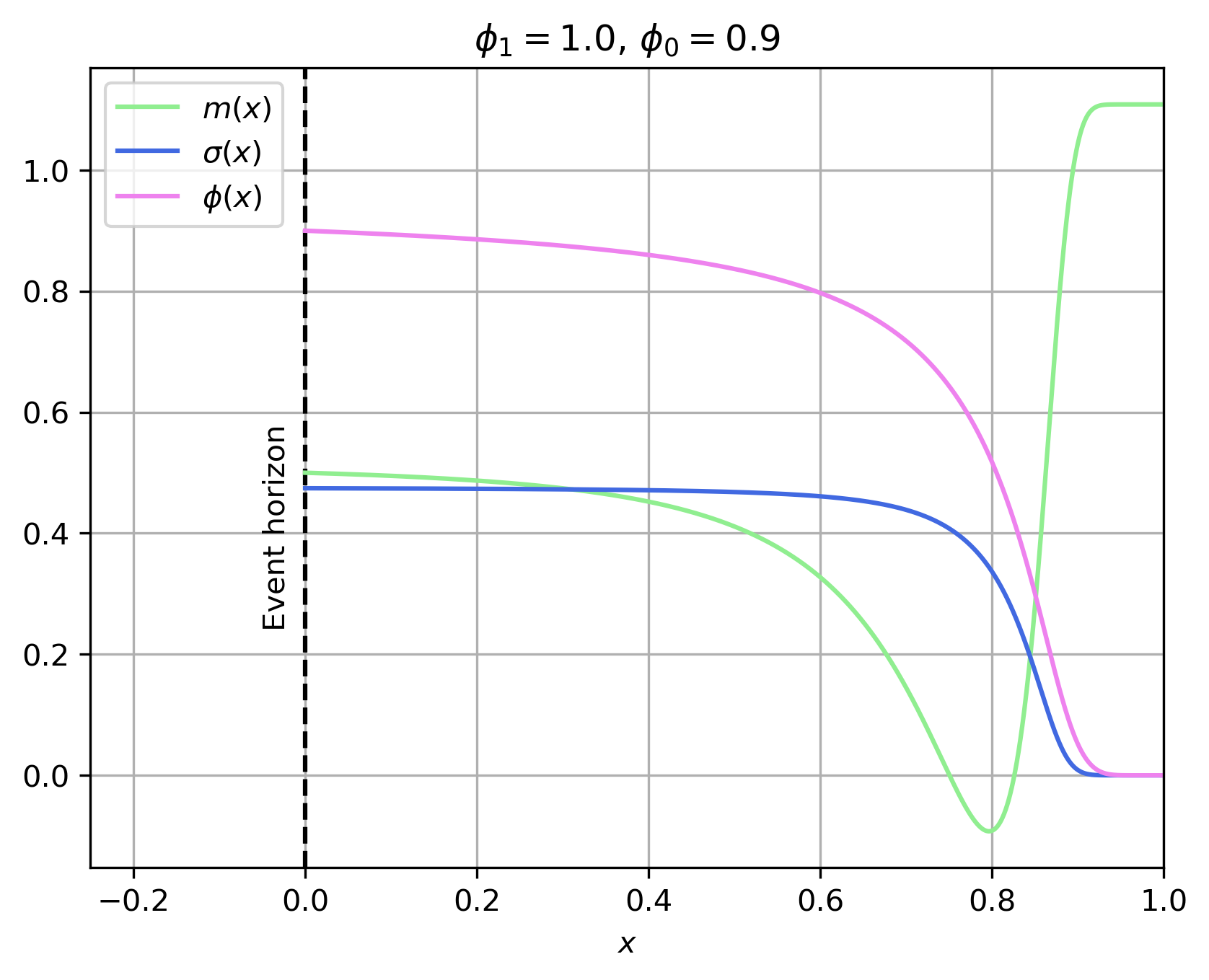}
    }
    \mbox{
    (c)
    \includegraphics[scale=0.55]{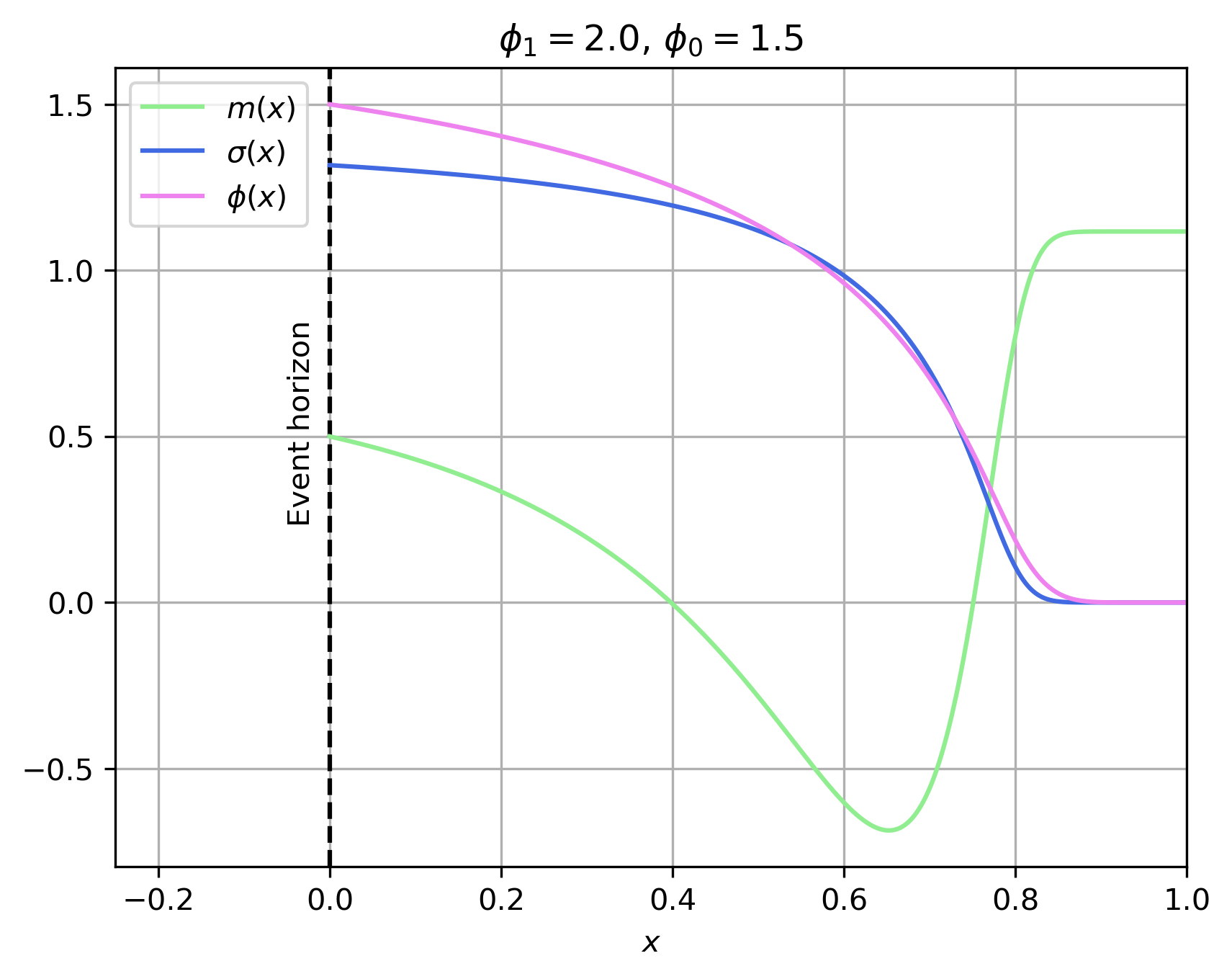}
    (d)
    \includegraphics[scale=0.55]{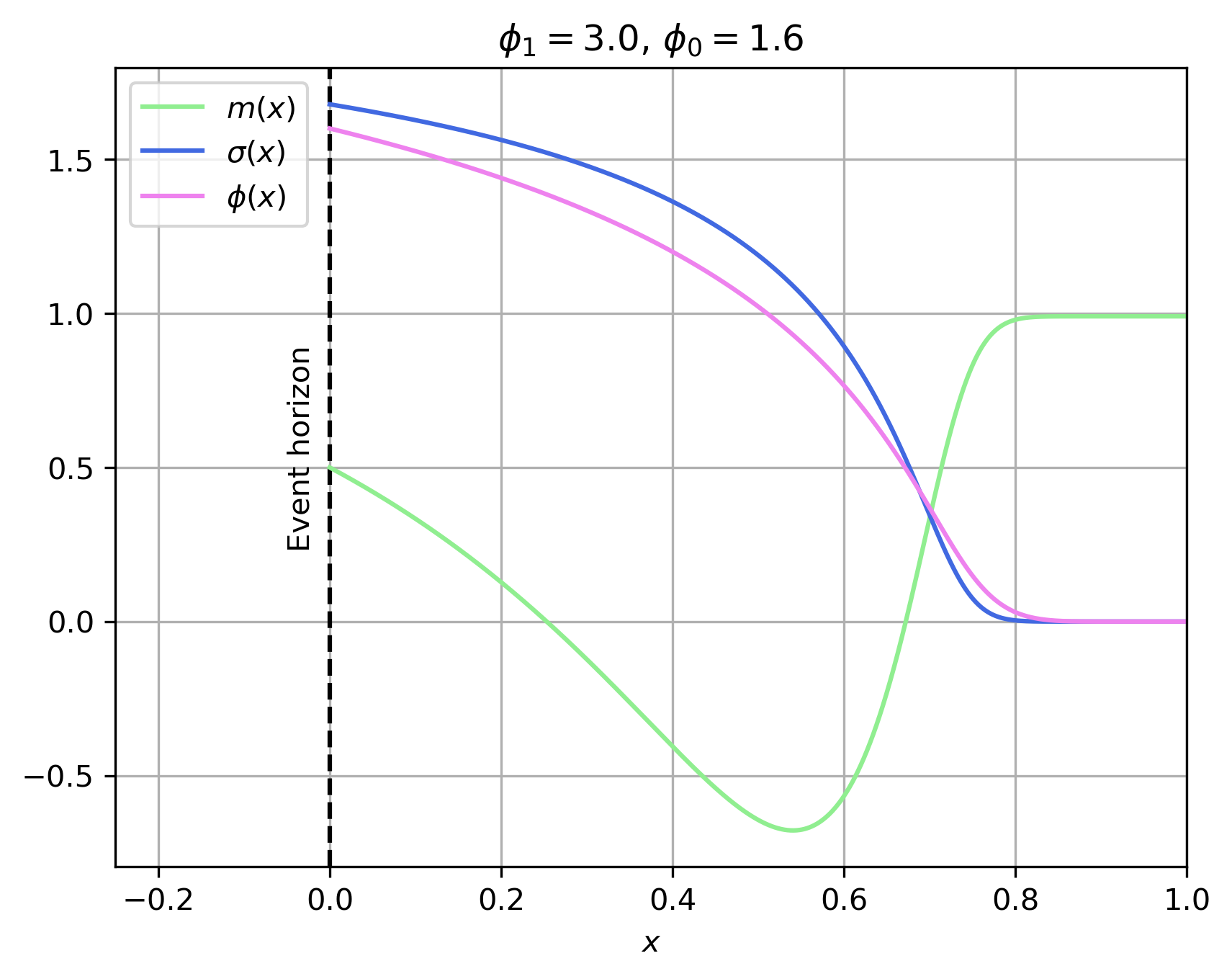}
    }
    \caption{Behavior of the functions $m(x)$, $\sigma (x)$ and $\phi (x)$ for different values of $\phi_1$ and $\phi_H$. Note that the event horizon $r_H$ is located at $x=0$, see equation \eqref{eq:randx}.}
    \label{Figure_1}
    \end{figure*}
Now, to obtain the SHBH solutions, we employ the static and spherically symmetric spacetime,
    \begin{equation}
    \label{eq:metric}
    ds^2 = g_{tt}dt^2+g_{rr}dr^2+r^2(d\theta^2+\sin^2\theta d\varphi^2),
    \end{equation}
where $g_{tt}$ and $g_{rr}$ are functions of the radial coordinate $r$ and given by the ansatz~\cite{Chew:2022enh}
    \begin{equation}
    \label{eq:LineElement}
    \begin{array}{ccc}
    g_{tt}=-N(r)e^{-2\sigma(r)}&\text{and}& g_{rr}=N^{-1}(r),
    \end{array}
    \end{equation}
where $N(r)=1-2 m(r)/r$ and $m(r)$ is the Misner-Sharp mass function. Note that this mass function gives the ADM mass of the black hole at infinity, i.e.  $m=M$ when $r\rightarrow\infty$. Hence, by substituting Eq.~\eqref{eq:LineElement} into the Einstein and the Klein-Gordon equations in~\eqref{einstein_KG}, one obtains a set of nonlinear ODEs for the metric functions~\cite{Chew:2022enh}
    \begin{align}
    \label{eq:dmdr}
    m' &= \frac{\kappa}{4} r^2 \left( N \phi'^2 + 2 V \right) \,, \\
    \sigma' &= - \frac{\kappa}{2} r \phi'^2 \,,  \label{eq:dsigmadr}  \\
    \phi'' &= -\frac{N - \kappa r^2 V +1}{r N} \phi' + \frac{V_0 \phi \left( \phi-\phi_0 \right) \left( \phi-\phi_1 \right)}{N} \,, \label{eq:dphidr}
    \end{align}
where the prime denotes the radial derivative of the functions. Therefore, the black hole solution is obtained by numerically solving the system~\eqref{eq:dmdr}-\eqref{eq:dphidr} with appropriate boundary conditions, see~\cite{Corichi:2005pa, Chew:2022enh} for details.

    In Fig.~\ref{Figure_1}, we show the behavior of the metric functions $m$, $\sigma$, and $\phi$. Note that we use the compactified radial coordinate, $x$, which is related to the radial coordinate, $r$, by the relation 
    \begin{equation} 
    \label{eq:randx}
    r = \frac{r_H}{1-x}, 
    \end{equation}
where $r_H$ is the radius of the horizon. Hence, Eq.~\eqref{eq:randx} allows us to see the behavior of the metric functions in the whole space-time; i.e. from $x=0$ ($r=r_H$) to $x=1$ ($r\rightarrow\infty$).

    \begin{figure*}
    \centering
    \mbox{
    (a)
    \includegraphics[scale=0.5]{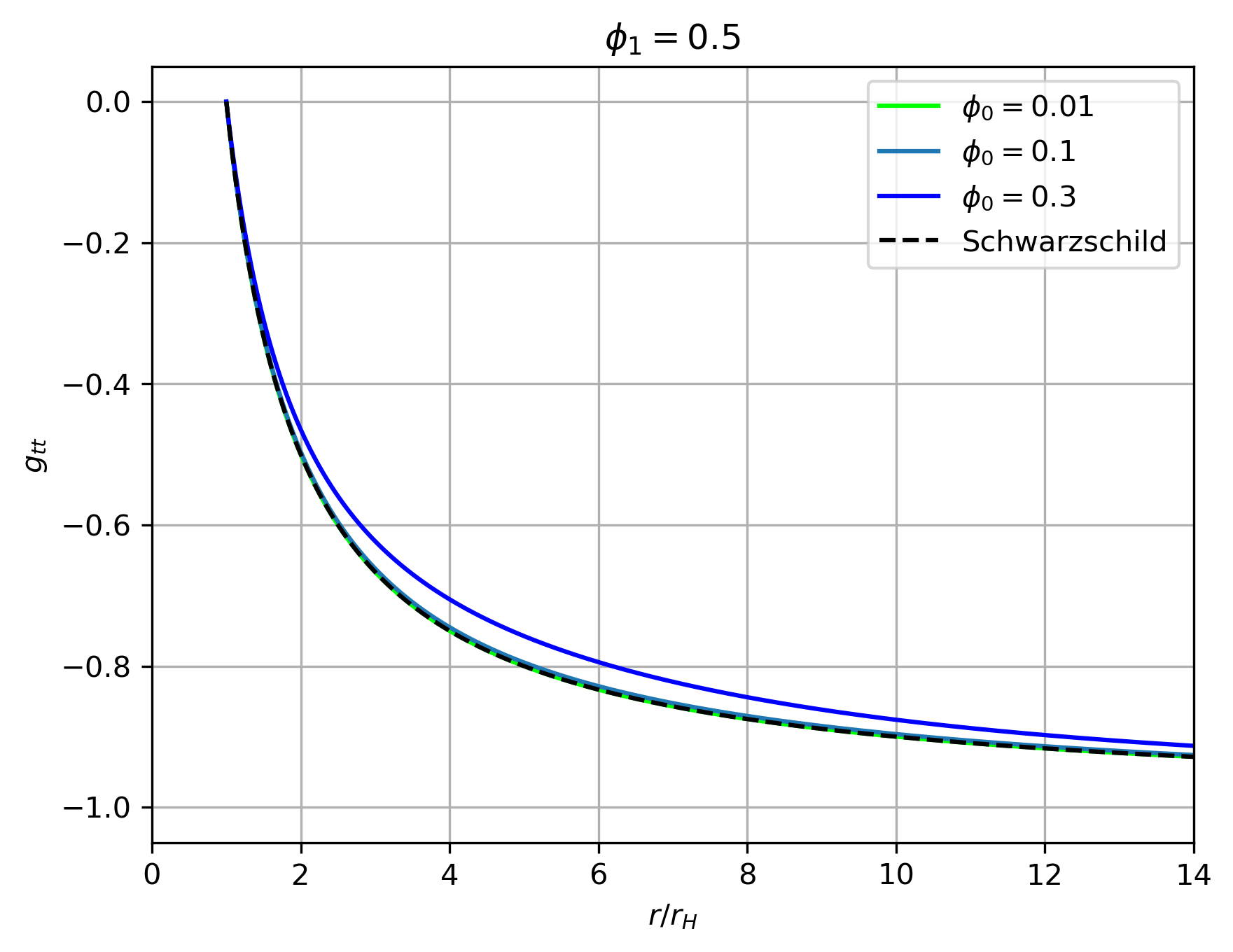}
    }
    \mbox{
    (b)
    \includegraphics[scale=0.5]{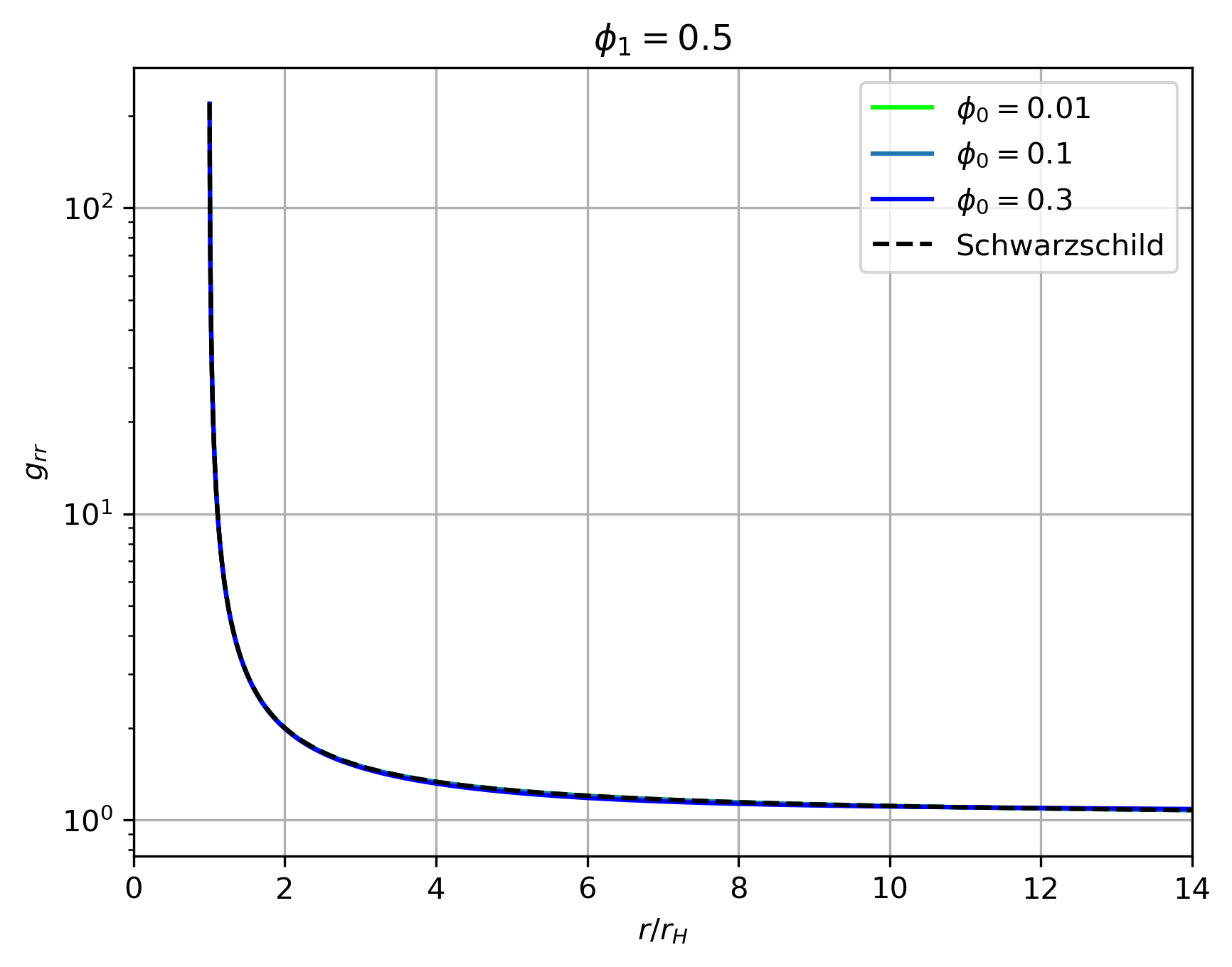}
    }
    \mbox{
    (c)
    \includegraphics[scale=0.5]{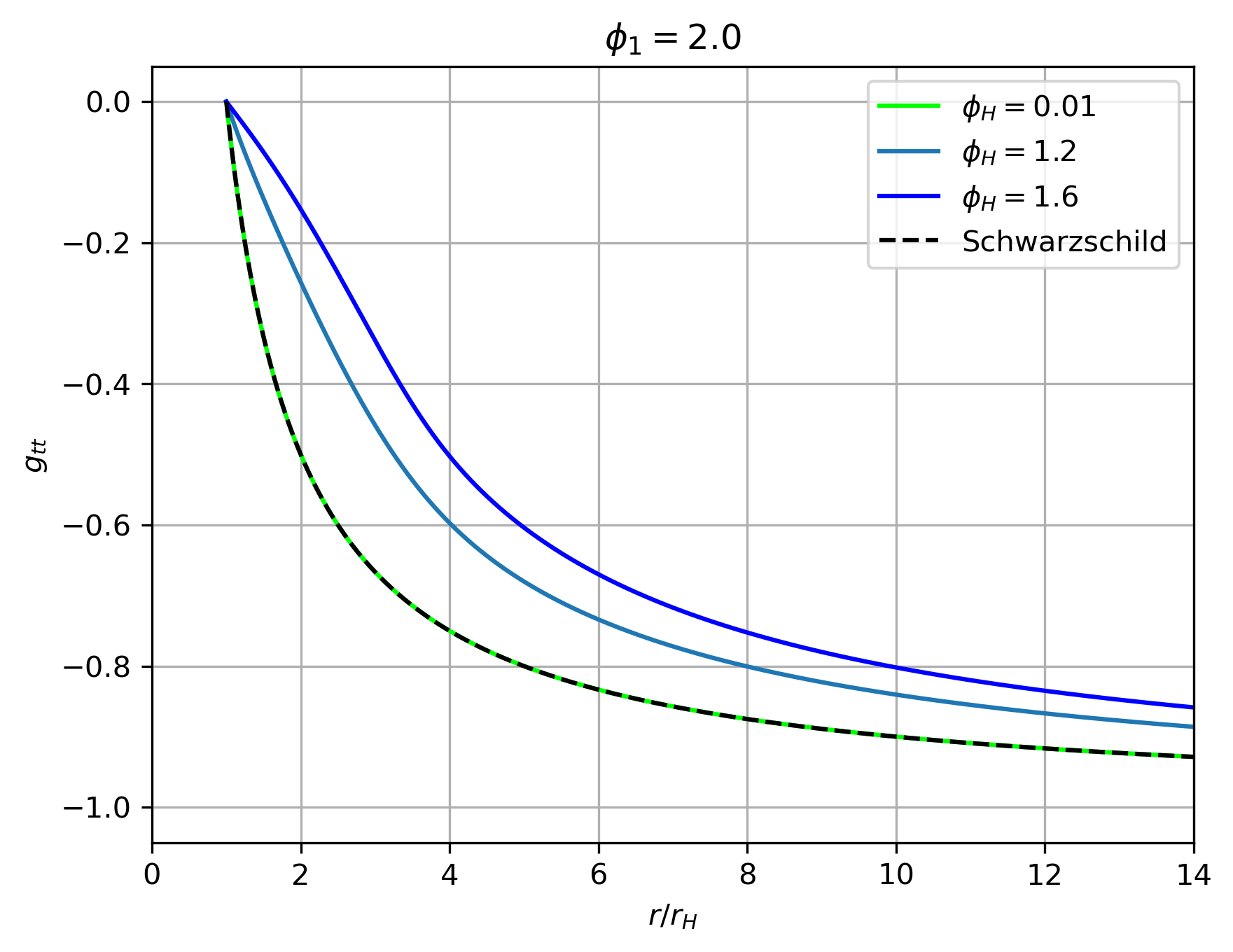}
    }
    \mbox{
    (d)
    \includegraphics[scale=0.5]{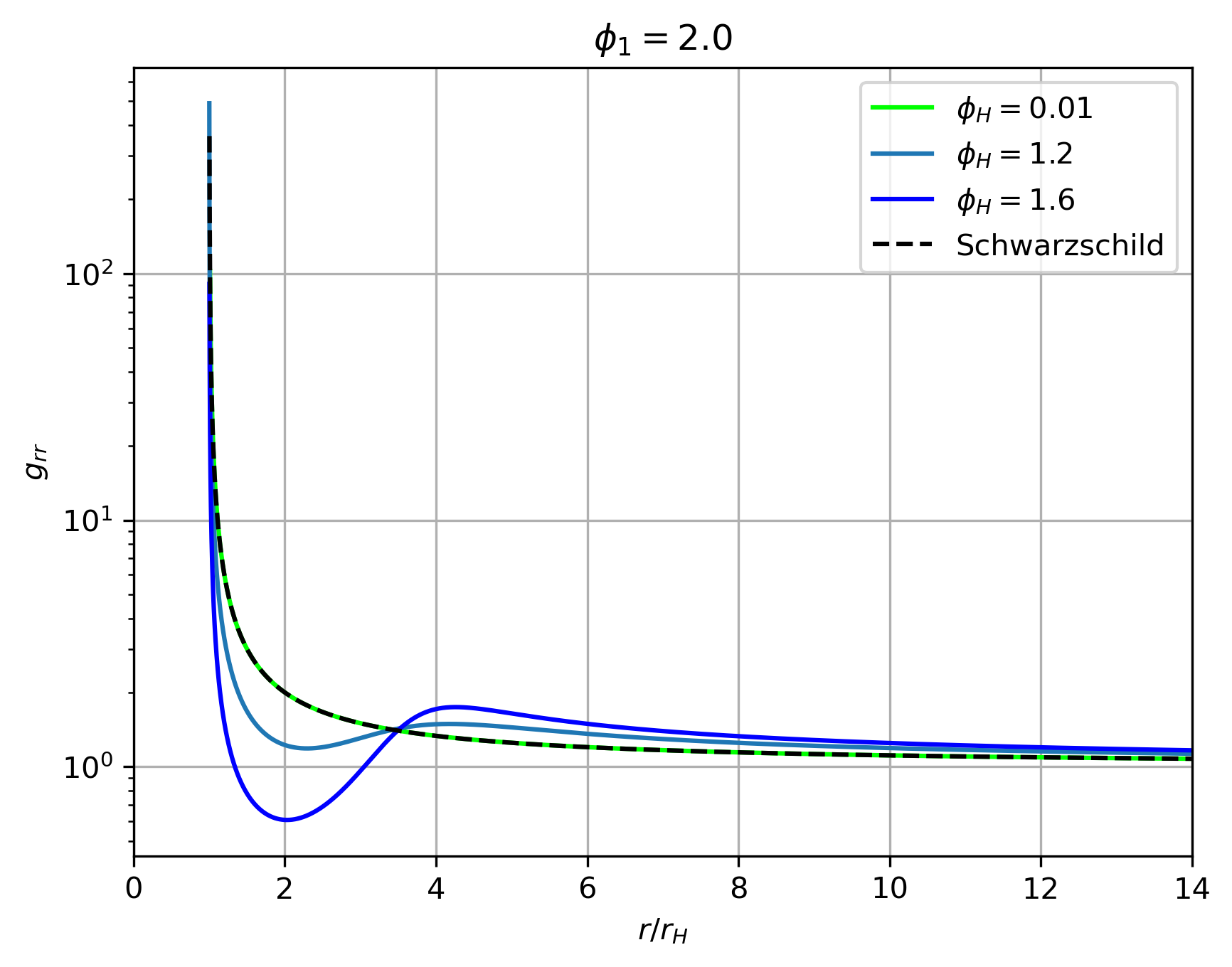}
    }
    \mbox{
    (e)
    \includegraphics[scale=0.5]{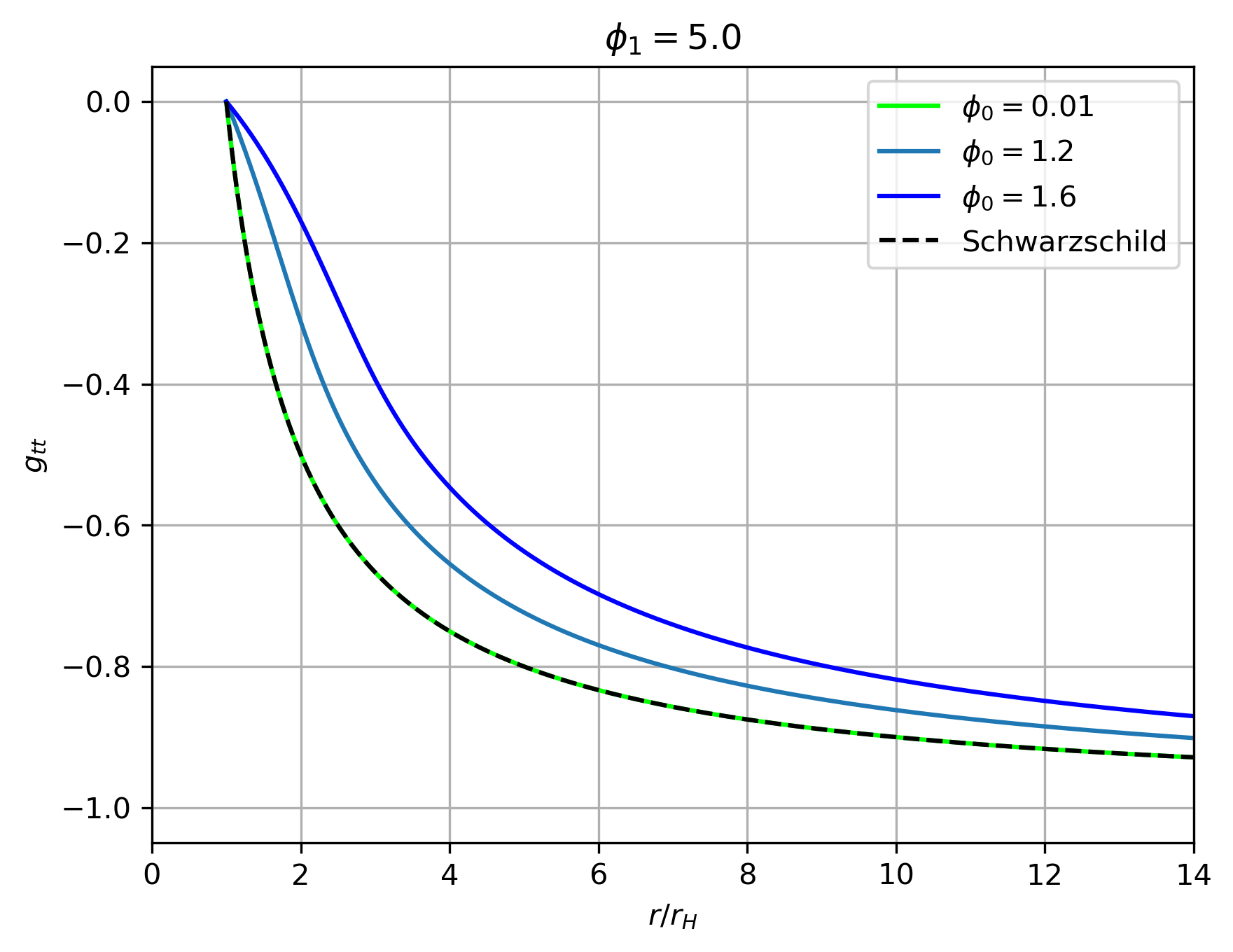}
    }
    \mbox{
    (f)
    \includegraphics[scale=0.5]{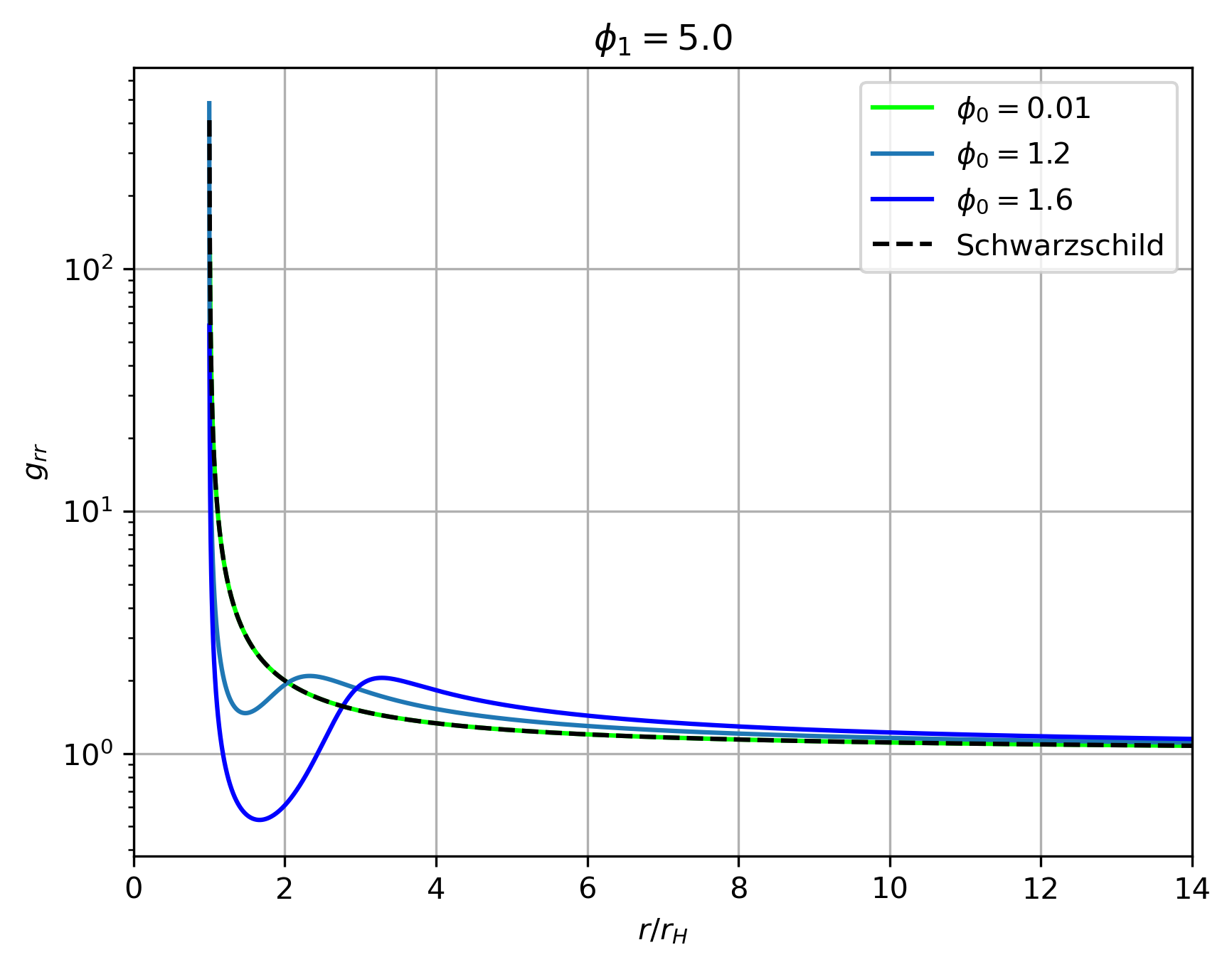}
    }
    \caption{Metric components $g_{tt}$ and $g_{rr}$ as function of the radial coordinate $r$ for a SHBH background with $\phi_1=0.5$, $2.0$ and $5.0$ and different values of $\phi_0$. The dashed-black curve corresponds to the Schwarzschild spacetime.}
    \label{fig:metric01}
    \end{figure*}
   For small values of $\phi_1$, the figure shows an interval where the metric functions have a constant value. See Fig.\ref{Figure_1}a when $\phi_1=0.5$. However, this behavior changes as the value of $\phi_1$ increases. For example, when $\phi_1=0.5$, the metric functions $m$, $\sigma$ and $\phi$ do not change their values from $x=0$ to $x\approx 0.7$, $x\approx 0.9$ and $x=0.8$, respectively, while for $\phi_1=1.0$, $2.0$, and $3.0$, the metric functions decrease as $x$ increases. Note that, in contrast to the metric functions $\sigma$ and $\phi$, where their values decrease to $\sigma=\phi=0$ for some value of $x$, the metric function $m$ reaches a minimum value and then increases up to a constant value $m=1$; the location of the minimum depending of $\phi_1$ and $\phi_0$.

    \begin{figure*}
    \centering
    \mbox{
    (a)
    \includegraphics[scale=0.5]{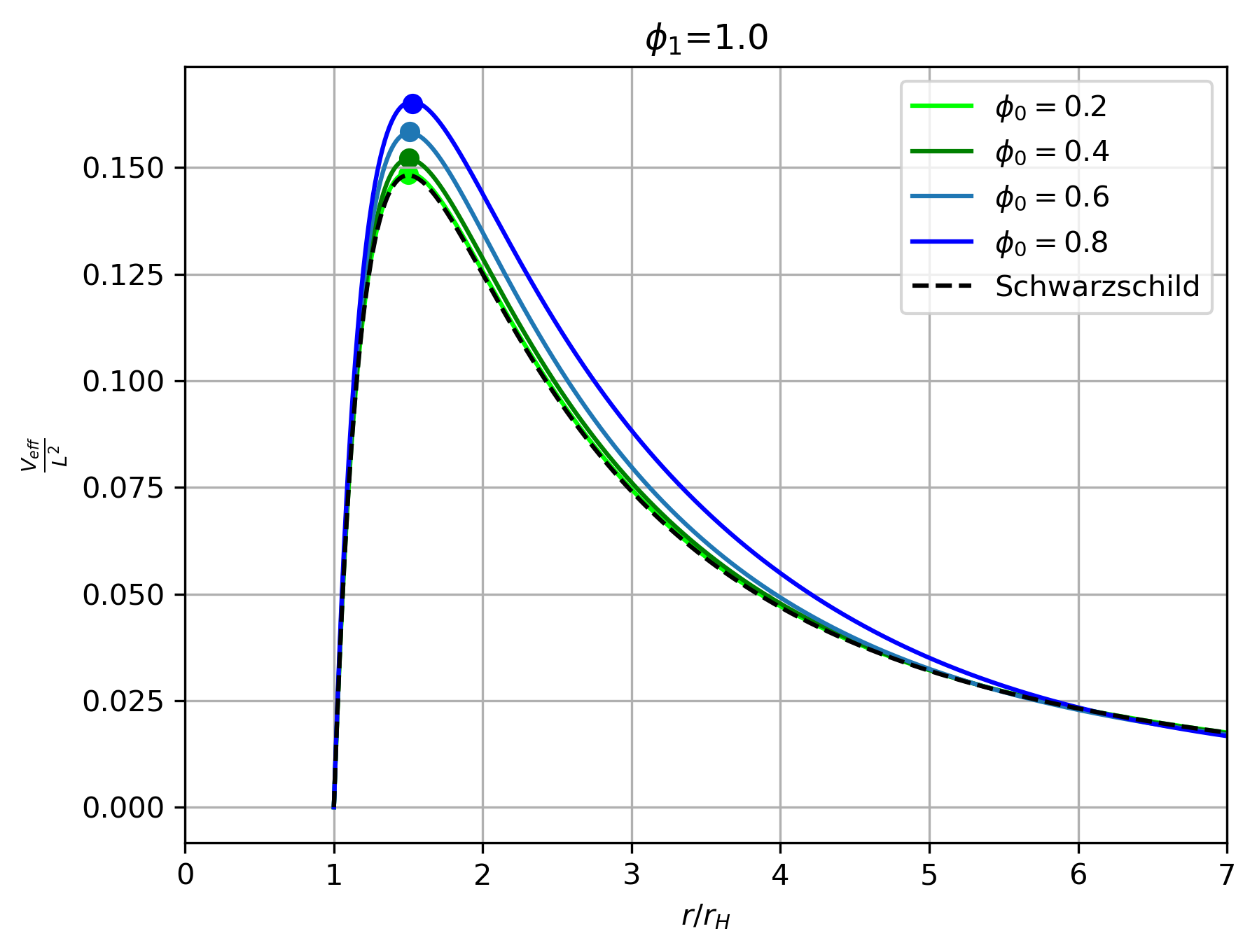}
    }
    \mbox{
    (b)
    \includegraphics[scale=0.5]{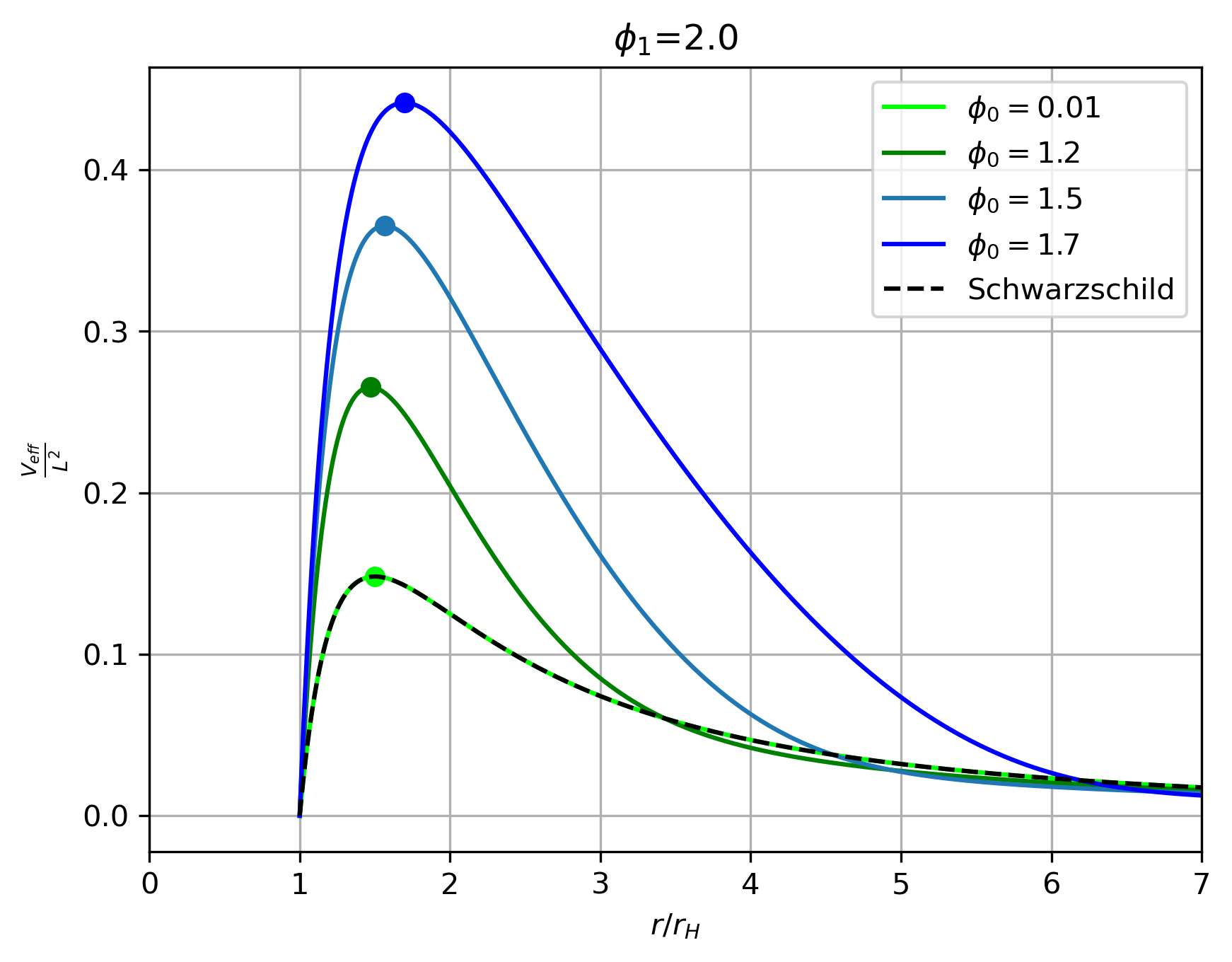}
    }
    \mbox{
    (c)
    \includegraphics[scale=0.5]{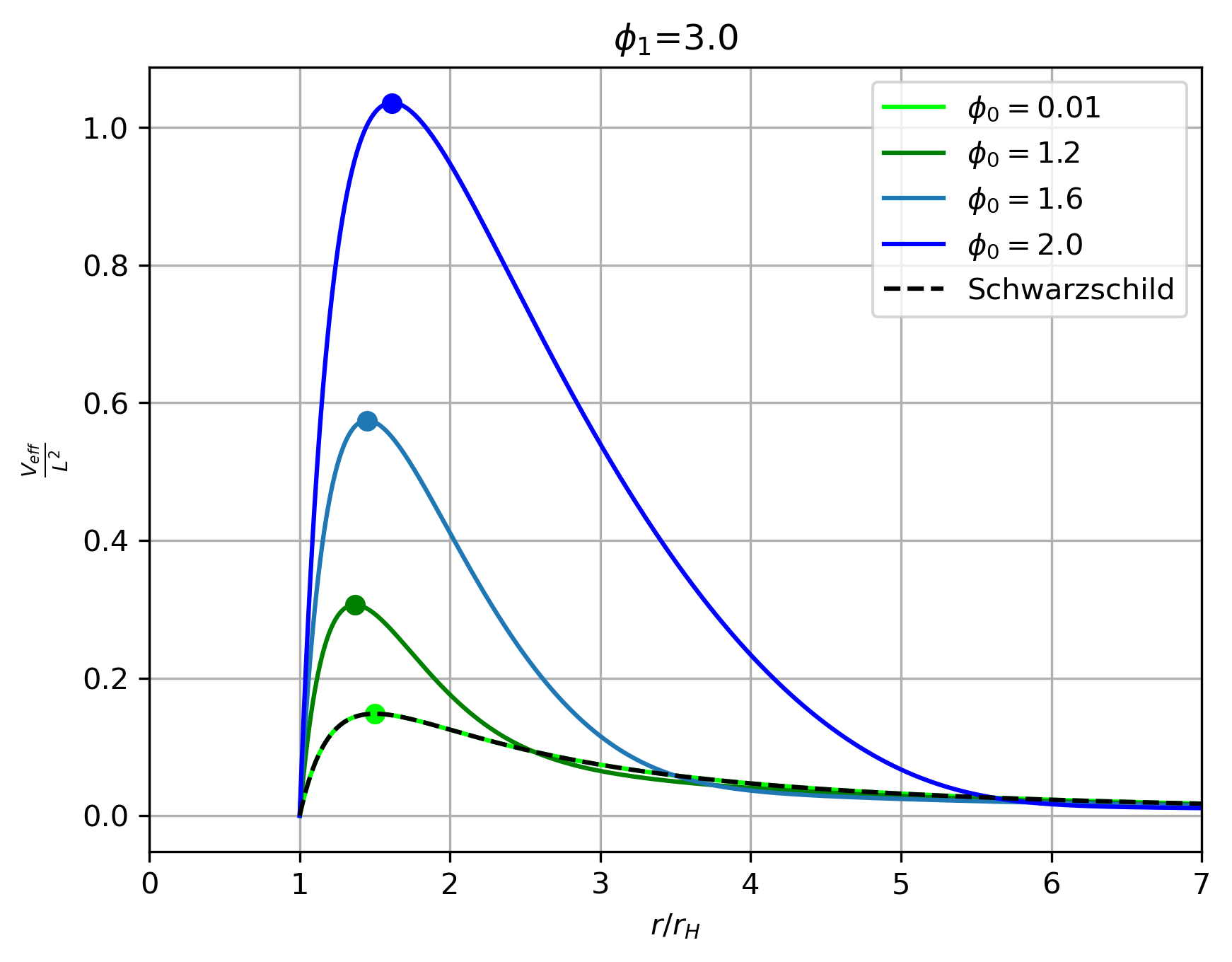}
    }
    \mbox{
    (d)
    \includegraphics[scale=0.5]{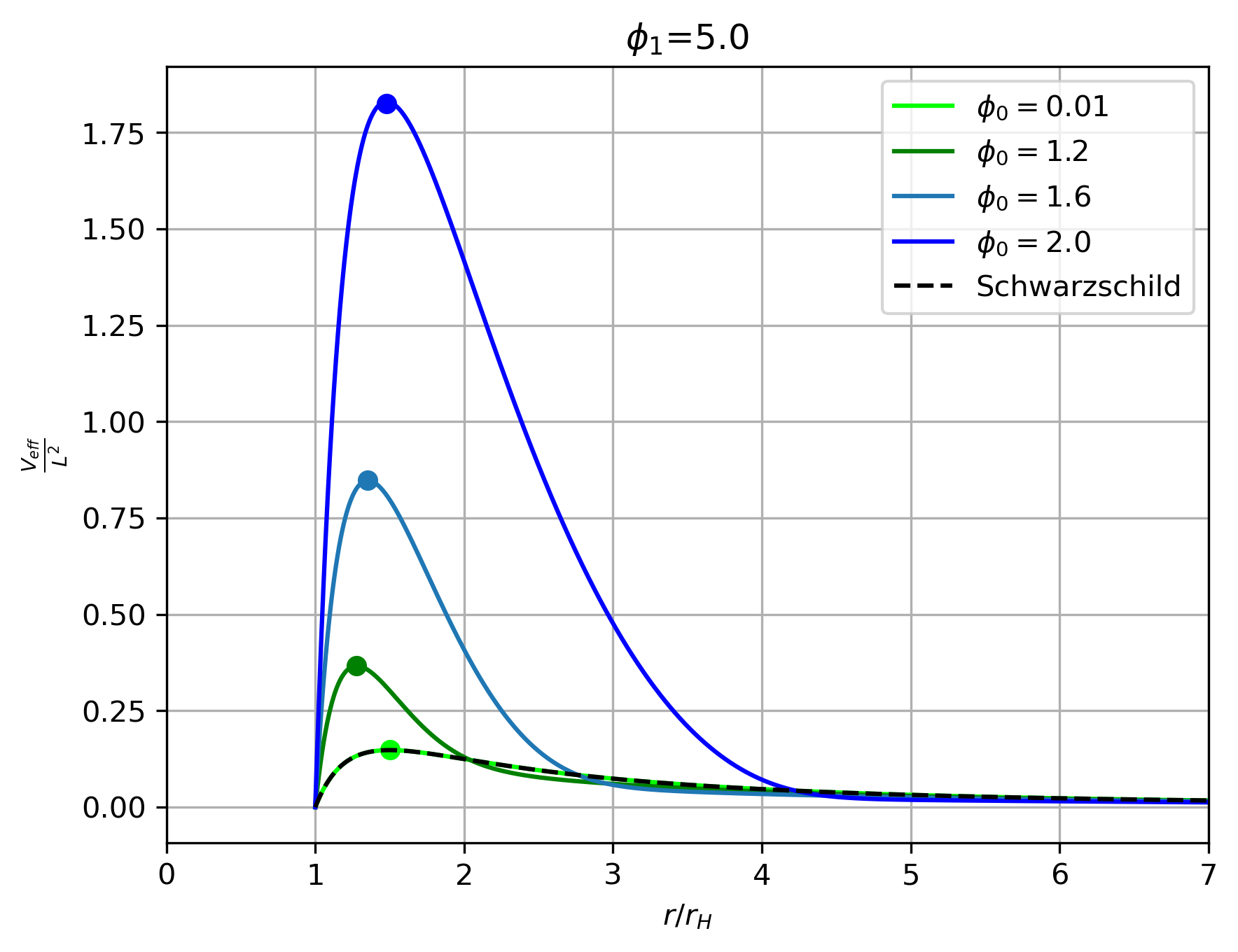}
    }
    \caption{Effective potential ($V_\text{eff}/L^2$) as a function of the radial coordinate $r$ for photons moving in the SHBH background with $\phi_1=1.0$, $2.0$, $3.0$ and $5.0$. The markers show the location of the photon sphere radius $r_\text{ps}/r_H$ and the dashed black curve corresponds to the effective potential for photons moving in the Schwarzschild spacetime.}
    \label{fig:effectivePotentialPhotons}
    \end{figure*}

    In figure \ref{fig:metric01}, we show the behavior of $g_{tt}$ (first column) and $g_{rr}$ (second column) as functions of the radial coordinate for different values of $\phi_1$ and $\phi_0$, contrasted with the Schwarzschild solution (dashed line). Note that for small values of $\phi_0$, both $g_{tt}$ and $g_{rr}$ are close to the Schwarzschild black hole solution. However, when we increase the value of $\phi_1$, some differences begin to appear. For example, the metric component $g_{tt}$ increases when we increase the value $\phi_1$. The increment is more visible near the black hole's center. Far from the black hole, $g_{tt}$ behaves similarly in all the cases, tending to $g_{tt}=-1$ when $r/r_H\rightarrow\infty$. In the case of $g_{rr}$, on the other hand, we also see some differences in the region near the black hole's center. Nevertheless, in contrast to the behavior of $g_{tt}$, $g_{rr}$ decreases to a minimum value. Then, it increases to a maximum and decreases again, tending to $g_{rr}=1$ when $r/r_H\rightarrow\infty$; see figure \ref{fig:metric01} panels $(\text{d})$ and $(\text{f})$.

    Different properties of the SHBH solutions, such as Hawking temperature $T_H$, the area of the horizon $A_H$, the Ricci and the Kretschmann scalars, $R$ and $K$, respectively, are investigated in detail in~\cite{Chew:2022enh}.  
    
	Finally, it is important to mention that the space-time studied in this paper corresponds to an example of an asymptotically flat black hole with scalar hair that is regular on and outside the event horizon~\cite{Herdeiro:2015waa}. The main idea of this solution, inspired by the isolated horizons (IH) formalism, is to consider space-times with an interior boundary (the horizon), satisfying quasi-local boundary conditions and ensuring that the horizon remains ``isolated''. While the boundary conditions are rooted in geometric considerations, they give rise to a well-defined action principle and Hamiltonian framework. Furthermore,  these boundary conditions suggest that specific ``quasi-local charges'' determined at the horizon remain constant ``over time''; therefore, one can interpret these quasi-local charges as similar to the global charges defined at infinity in the context of asymptotically flat space-times~\cite{Corichi:2005pa}.

	It is well-known that Bekenstein developed a method for proving the inexistence of scalar hair~\cite{Bekenstein:1972ny, Bekenstein:1971hc, Bekenstein:1972ky}, which contains three key assumptions: 1) the type of scalar field equation; 2) the space-time symmetry inheritance by the scalar field and 3) an energy condition~\cite{Herdeiro:2015waa}. Although violating one of the assumptions does not guarantee the existence of a regular BH with scalar hair, one can find some explicit solutions in Ref.~\cite{Herdeiro:2015waa}. In particular, the solution proposed by Corichi et al. violates the third assumption, i.e.,  the weak energy condition (WEC). From the physical point of view, the violation of the WEC means that the asymmetric potential is not positive-define, ``\textit{providing an extra repulsive interaction and, therefore, preventing the scalar field from infinitely piling up at the (would be) horizon when one requires $\phi$ to be non-trivial}~\cite{Herdeiro:2015waa}. In this sense, the violation of the WEC is necessary in the horizon to ensure a regular solution on and outside it.

	On the other hand, as pointed out by C.~Hereido~\cite{Herdeiro:2015waa}, although there is not yet a clear physical setting for the WEC violation on the horizon, at least in the asymptotically flat case, the assumption of it leads to BHs with scalar hair possessing distinct properties. Hence, from the astrophysical perspective, it is worth exploring and investigating these solutions from the observational point of view. In particular, as we show in this work, we explore the optical appearance of a numerical SHBH solution, investigate the shadow and the profile of the specific intensity, and use the EHT observation to constrain the parameter $\phi_0$. 
    

\section{The Hamiltonian formalism and the Equation of motion\label{secIII}}

    To generate and investigate the image of scalar hairy black holes (SHBHs), we need to describe the motion of massive and massless particles in this spacetime. Usually, one models the particle motion by a system of differential equations obtained using the Hamiltonian Formalism (HF) because, according to Ref.~\cite{Levin:2008mq}, numerical difficulties such as error propagation when solving the equations of motion, can be avoided naturally in this formalism. 
    
    We begin by considering a general static and spherically symmetric spacetime, given by the line element
    \begin{equation}
    ds^2= g_{tt}dt^2+g_{rr}dr^2+g_{\theta\theta}d\theta^2+g_{\varphi\varphi}d\varphi^2,\label{s3e1}
    \end{equation}
where we assume that the metric components only depend on the coordinates $r$ and $\theta$. The Lagrangian describing the dynamics of a test particle moving in this spacetime is given by
    \begin{equation}
    \mathcal{L} = \frac{1}{2} g_{\mu\nu}\dot{x}^\mu \dot{x}^\nu = -\frac{\epsilon}{2}, \label{eq:Lagrangian}
    \end{equation}
where $\epsilon=0$ and $\epsilon=1$ refer to massless and massive particles, respectively. In this expression, we use a dot to denote the derivative with respect to the affine parameter $\lambda$. The canonical 4-momentum is defined as
    \begin{equation}
    \label{s3e3}
    p_\alpha\equiv\frac{\partial \mathcal{L}}{\partial{\dot{x}^\alpha}}=g_{\alpha\beta}\dot{x}^\beta=g_{\alpha\beta}p^\beta.
    \end{equation}
    
    For the spacetime in Eq.~\eqref{s3e1}, the components of the canonical momentum are
    \begin{equation}
    \label{s3e4}
    \begin{aligned}
    p_t&=\frac{\partial \mathcal{L}}{\partial{\dot{t}}}=g_{tt}\dot{t}\\
    p_r&=\frac{\partial \mathcal{L}}{\partial{\dot{r}}}=g_{rr}\dot{r}\\
    p_\theta&=\frac{\partial \mathcal{L}}{\partial{\dot{\theta}}}=g_{\theta\theta}\dot{\theta}\\
    p_\varphi&=\frac{\partial \mathcal{L}}{\partial{\dot{\varphi}}}=g_{\varphi\varphi}\dot{\varphi}.
    \end{aligned}
    \end{equation}

    It is straightforward to show that the Hamiltonian and the Lagrangian are equal, 
    \begin{equation}
    \label{s3e5}
    \mathcal{H}=p_\mu\dot{x}^\mu-\mathcal{L}=\frac{1}{2}g^{\alpha\beta}p_\alpha p_\beta=\mathcal{L},
    \end{equation}
and constant~\cite{Chandrasekhar:1998}. Hence, the equations of motion for the coordinates $x^\mu$ and the momenta $p_\mu$, in the Hamiltonian formalism, are given by the relations
   \begin{equation}
   \label{s3e6}
   \begin{array}{ccc}
   \dot{x}^\mu=\frac{\partial \mathcal{H}}{\partial p_\mu}& \text{ and }&
   \dot{p}_\mu=-\frac{\partial \mathcal{H}}{\partial x^\mu}.
   \end{array}
   \end{equation}
For the general line element in Eq.~\eqref{s3e1}, we obtain the following system of eight differential equations:
    \begin{equation}
    \label{s3e7}
    \begin{aligned}
    \dot{t}=&g^{tt}p_t\\
    \dot{r}=&g^{rr}p_r\\
    \dot{\theta}=&g^{\theta\theta}p_\theta\\
    \dot{\varphi}=&g^{\varphi\varphi}p_\varphi \\
    \dot{p}_t=&0\\
    \dot{p}_r=&-\frac{1}{2}\big(p^2_t\partial_r g^{tt}+p^2_r\partial_rg^{rr}+p^2_\theta\partial_rg^{\theta\theta}+p^2_\varphi \partial_rg^{\varphi\varphi}. \big)\\
    \dot{p}_\theta=&-\frac{1}{2}\big(p^2_t\partial_\theta g^{tt}+p^2_r\partial_\theta g^{rr}+p^2_\theta\partial_\theta g^{\theta\theta}+p^2_\varphi \partial_\theta g^{\varphi\varphi}\big)\\
    \dot{p}_\varphi=&0.\\
    \end{aligned}
    \end{equation}
Note that $p_t$ and $p_\varphi$ are constants of motion associated with the energy, $E$, and the angular momentum, $L$, respectively. Therefore, there are two Killing vector fields, $\xi=\partial_t$ and $\zeta=\partial_\varphi$ from which 
     \begin{equation}
     \begin{aligned}
     -E = p_t &=g_{tt}\dot{t}\\
      L =p_\varphi &=g_{\varphi\varphi}\dot{\varphi},\label{eq:ConservedQuantities}
     \end{aligned}
     \end{equation}
    and, $\dot{t}$ and $\dot{\varphi}$ can be written in the form
    \begin{equation}
    \label{s3e9}
    \begin{aligned}
    \dot{t}&=-g^{tt}E\\
    \dot{\varphi}&=g^{\varphi\varphi}L.
    \end{aligned}
    \end{equation}
In the case of SHBH given by Eq.~\eqref{eq:LineElement} and restricting the analysis to the equatorial plane, $\theta=\pi/2$, so that $\dot{\theta}$ and $\dot{p}_\theta$ vanish, the system \eqref{s3e7} reduces to 
    \begin{equation}
    \label{s3e10}
    \begin{aligned}
    \dot{t}=&e^{2 \sigma}N^{-1}E\\
    \dot{r}=&Np_r\\
    \dot{\varphi}=&\frac{L}{r^2} \\
    \dot{p}_r=&\frac{L^2}{r^3}-\frac{1}{2} p^2_r N'-\frac{E^2 e^{2 \sigma} \left[N'-2 N \sigma'\right]}{2 N^2}
    \end{aligned}
    \end{equation}
while the conserved quantities become 
    \begin{equation}
     E = e^{-2 \sigma} N \dot{t} \,, \quad  L =   r^2 \dot{\varphi} \,. 
    \end{equation}

\subsection{The effective potential}

    Using the Lagrangian (\ref{eq:Lagrangian}) and the fact that our analysis is restricted to the equatorial plane, we can define the effective potential, $V_\text{eff}(r)$, from the radial equation of motion
    \begin{equation}
    \dot{r}^2 = e^{2\sigma}E^2 - V_{\text{eff}} (r)  \,,\label{eq:radialEoM} 
    \end{equation}
    with 
    \begin{equation}
    V_{\text{eff}} (r) = N(r) \left(\frac{L^2}{r^2} + \epsilon \right) \,. \label{eq:effectivePotential}
    \end{equation}

\subsubsection{The Photon Sphere}
 
    In Fig.~\ref{fig:effectivePotentialPhotons}, we show the behavior of the effective potential (\ref{eq:effectivePotential}) for photons ($\epsilon = 0$) as a function of the radial coordinate $r/r_H$. Note that the quantity $V_\text{eff}/L^2$ vanishes at the horizon; then, it reaches its maximum value at the unstable circular orbit called the \textit{photon sphere} (which has a radius $r_\text{ps}/r_H$) and later the potential decreases, vanishing again asymptotically. This behavior is similar for all the values of $\phi_1$ and $\phi_0$; nevertheless, when increasing the parameter $\phi_0$, the maximum value of the effective potential increases (clearly seen when $\phi_1 = 1.0$, $2.0$, and $5.0$). Furthermore, note that the maximum changes its position for different values of $\phi_0$, shifting to the left for small values and to the right for greater values of $\phi_0$. The figure also shows that $V_\text{eff}/L^2$ behaves quite close to that of the Schwarzschild BH solution for small values of $\phi_1$ and the differences begin to surface for greater values of $\phi_1$.

    \begin{figure}
    \centering
    \includegraphics[width=0.95\linewidth]
    {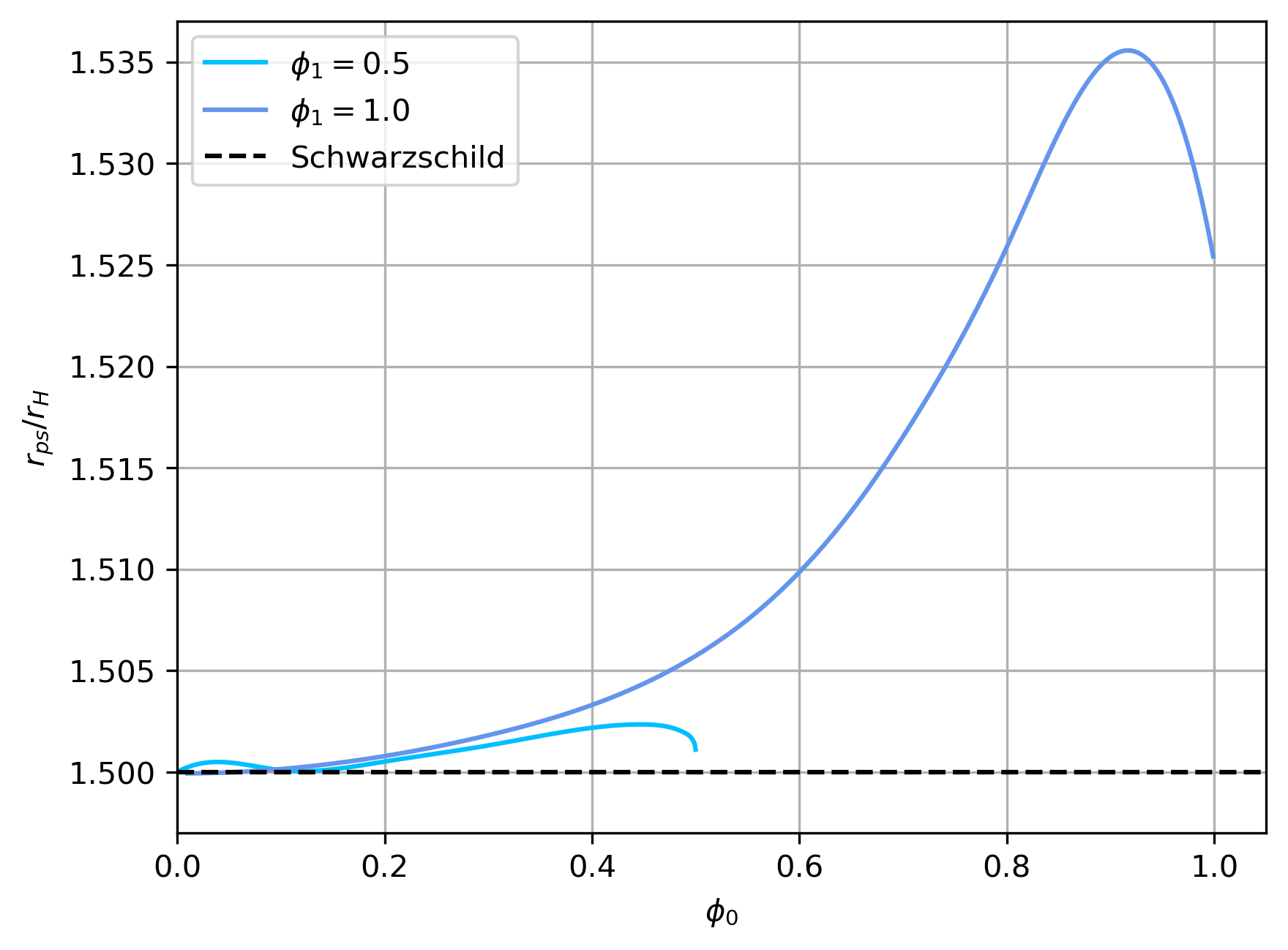}
    \includegraphics[width=0.95\linewidth]
    {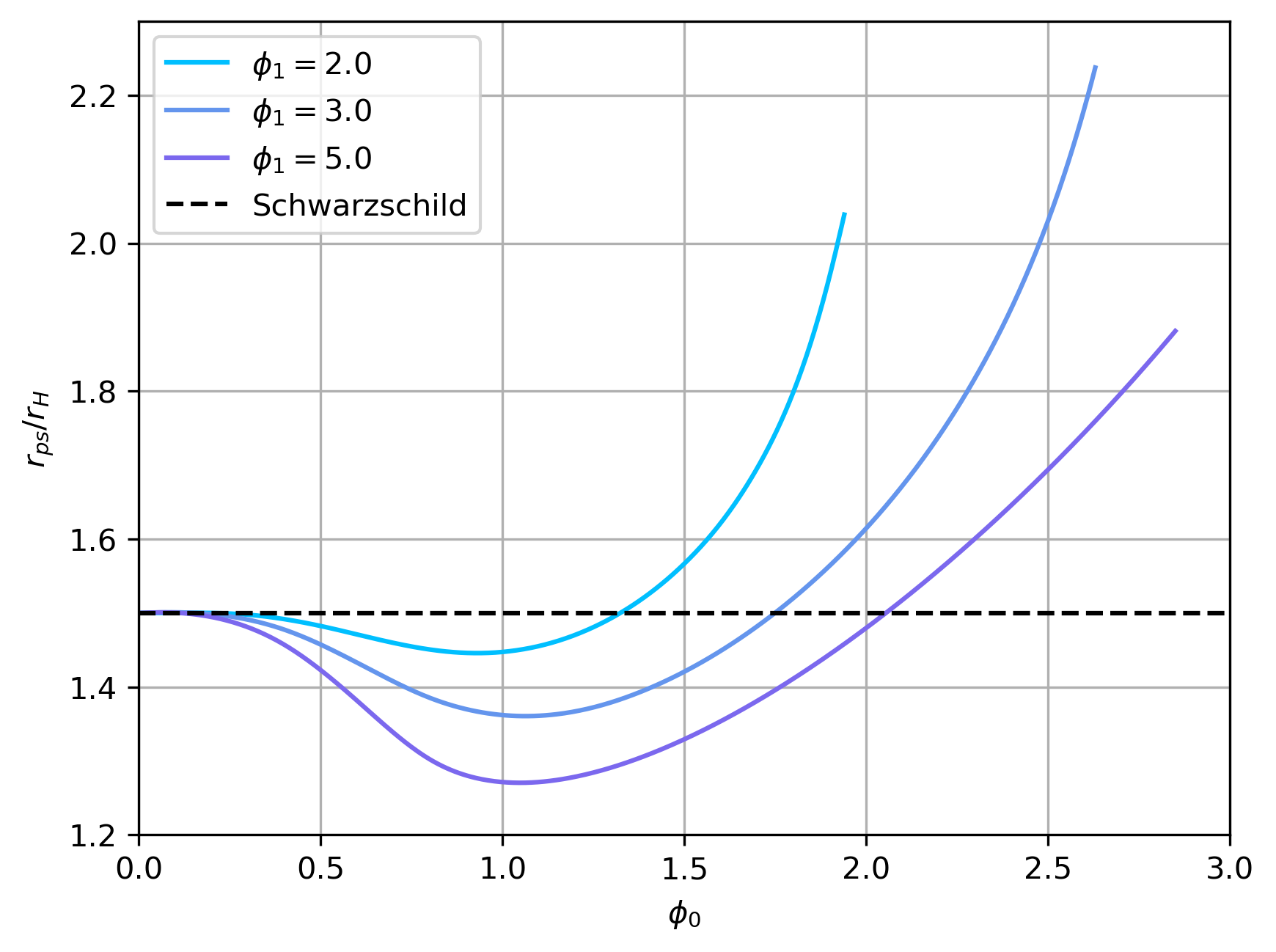}
    \caption{Radius of the photon sphere $r_{\text{ps}}/r_H$ depending on the parameter $\phi_0$ for SHBHs with the values $\phi_1= 0.5$, $1.0$, $2.0$, $3.0$ and $5.0$. The dashed black line represents the radius of the photon sphere around a Schwarzschild black hole.}
    \label{fig:plot_rph}
    \end{figure}


    \begin{figure*}[t]
    \centering
    \mbox{
    (a)
    \includegraphics[scale=0.45]{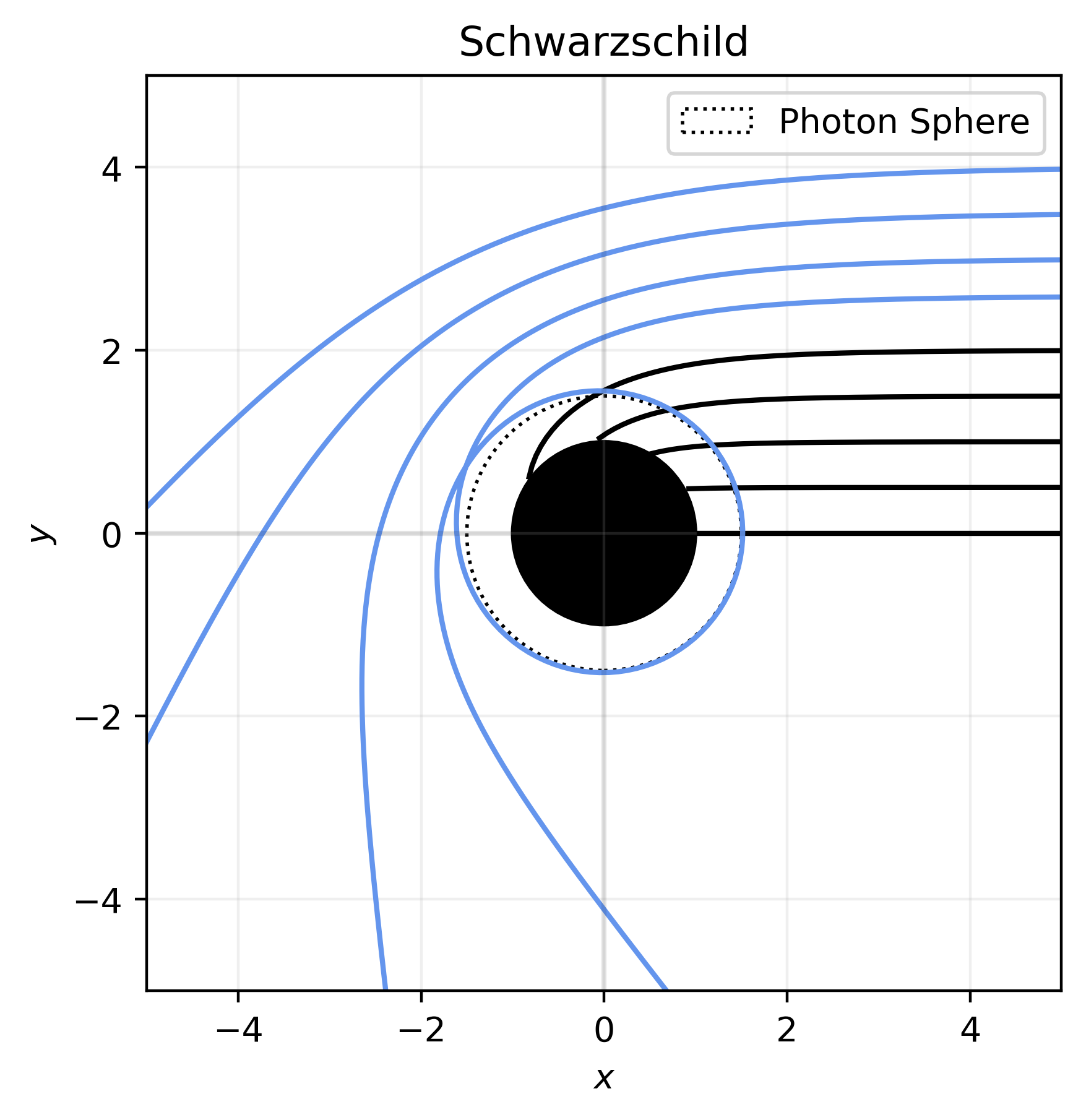}
    (b)    
    \includegraphics[scale=0.45]{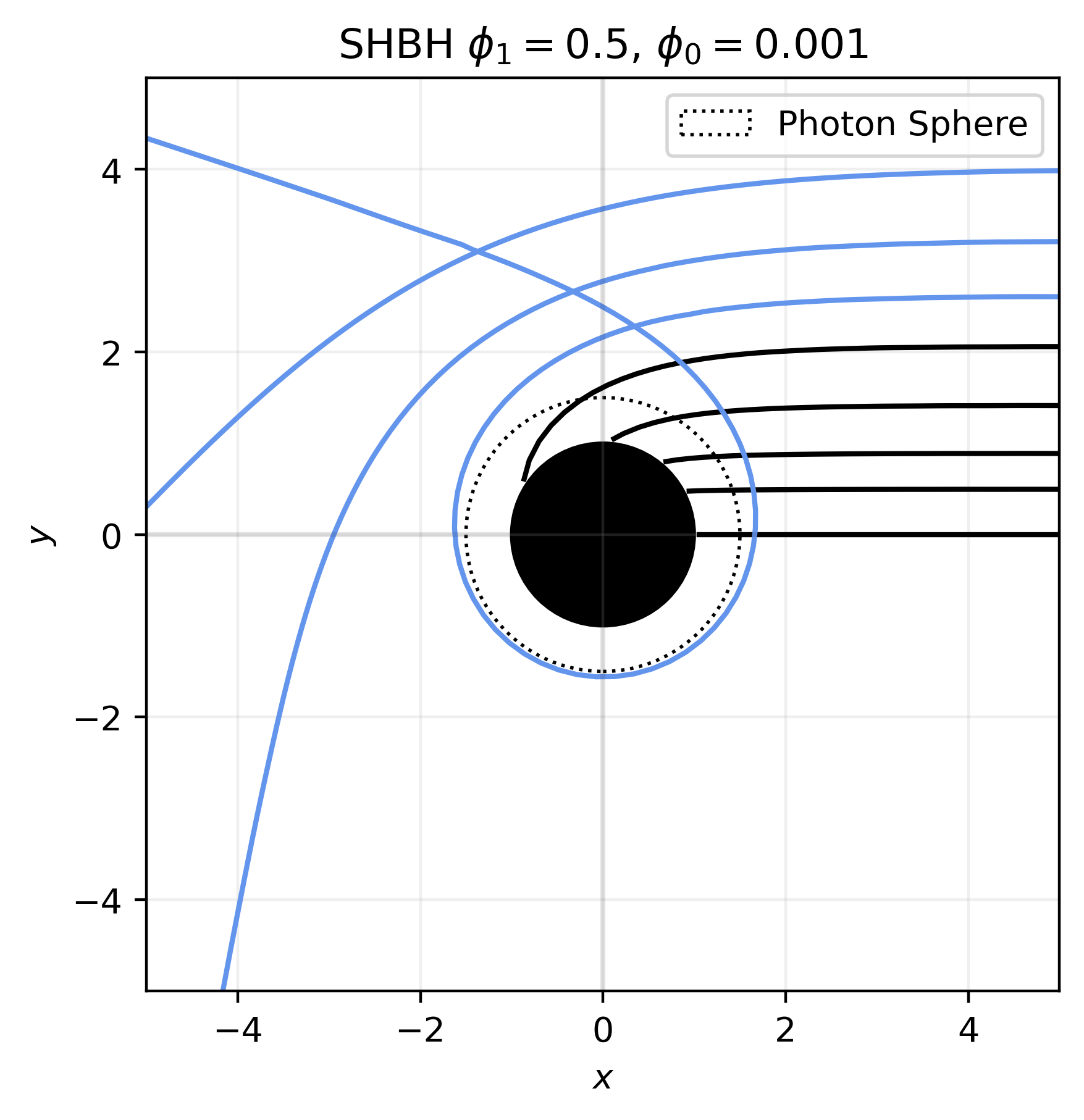}
    (c)
    \includegraphics[scale=0.45]{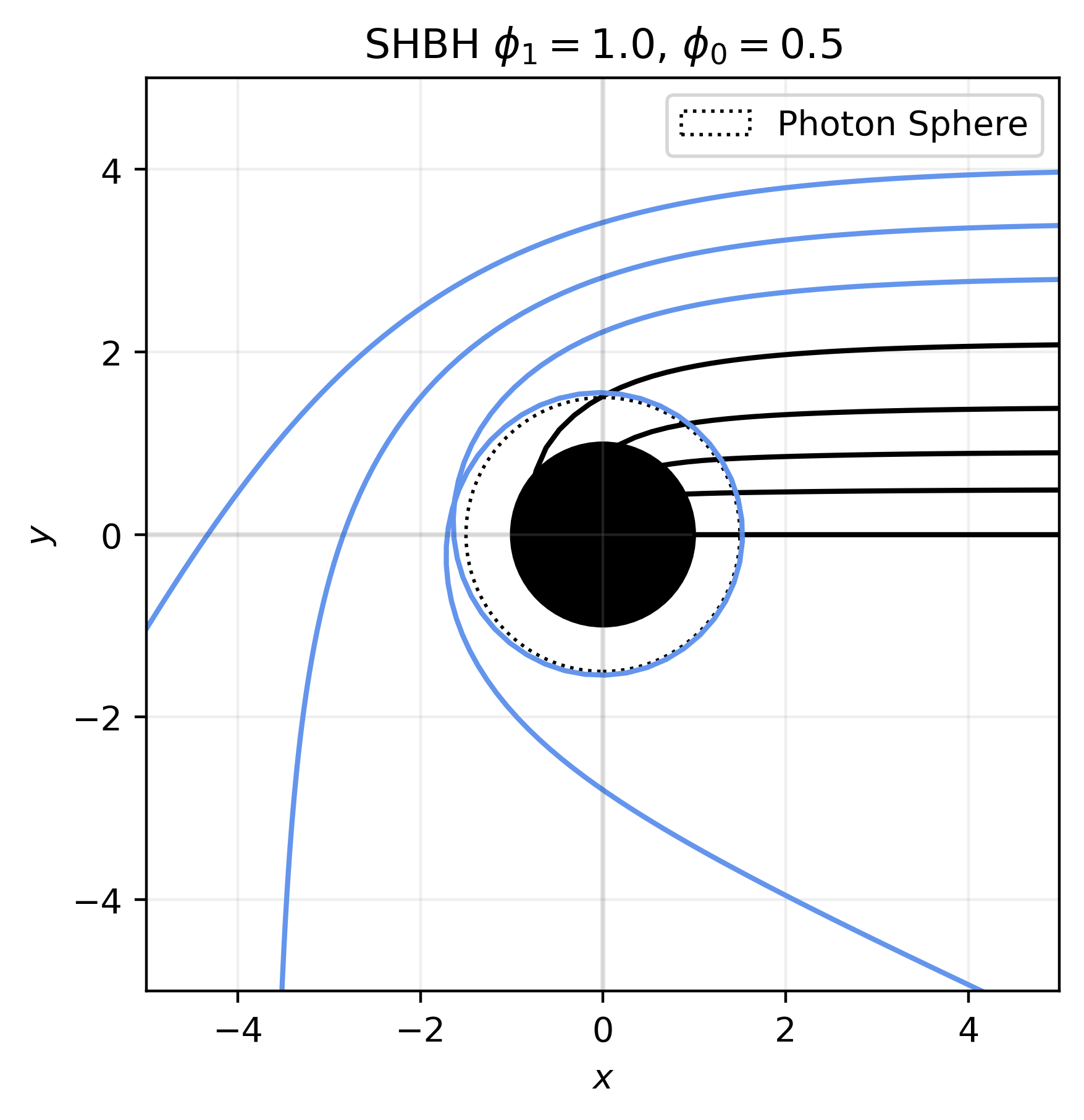}
    }
    \mbox{
    (d)
    \includegraphics[scale=0.45]{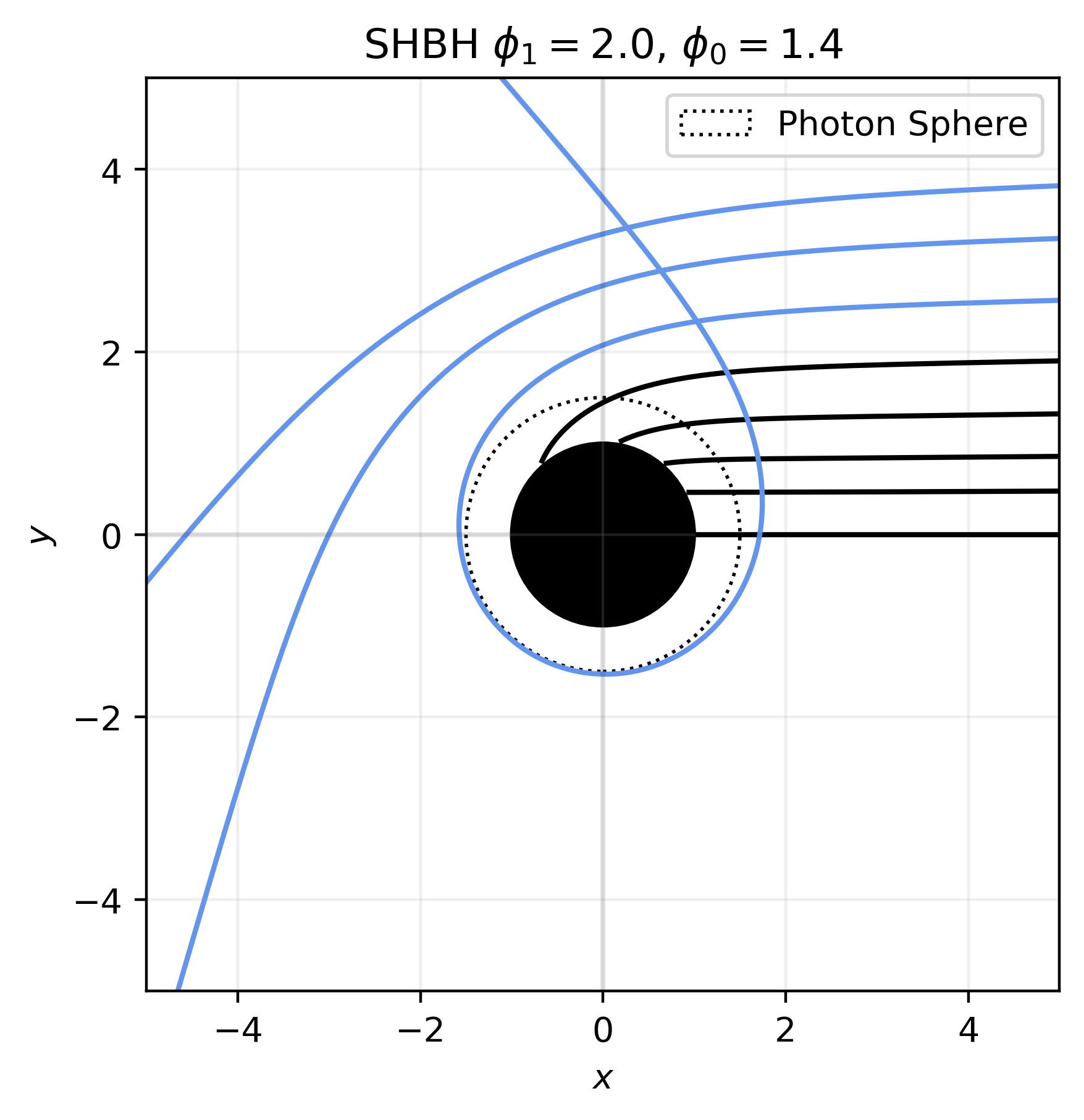}
    (e)
    \includegraphics[scale=0.45]{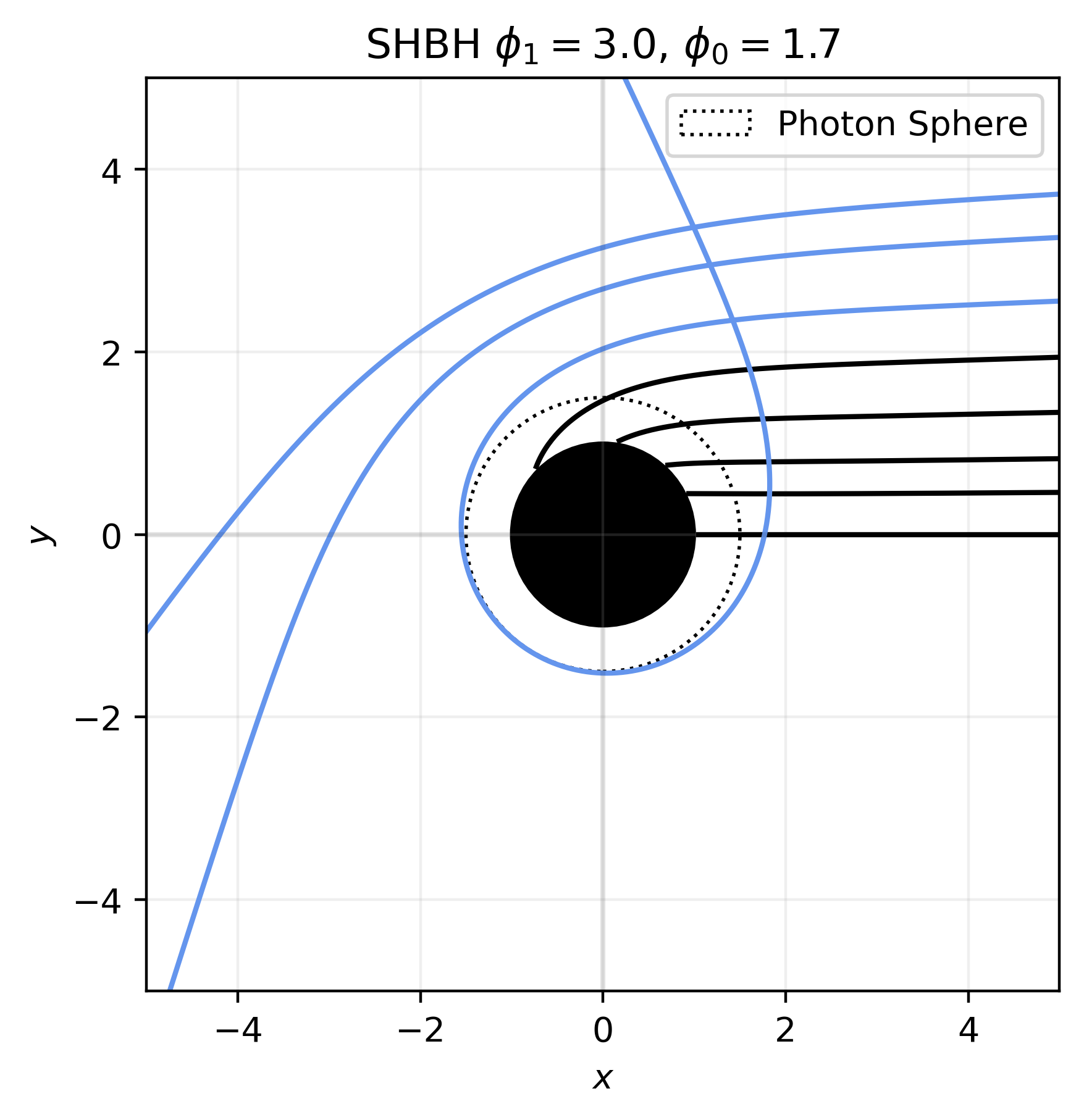}
    (f)
    \includegraphics[scale=0.45]{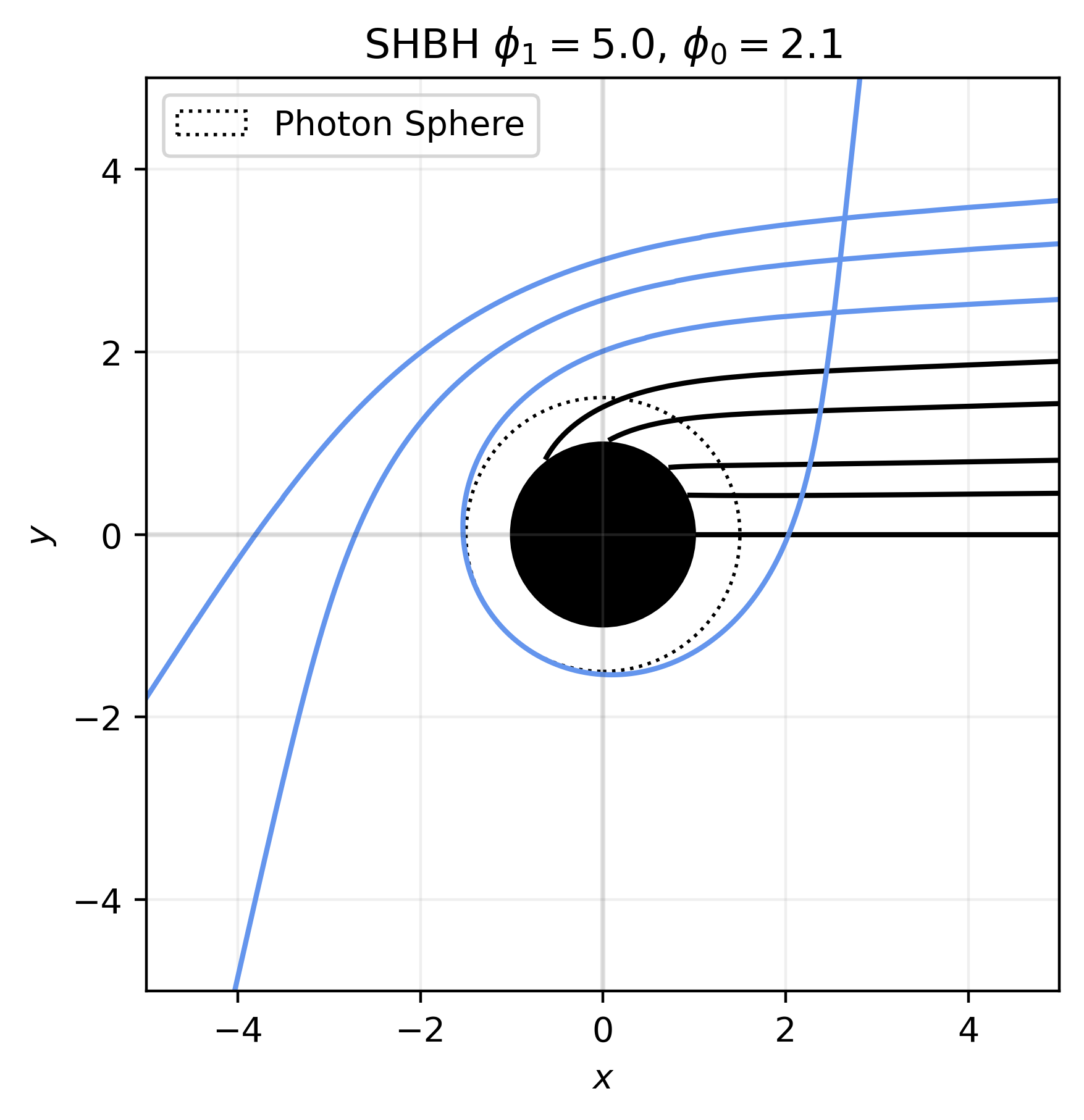}
    }
    \caption{Photon trajectories with different values of $\phi_1$ ($0.001$, $1.0$, $2.0$, $3.0$, and $5.0$), $\phi_0$ ($0.001$, $0.5$, $1.4$, $1.7$, and $2.1$) for different impact parameters. We choose the values of $\phi_0$ in such a way that $r_\text{ps}/r_H\approx 1.5$ (see figure~\ref{fig:plot_rph}) to compare with the Schwarzschild black hole solution shown in the first panel of the figure.}
    \label{fig:rays1}
    \end{figure*}
    
    The conditions determining the photon sphere radius, $r_\text{ps}$, is the vanishing of the derivative of the effective potential, $\left.\frac{d  V_{\text{eff}}}{d r}  \right|_{r=r_\text{ps}} = 0$. For the SHBH, this condition reads 
    \begin{equation}
    r_\text{ps}N'(r_\text{ps})-2N(r_\text{ps})=0 . \label{eq:photon_sphere_condition}
    \end{equation}


    \begin{figure*}[t]
    \centering
    \mbox{
    (a)
    \includegraphics[scale=0.5]{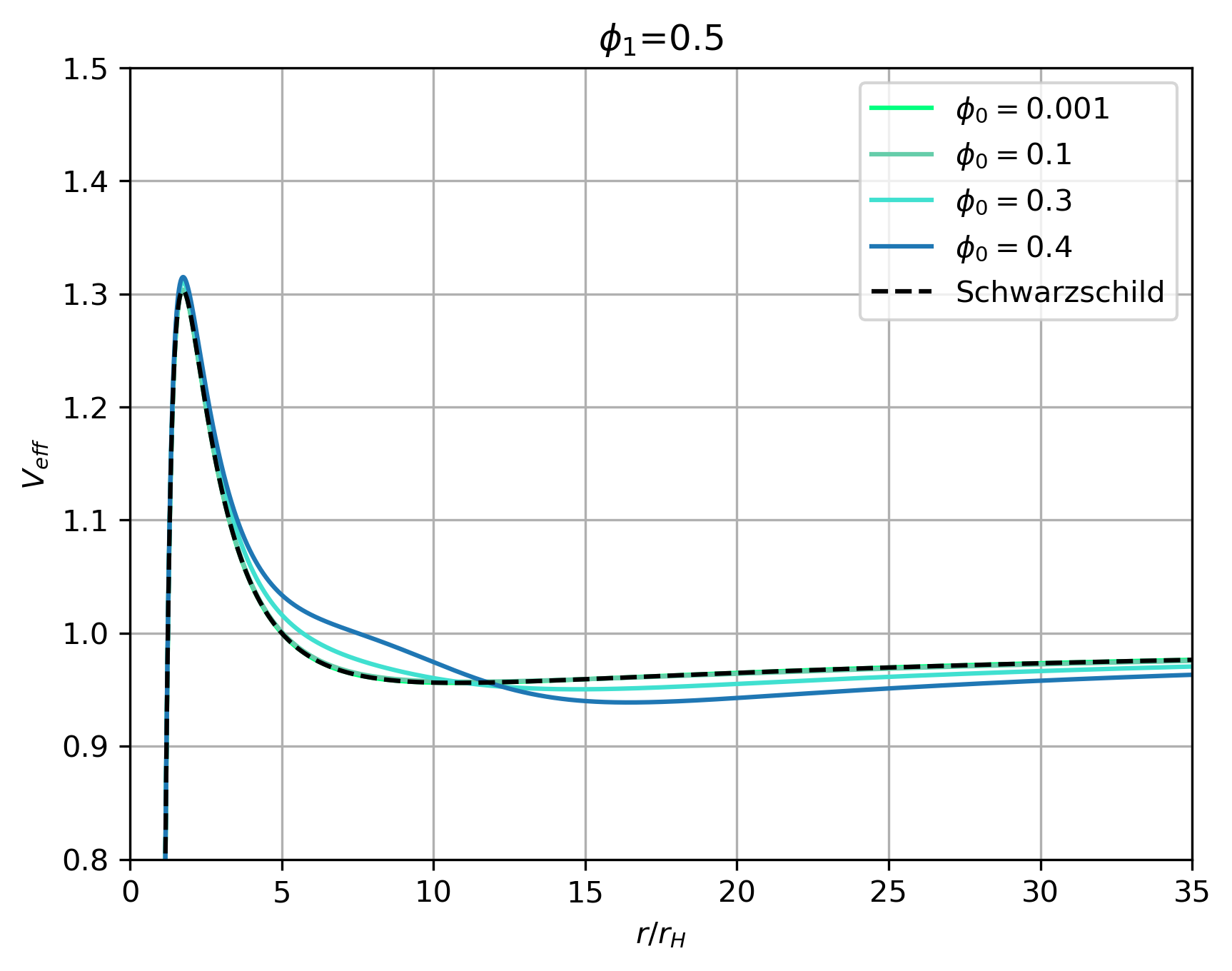}
    }
    \mbox{
    (b)
    \includegraphics[scale=0.5]{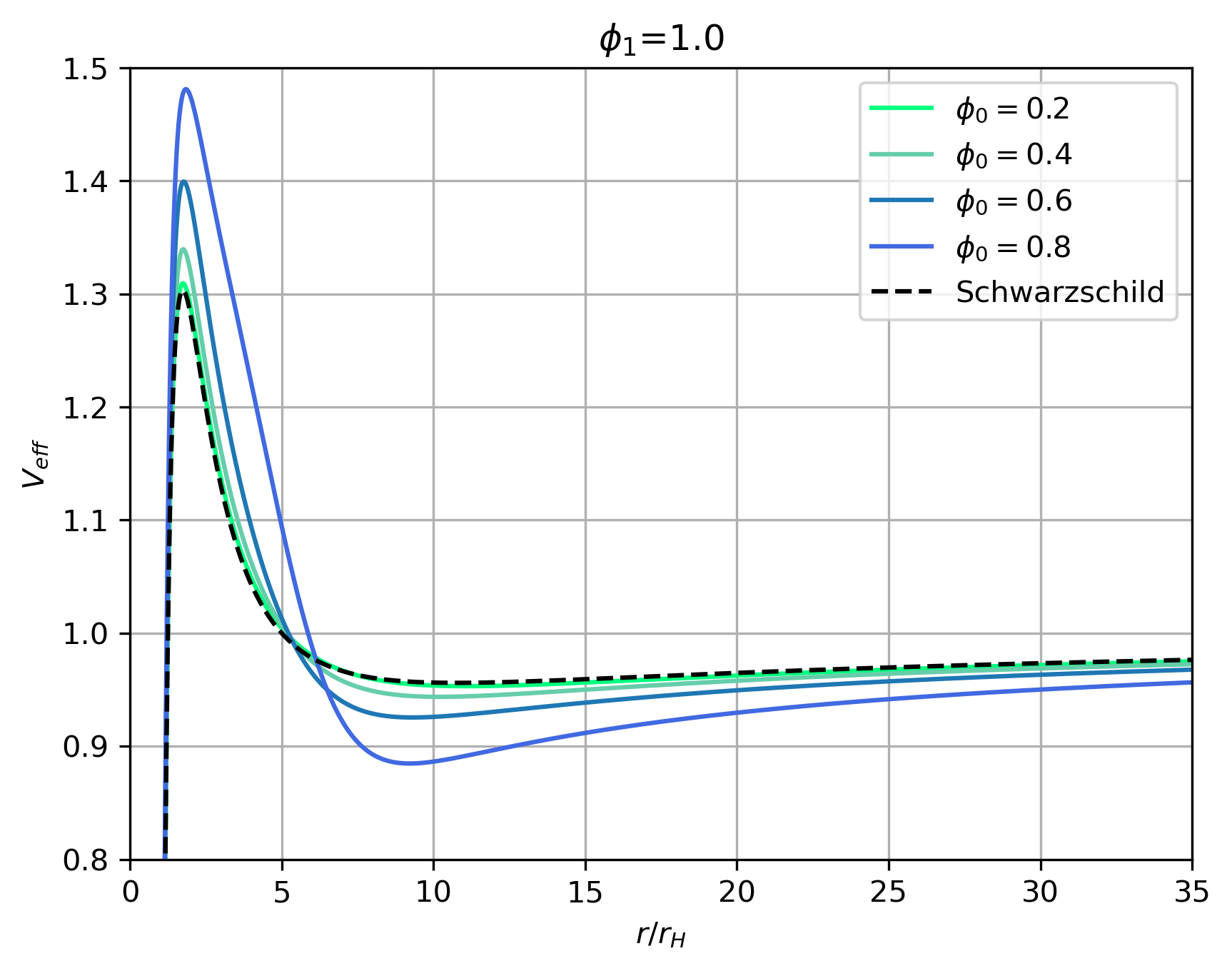}
    }
    \mbox{
    (c)
    \includegraphics[scale=0.5]{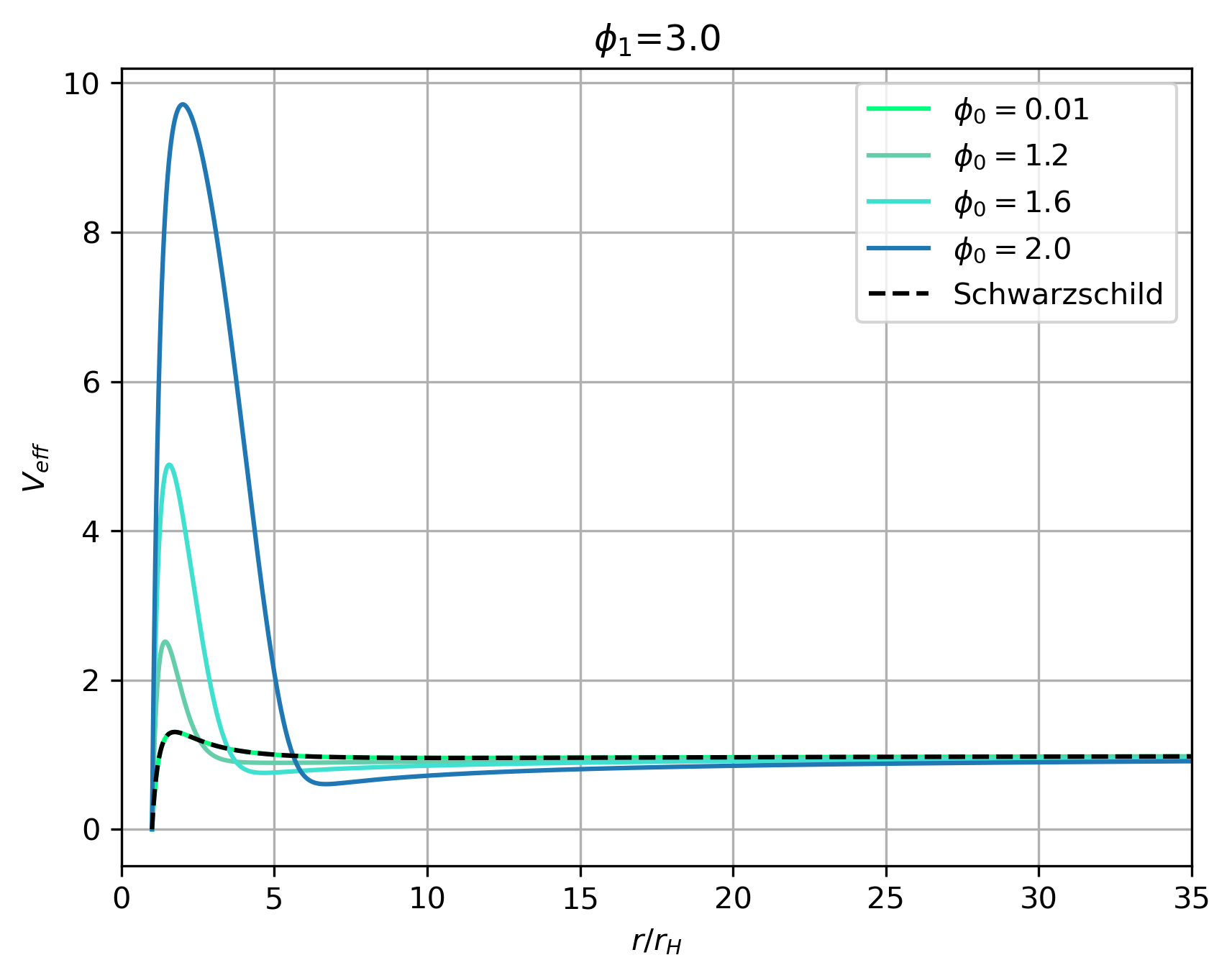}
    }
    \mbox{
    (d)
    \includegraphics[scale=0.5]{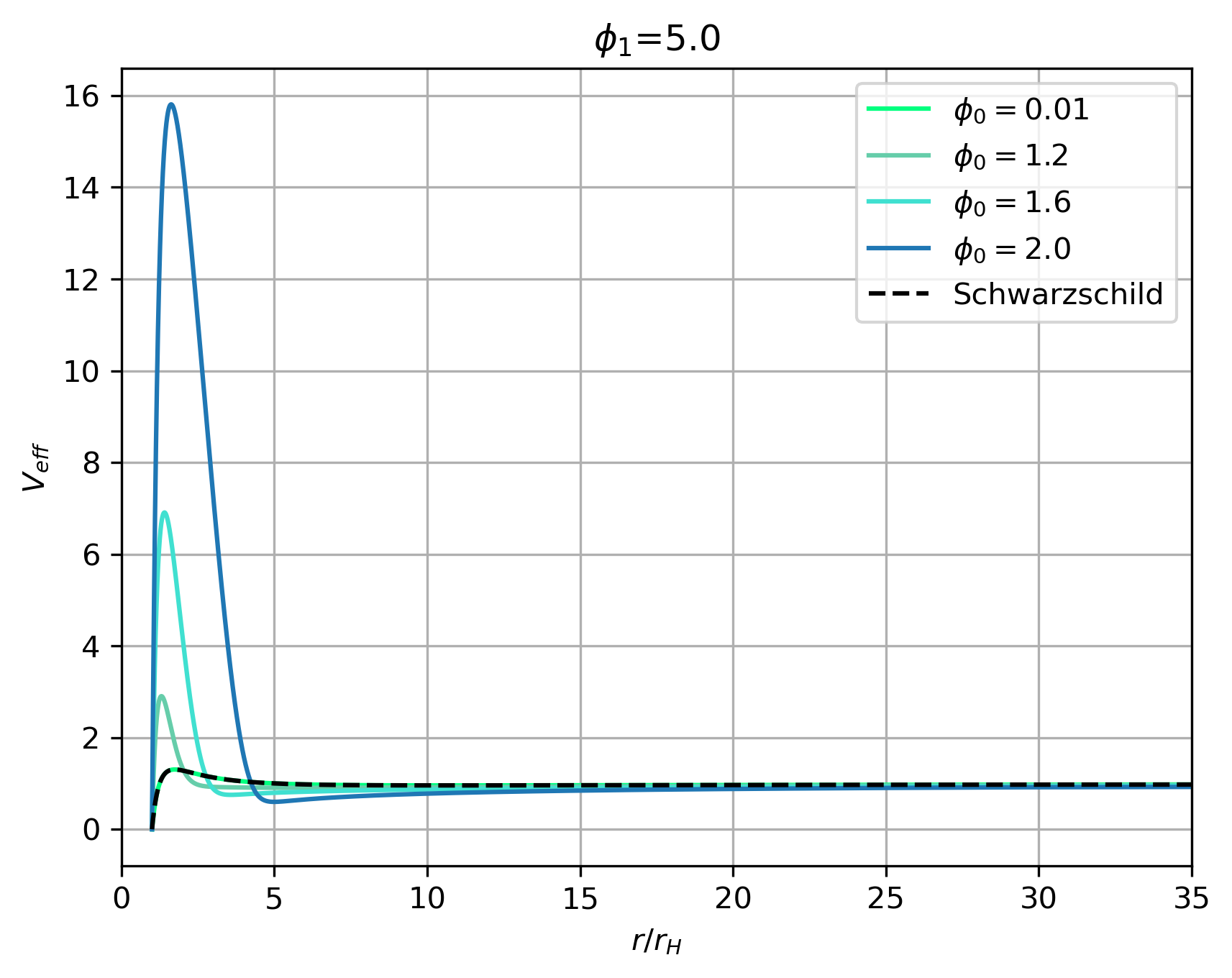}
    }
    \caption{Effective potential as a function of the radial coordinate $r/r_H$ for massive particles moving around SHBHs with $\phi_1=0.5$, $1.0$, $3.0$ and $5.0$. In all the cases, we consider $L=2.5$ and the dashed black curve corresponds to the effective potential for test particles moving in Schwarzschild spacetime.}
    \label{fig:effectivePotentialParticles}
    \end{figure*}
    
    We solve Eq.~\eqref{eq:photon_sphere_condition} numerically to obtain the photon sphere radius, $r_\text{ps}/r_H$, as a function of $\phi_0$ and characterized by different values of $\phi_1$; see Fig.~\ref{fig:plot_rph}. Note that the radius of the photon sphere for $\phi_1 = 0.5$ along the interval $0\leq \phi_0 < 0.5$ has small deviations from that of the Schwarzschild BH solution; i.e., $r_\text{ps}\approx r_\text{ps}^{(\text{Sch})} = 3 r_H/2$, in contrast to the case  $\phi_1 = 1.0$ where $r_\text{ps}/r_H$  has larger variations with the parameter $\phi_0$; see the upper panel in the figure. On the other hand, when considering greater values of $\phi_1$ (i.e.,  $2.0$, $3.0$, and $5.0$), the photon sphere radius changes notoriously. In all the cases, $r_\text{ps}/r_H$ tends to have a constant value for some interval of $\phi_0$. Then, $r_\text{ps}/r_H$ decreases to a minimum value from which it increases again, in agreement with the left and right shifting of $r_\text{ps}/r_H$ described in Fig.~\ref{fig:effectivePotentialPhotons}. Note that the minimum value of $r_\text{ps}/r_H$ decreases as $\phi_1$ goes from $2.0$ to $5.0$; see the lower panel in Fig.~\ref{fig:plot_rph}.
    
    Finally, in Fig~\ref{fig:rays1}, we show a few examples of photon trajectories calculated from a ray-tracing code. We obtained these figures by solving numerically the system of equations~\eqref{s3e10}. To solve the system, we constrain the motion of the photons to the equatorial plane, i.e., $\theta_0=\pi/2$. Hence, because $\dot{\theta}=0$, the photon remains on the equatorial plane along its trajectory. There are two possible trajectories: 1) the photon is caught by the black hole (black continuous line in the figure), or 2) it can escape to infinity after changing its direction (see the continuous cornflower blue). For the figure, we selected the values of $\phi_0$ so that the photon sphere in each case was close to that of the Schwarzschild black hole, i.e., $r_\text{ps}/r_H\approx 1.5$. In all the figures, the initial condition for the radial coordinate, $r_0=\sqrt{x^2_0+y^2_0}$, was chosen in such a way that $x_0=10$ and changing $y_0$ from $y_0=0.0$ (the horizontal line on the x-axis) to $y_0=4.0$.

\subsubsection{Circular Orbits for Massive Test Particles}

    In Fig.~\ref{fig:effectivePotentialParticles}, we show the behavior of the effective potential for massive particles ($\epsilon = 1$) as a function of the radial coordinate, $r/r_H$, for different values of $\phi_1$ and $\phi_0$. From this figure, it is possible to see that $V_\text{eff}$ shows a maximum and a minimum, corresponding to an unstable and a stable circular orbit, respectively. Note that, in all cases, the effective potential tends to $V_\text{eff}=1$ when $r/r_H\rightarrow \infty$, just as in the Schwarzschild case. 

    Circular trajectories are characterized by:
    \begin{equation}
    \left. \frac{d V_{\text{eff}} }{d r} \right|_{r = r_c} =\left(1+\frac{L^2}{r_c^2}\right)N'(r_c)-\frac{2L^2}{r_c^3}N(r_c) = 0 \, ,
    \end{equation} 
    where $r_c$ is the radius of the orbit. From this expression, it is possible to obtain the angular momentum of the test particle when moving in the circular orbit as
    \begin{equation}
     L_c = \frac{r_c \sqrt{r_c N'(r_c)}}{\sqrt{2 N(r_c) - r_c N'(r_c)}}. \label{eq:AngMomentumCircular}
     \end{equation}
    Using this relation, together with Eq. (\ref{eq:radialEoM}) and the definition $\Omega = \frac{p^\varphi}{p^t}$, we obtain the energy and the angular frequency of the test particle at the circular orbit,
    \begin{align}
    E_c = &\frac{e^{-\sigma (r_c)}N(r_c)}{\sqrt{N(r_c) - \frac{r_c}{2} N'(r_c)}} \label{eq:EnergyatCirc}\\
    \Omega_c = & e^{-\sigma (r_c)}\sqrt{\frac{N'(r_c)}{2r_c}}.\label{eq:OmegaatCirc}
    \end{align}
    
    On the other hand, the effective potential \eqref{eq:effectivePotential} also determines the existence of the Innermost Stable Circular Orbit (ISCO) through the  conditions 
    \begin{equation}
    \left.\frac{d V_{\text{eff}} }{d r}\right|_{ISCO} =   \left.\frac{d^2 V_{\text{eff}} }{d r^2}\right|_{ISCO} = 0 \,. 
    \end{equation}    
    The explicit form of these expressions for the SHBH are
    \begin{equation}
    2L_{ISCO}^2 N -
     \left(L_{ISCO}^2 + r_{ISCO}^2 \right)r_{ISCO}N' = 0
    \end{equation}
    and
    \begin{align}
    \left[2 (N')^2 - N N'' \right]
    \left(L_{ISCO}^2+r_{ISCO}^2\right) r_{ISCO}^2 & \notag \\ 
    - 2 r_{ISCO} N N'\left(L_{ISCO}^2+r_{ISCO}^2\right) -2L_{ISCO}^2 N^2&=0
    \end{align}
    with $N$, $N'$ and $N''$ evaluated at $r_\text{ISCO}$. Eliminating the angular momentum from these equations, we obtain the equation
    \begin{equation}
    \left[2 (N')^2 - N N'' \right] r_{ISCO} - 3NN' = 0. \label{eq:ISCO_condition}
    \end{equation}
    We solve Eq.~\eqref{eq:ISCO_condition} numerically to obtain the radius of the ISCO and the corresponding angular momentum, $L_\text{ISCO}$, from Eq.~\eqref{eq:AngMomentumCircular}.

    \begin{figure}
    \centering
    \mbox{
    \includegraphics[width=0.95\linewidth]{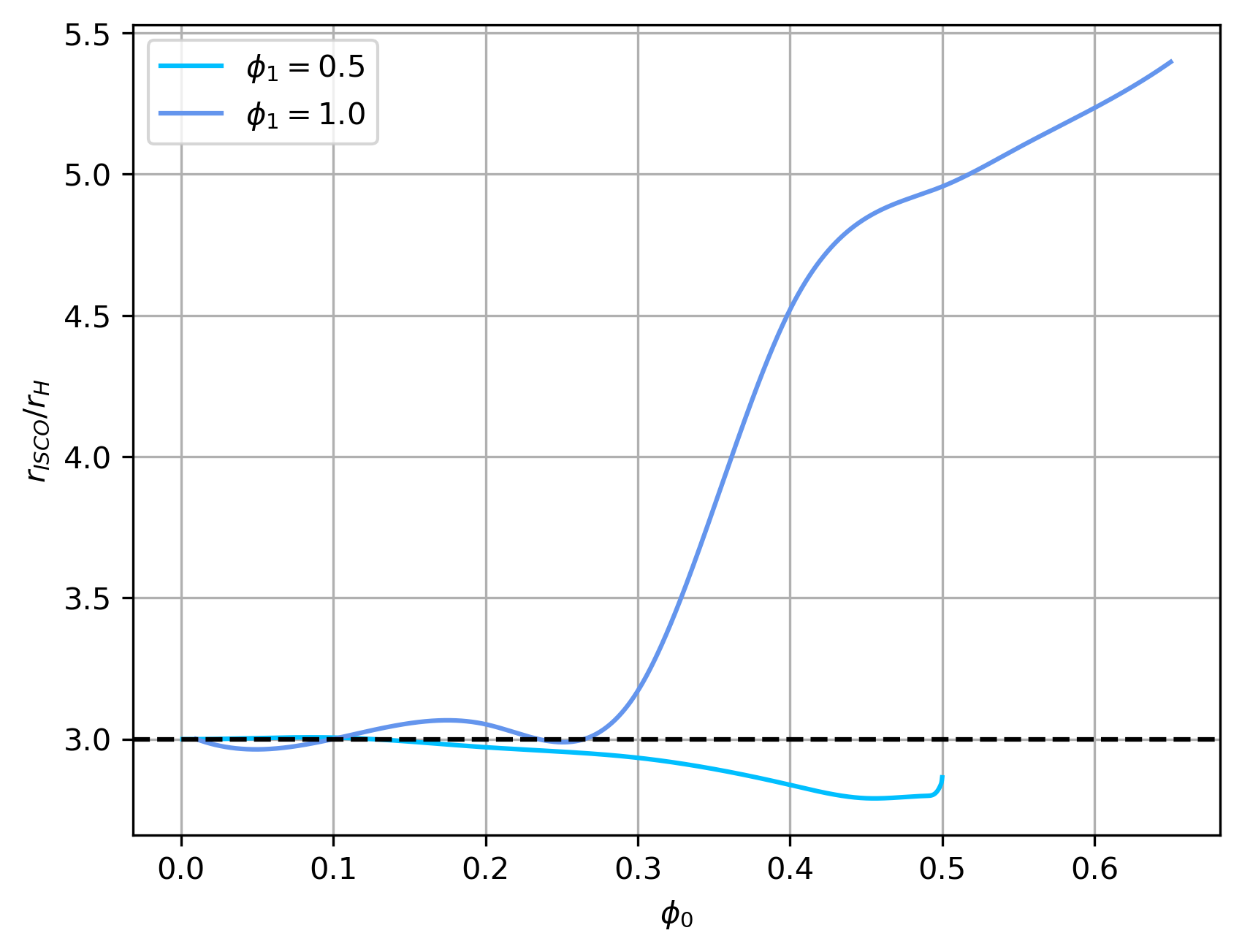}}	
    \mbox{\includegraphics[width=0.95\linewidth]{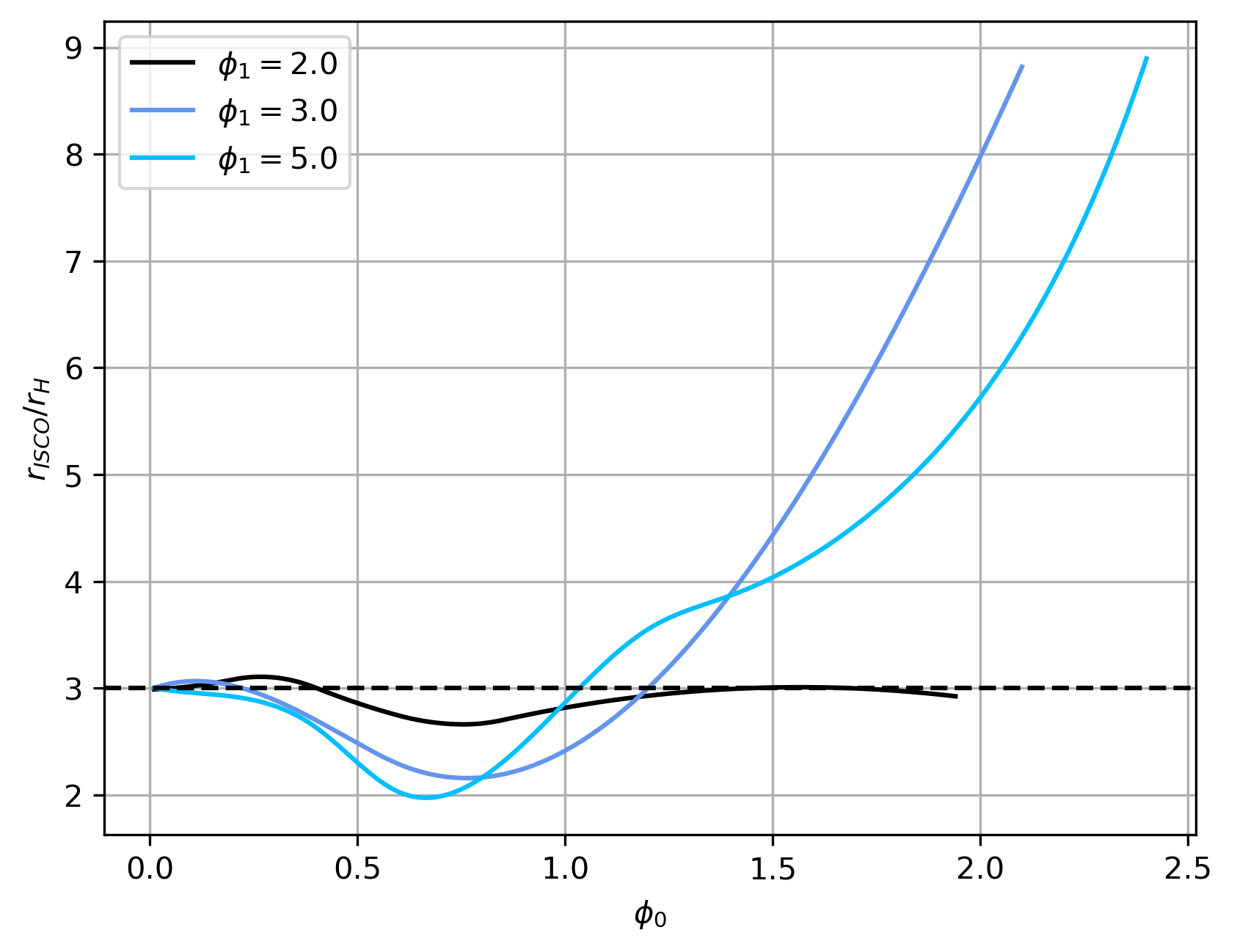}}
    \caption{Radius of the ISCO depending on the parameter $\phi_0$ for SHBHs with the values $\phi_1= 0.5, 1.0, 2.0, 3.0$ and $5.0$, upper and 	lower panels respectively. The dashed black line represents the radius of the photon sphere around a Schwarzschild black hole.}
    \label{fig:rISCO}
    \end{figure}
    
    Figure~\ref{fig:rISCO} shows the general behavior of the ISCO radius for the SHBHs as a function of the parameter $\phi_0$  for different values of $\phi_1$. Note that, in both figures (upper and lower panels), the behavior of the ISCO is close to that of a Schwarzschild black hole, $r_\text{ISCO}^\text{(Sch)} = 3 r_H $, for small values of $\phi_0$ ($\phi_0=0.5$ being the closest case); however, for higher values of $\phi_0$, the figure shows a variation in the size of the ISCO. For example, when $\phi_1=1.0$, it is possible to see that the ISCO is close to the Schwarzschild case, and then (around $\phi_0=1.5$) the ISCO increases. On the other hand, when $\phi_1=2.0$, $3.0$, and $5.0$, the figure shows a different behavior. Although the ISCO is still close to that of Schwarzschild for small values of $\phi_0$, there is a value of the parameter  $\phi_0$ for which the ISCO decreases, reaching a minimum value at some point $\phi^\text{min}_0$; then it begins to increase again. Note that the location of the minimum decreases as $\phi_1$ changes from $2.0$ to $5.0$.

\begin{figure*}
    \centering
    \mbox{
    (a)
    \includegraphics[scale=0.5]{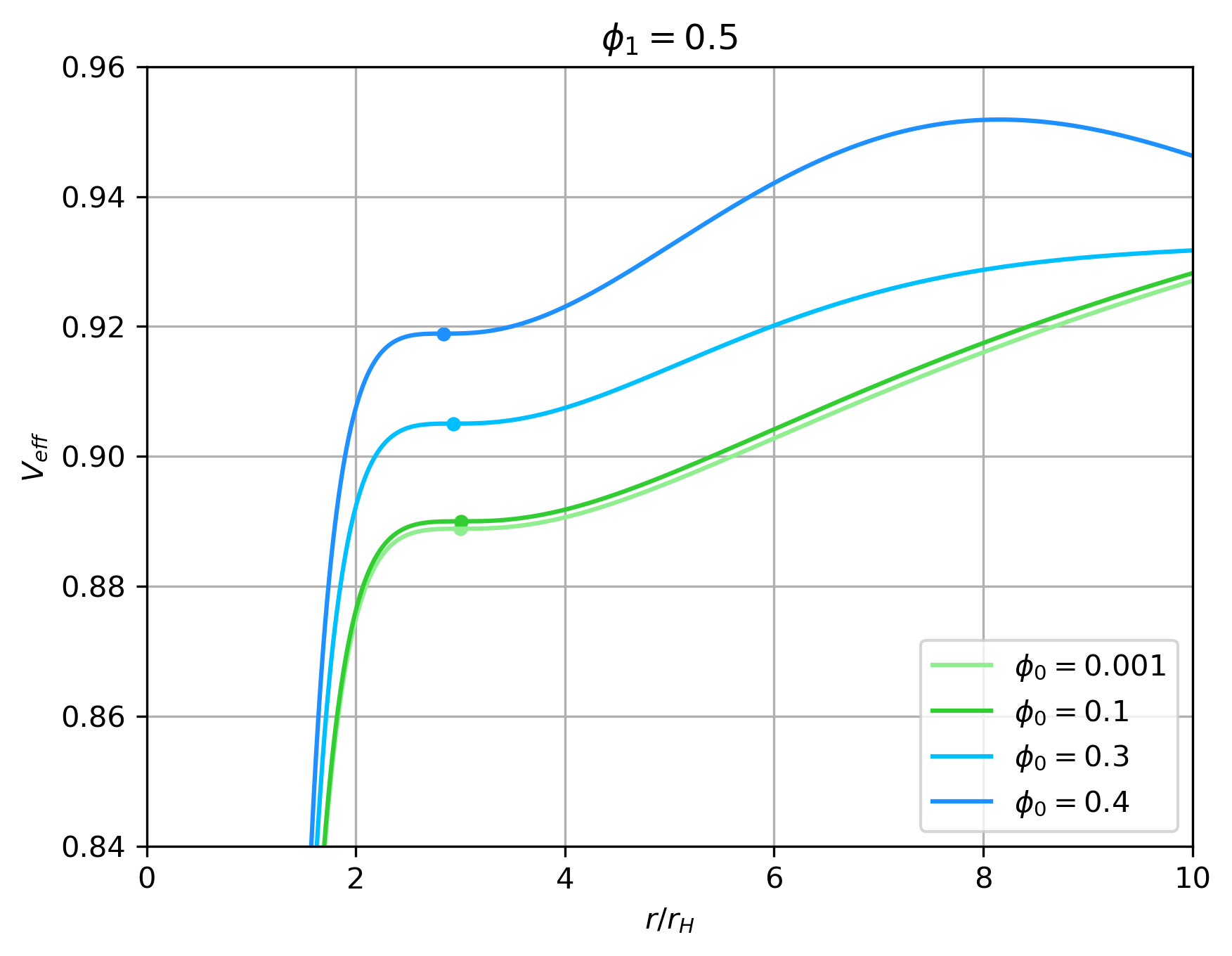}
    }
    \mbox{
    (b)
    \includegraphics[scale=0.5]{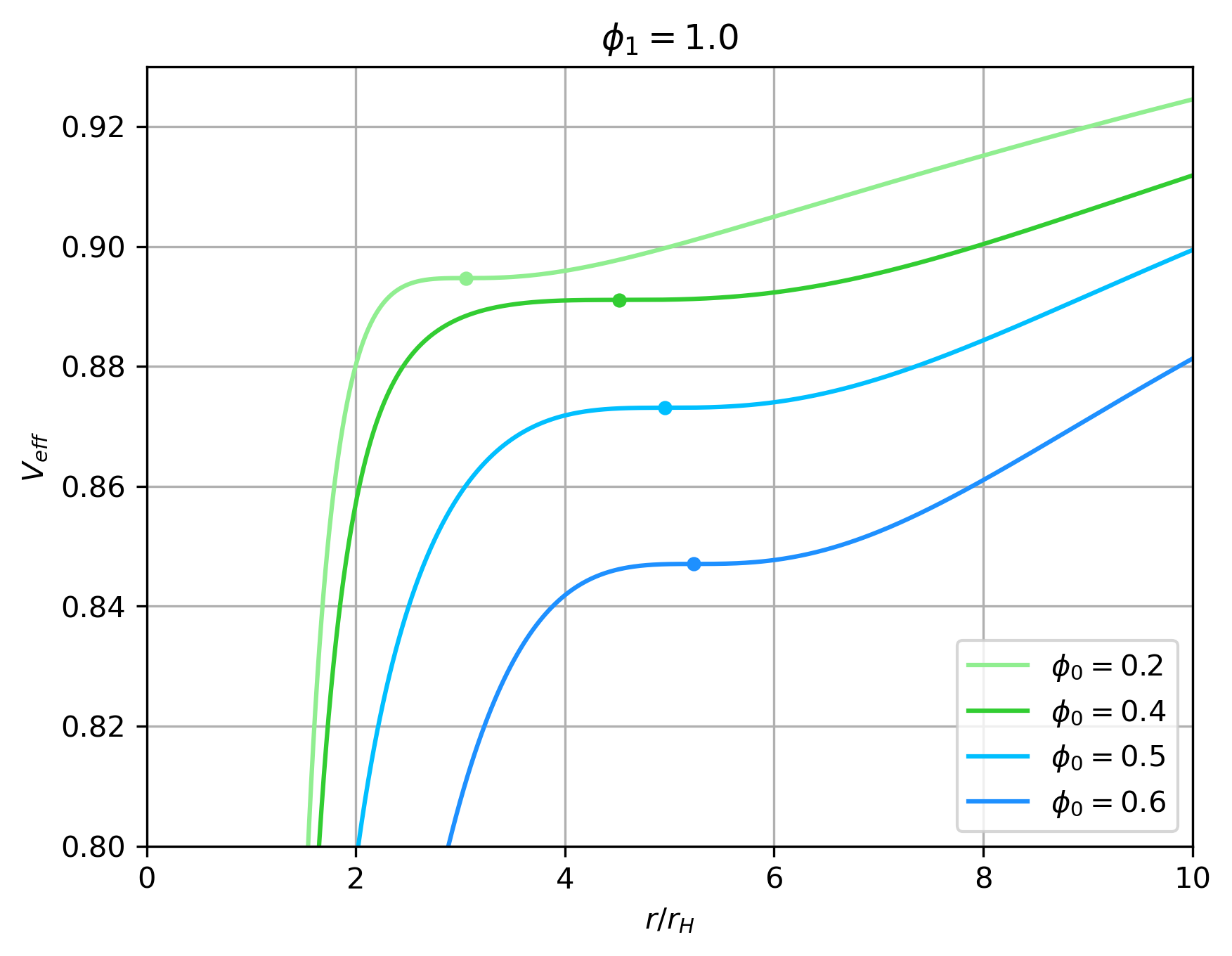}
    }
    \mbox{
    (c)
    \includegraphics[scale=0.5]{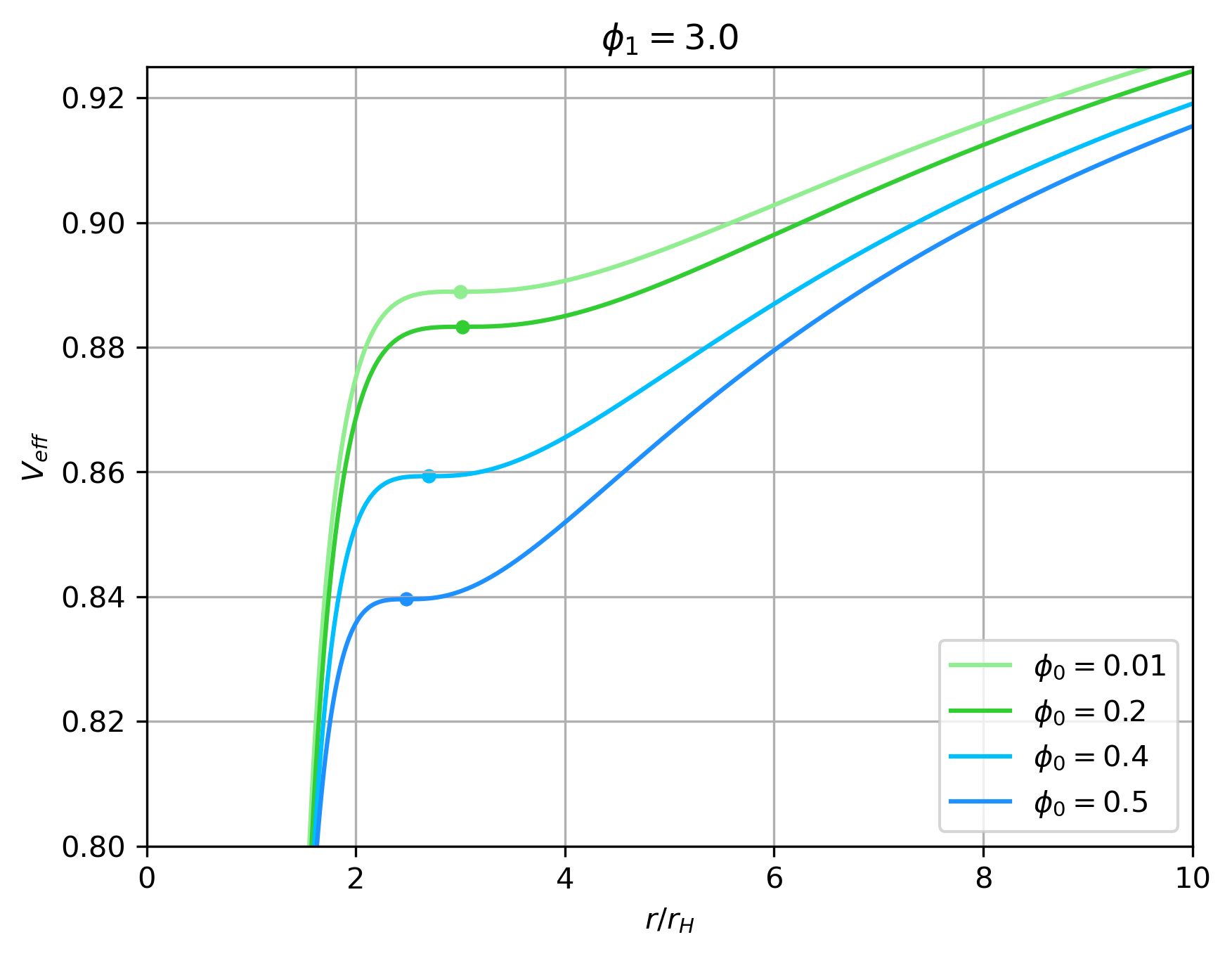}
    }
    \mbox{
    (d)
    \includegraphics[scale=0.5]{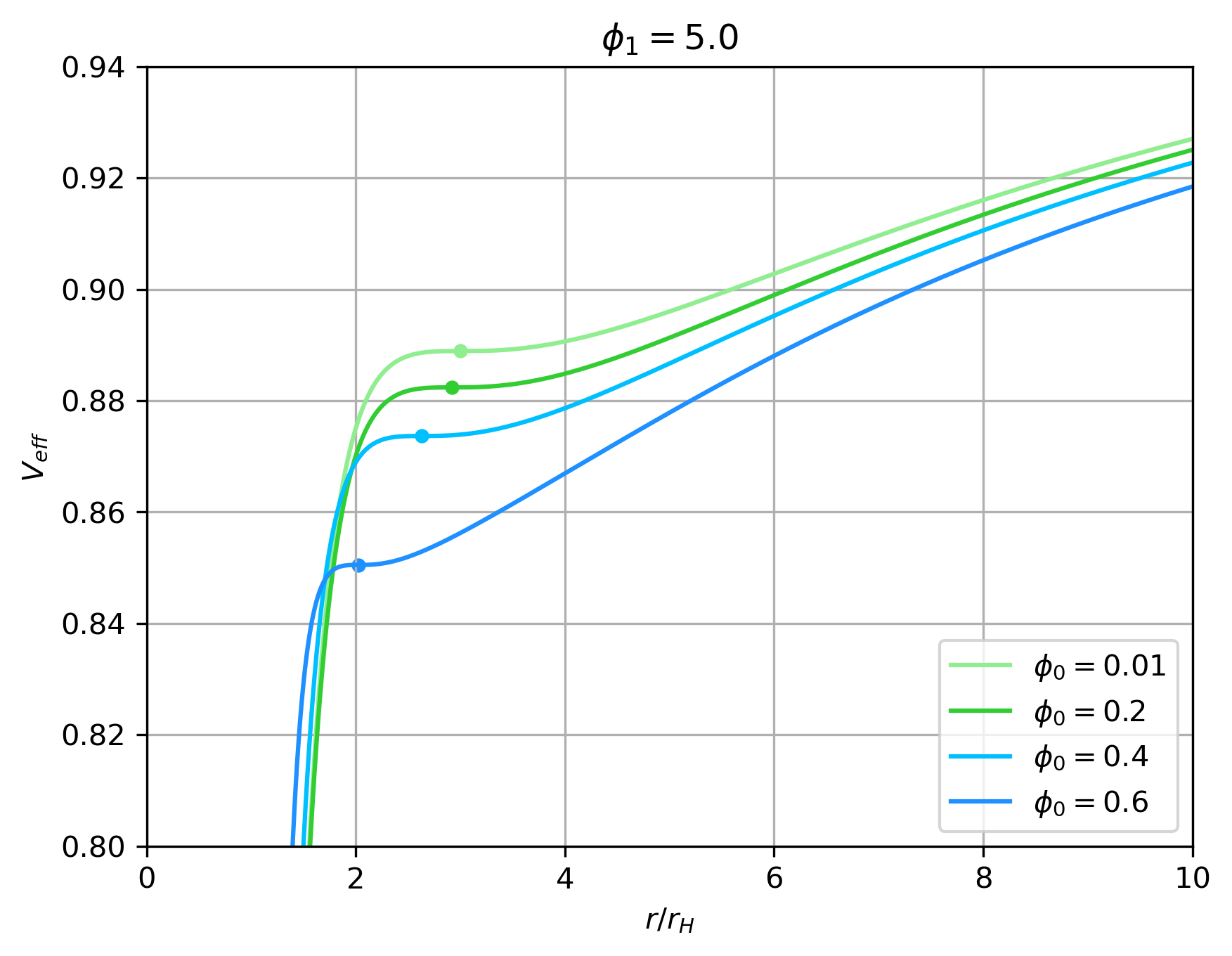}
    }
    \mbox{
    (e)
    \includegraphics[scale=0.5]{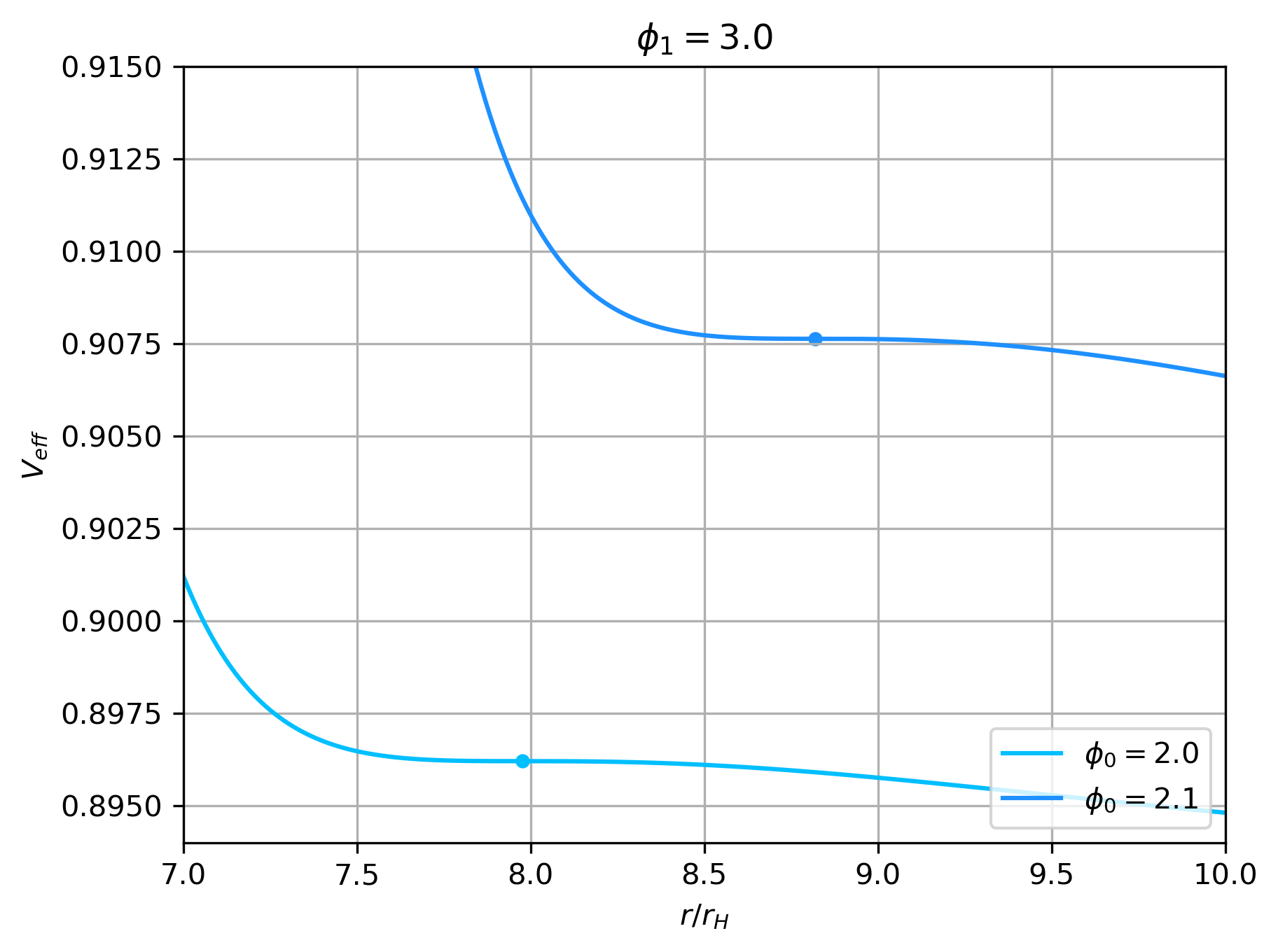}
    }
    \mbox{
    (f)
    \includegraphics[scale=0.5]{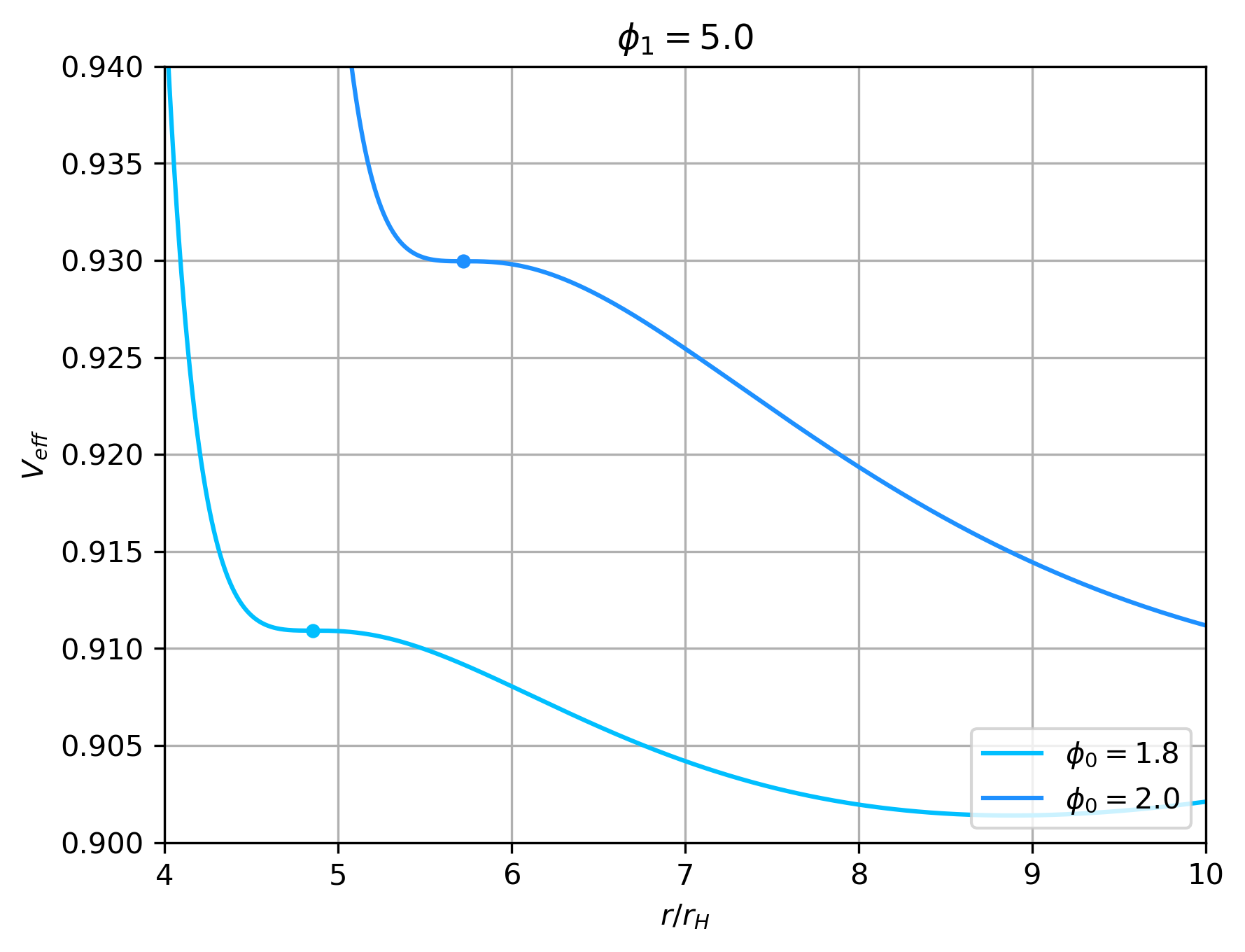}
    }
    \caption{Effective potential for a test particle moving with the angular momentum required at the ISCO as a function of the radial coordinate $r/r_H$. The marker shows the position of the ISCO.}
    \label{fig:effectivePotentialISCO}
    \end{figure*}

    In Fig.~\ref{fig:effectivePotentialISCO}, we show the effective potential vs. $r/r_H$ for different values of $\phi_0$. We represent the location of the ISCO using a dot as a marker. In the case of $\phi_1=0.5$, see Fig.~\ref{fig:effectivePotentialISCO} (a), the location of the ISCO does not change considerably as $\phi_0$ increases from $0.001$ to $0.4$. In the case of $\phi_1=1.0$, the change is more evident as we increase the value of $\phi_0$, in agreement with the upper panel of Fig.~\ref{fig:rISCO}. In panels (c) and (d), we show the behavior of $V_\text{eff}$ when $\phi_1=3.0$ and $\phi_1=5.0$, respectively. We consider small values of $\phi_0$. Note that the profile of the effective potential decreases its value as $\phi_0$ increases, i.e., green lines are higher than blue lines, in contrast to the behavior shown in panel (a). We observe the same behavior when $\phi_1=1.0$; see panel (b). On the other hand, when $\phi_0$ increases, the ISCO radius decreases. For example, for $\phi_1=3.0$, $r_\text{ISCO}/r_\text{H}$ changes from $3.00$ to $2.48$ as $\phi_0$ goes from $0.01$ to $0.5$, in agreement with the behavior shown in the lower panel of Fig.~\ref{fig:rISCO}. For greater values of $ph_0$, the profile of the effective potential changes. In panels (e) and (f) of Fig.~\ref{fig:effectivePotentialISCO}, we consider the cases $\phi_1=3.0$ and $5.0$ with $\phi_0=2.0$, $2.1$ and $\phi_0=1.8$, $2.0$, respectively. In both cases, the value of $V_\text{eff}$ increases as $\phi_0$ increases. Moreover, in agreement with the behavior shown in the lower panel of Fig.~\ref{fig:rISCO}, the location of $r_\text{ISCO}$ also increases, changing from $7.97$ to $8.82$, and from $4.85$ to $5.72$ in the case of $\phi_1=3.0$ and $\phi_1=5.0$, respectively. The panels also show how the effective potential decreases its value as $r/r_H$ increases. Nevertheless, it will reach a second minimum from which its value increases again, approaching $V_\text{eff}=1.0$ as $r/r_H\rightarrow \infty$. For example, in the case of $\phi_1=3.0$ with $\phi_0=2.0$ and $2.1$, the second minima are located at $r/r_H\approx12.5$ and $15$, respectively. We do not show these minima in the figure since our main interest is the behavior of the ISCO.
    
\section{The Image of Scalar Hairy Black Holes\label{secIV}}

\subsection{Angular Diameter of the Shadow}
\begin{figure}
    \centering
    \includegraphics[width=0.9\linewidth]{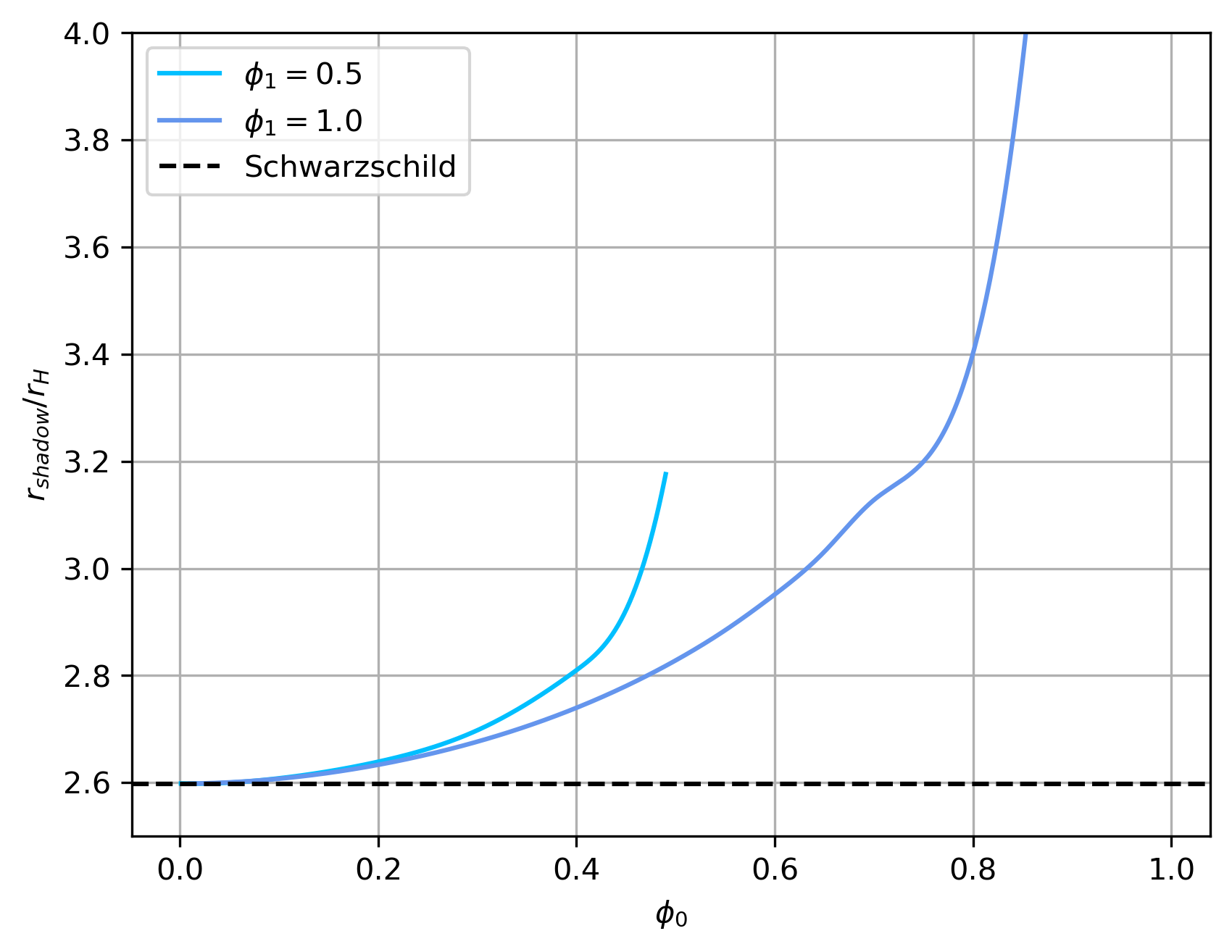}
    \includegraphics[width=0.9\linewidth]{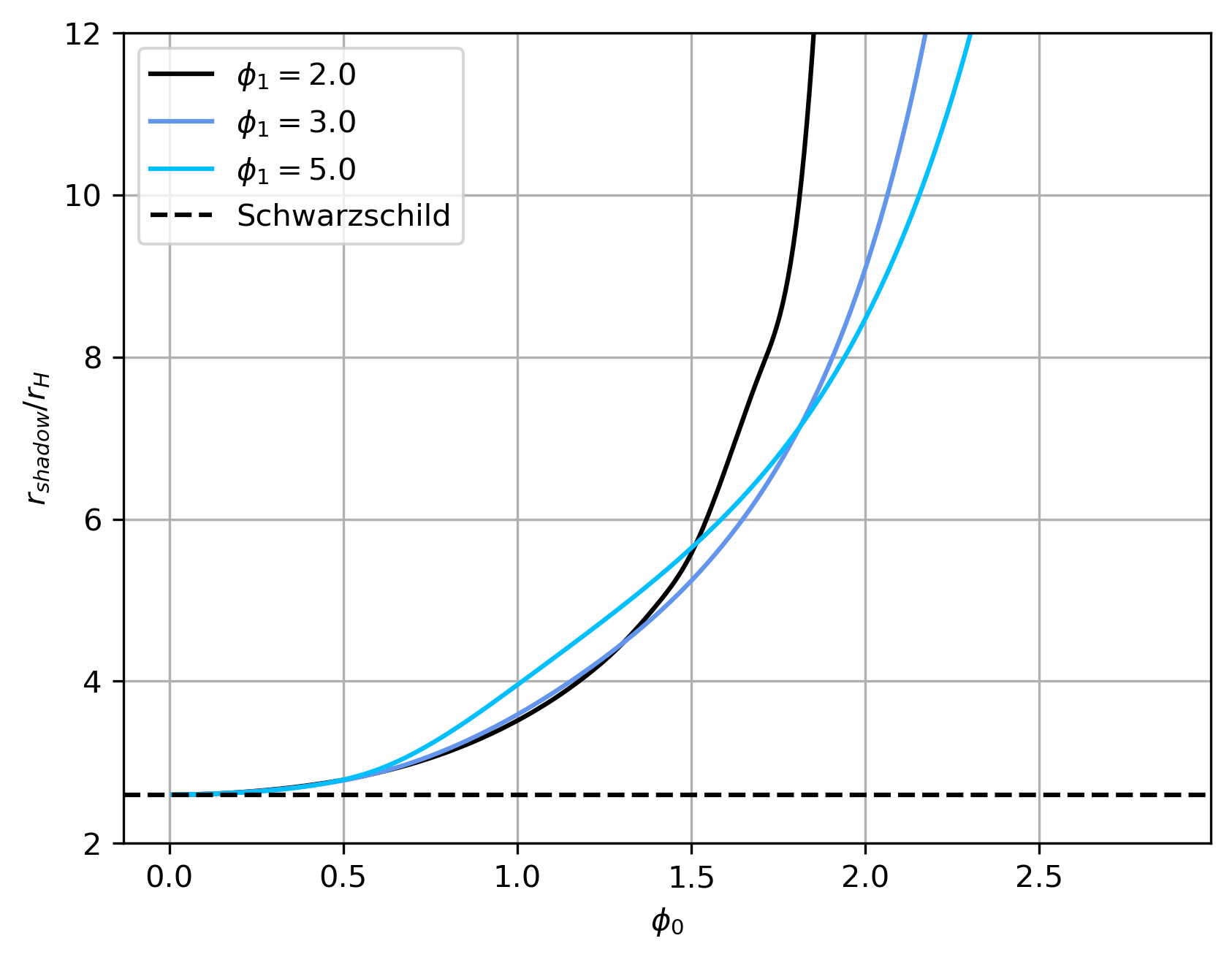}
    \caption{Size of the black hole shadows perceived by an asymptotic observer. The dashed black line represents the radius of the shadow of a  Schwarzschild black hole.}
\label{fig:r_shadow}
\end{figure}

    To investigate the shadow of hairy black holes, we start by defining the impact parameter of photons as
\begin{equation}
\label{eq:impactParameter}
b = \frac{L}{E}. 
\end{equation}
Replacing $\dot{r} = 0$ in Eq.~\eqref{eq:radialEoM} gives
\begin{equation}
b = \frac{re^{\sigma(r)}}{\sqrt{N(r)}}.
\end{equation}
The radius of the shadow of the black hole, as seen by a distant observer, is obtained by evaluating this impact parameter at $r=r_\text{ps}$, i.e.
\begin{equation}
r_\text{shadow} = b(r_\text{ps}) = \frac{r_\text{ps} e^{\sigma(r_\text{ps})}}{\sqrt{N(r_\text{ps})}}.
\end{equation}
Figure \ref{fig:r_shadow} shows the radius of the shadow for the SHBHs compared with the expected size for the shadow of a Schwarzschild black hole. Notice that increasing the value of the parameter $\phi_0$ produces an increment in the radius of the shadow.
    
Distant observers usually measure the angular diameter of the shadow, defined by the relation
    \begin{equation}
    \Theta = \frac{2r_{\text{shadow}}}{D},
    \end{equation}
where $D$ is the distance between the black hole and the observer. Giving $D$ in units of [Mpc] and introducing the mass of the black hole measured in solar masses, we obtain the angular diameter of the shadow in units of [$\mu$as] as
     \begin{equation}
    \frac{\Theta}{\text{ $\mu$as} }= \frac{6.191165 \times 10^{-8}}{\pi} \left(\frac{M}{M_\odot }\right) \left(\frac{D}{\text{Mpc}} \right)^{-1} \left(\frac{r_\text{shadow}}{M}\right).
    \end{equation}

\begin{figure}[t]
    \centering
    \includegraphics[width=0.9\linewidth]{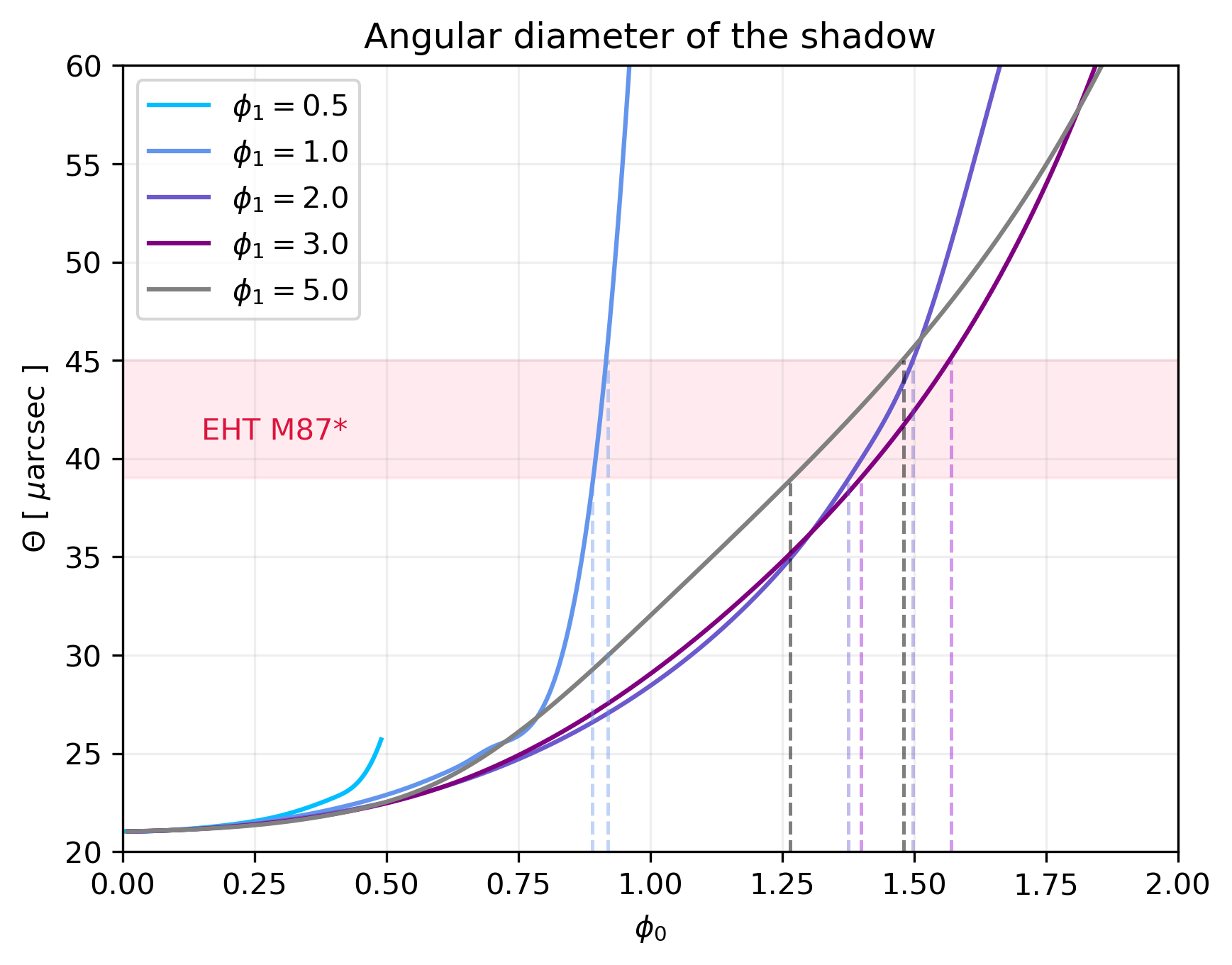}
    \includegraphics[width=0.9\linewidth]{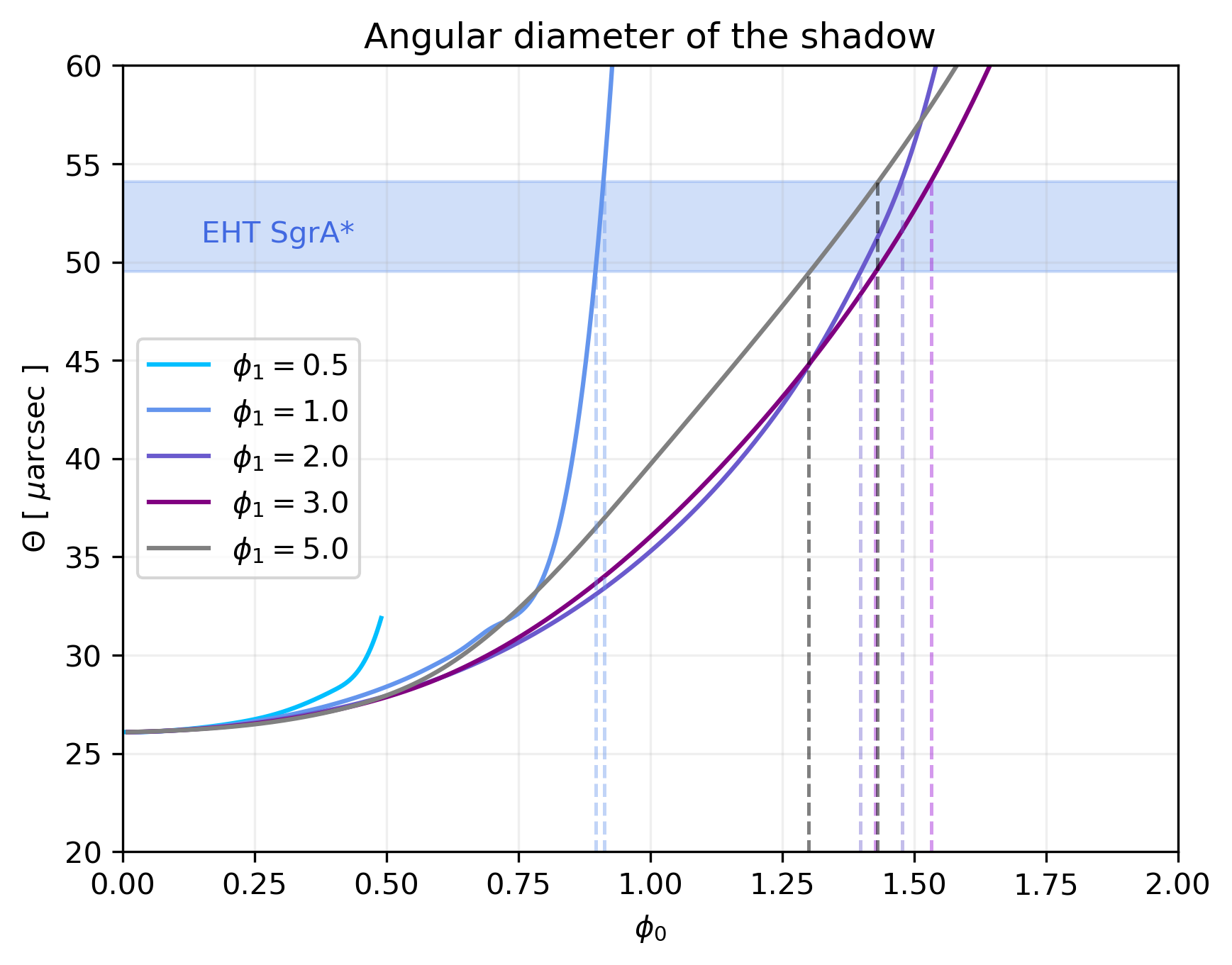}
    \caption{Angular diameter of the scalar hairy black hole compared with the angular sizes of the shadow reported by the Event Horizon Telescope for Sagittarius A* and M87* black holes. The vertical dashed lines show the constraints on the parameter $\phi_0$ to each of the SHBHs imposed by the comparison.}
    \label{fig:EHTshadows2}
    \end{figure}

    Two interesting black holes may help to study the shadow size. The first one is Sgr A*, located at a distance of $D=8.127 \text{ kpc}$ and with a mass of $M = 4.14 \times 10^6 M_\odot$. The second one is M87*, located at a distance $D=16.8 \text{ Mpc}$ with a mass of $M = 6.2 \times 10^9 M_\odot$. The angular diameter of the shadows reported by the Event Horizon Telescope (EHT) collaboration \cite{EHT1, EHT2} are  $51.8 \pm 2.3 \text{$\mu$as}$ and $42 \pm 3 \text{$\mu$as}$ for Sgr A* and M87*, respectively. With these values, we may restrict the possible values of the parameters $\phi_1$ and $\phi_0$ of the SHBH. In Fig.~\ref{fig:EHTshadows2}, we compare the predicted value of the angular size of the shadows for SgrA* and M87* when considered as SHBHs with the reported values from the EHT. 
    
    Our results show it is impossible to constrain the data obtained for $\phi_1=0.5$ since the behavior of the angular diameter does not reach the limits established for SgrA* and M87* by the EHT observations. However, when considering $\phi_1=1.0$, $2.0$, $3.0$, and $5.0$, we see a minimum and a maximum value for $\phi_0$. Note that in these cases, $\theta$ increases as $\phi_0$ increases, reaching the limits established by the EHT. For example, for $\phi_1=1.0$, the possible values for $\phi_0$ belong to the interval $0.89\leq\phi_0\leq0.92$ and $\leq\phi_0\leq$ for M87* and Sgr A*, respectively. In the case of $\phi_1=2.0$, the constraints are given by $0.89\leq\phi_0\leq0.92$ for M87* and $0.89\leq\phi_0\leq0.92$ for Sgr A*. In Fig.~\ref{fig:EHTshadows2}, the vertical dashed lines indicate the minimum and maximum values for the parameter $\phi_0$ allowed by the observational results for each value of $\phi_1$. The numerical values for these bounds are reported in Table \ref{table:constrainsSgrA}. Note that the interval in the case of $\phi_1=1.0$ is smaller when compared to the other cases, which increases as $\phi_1$ changes from $1.0$ to $5.0$. Finally, in Table \ref{table:ISCO}, we show some values of $r_\text{ps}$, $r_\text{shadow}$, $r_\text{ISCO}$, and $L_\text{ISCO}$ for some specific SHBHs solutions.  
    
    \begin{center}
    \begin{table}
    \begin{tabular}{||c | c | c | c | c ||} 
    \hline
    & \multicolumn{2}{|c|}{SgrA*} & \multicolumn{2}{|c||}{M87*} \\ [0.5ex]
    \hline
    $\phi_1$ &  $\phi^\text{min}_0$  & $\phi^\text{max}_0$ &$\phi^\text{min}_0$ & $\phi^\text{max}_0$ \\ [0.5ex]  
    \hline\hline \rule{0pt}{3ex}  
    1.0 & 0.8967 & 0.9131 & 0.89 & 0.92\\ [1.5ex]
    \hline \rule{0pt}{3ex}  
    2.0 & 1.39826 & 1.4775 & 1.375 & 1.498  \\ [1.5ex]
    \hline  \rule{0pt}{4ex}  
    3.0 & 1.42694 & 1.532 & 1.4 & 1.57  \\ [1.5ex]
    \hline  \rule{0pt}{4ex}  
    5.0 & 1.3 & 1.43 & 1.265 & 1.48 \\ [1.5ex] 
    \hline  
    \end{tabular}
    \caption{Constraints of the parameter $\phi_0$ arising from comparing the predicted angular diameter of the shadow with the values reported by the Event Horizon Telescope cooperation for the black holes SgrA* and M87*.}
    \label{table:constrainsSgrA}
    \end{table}
    \end{center}

    \begin{center}
    \begin{table}
    \begin{tabular}{||c | c | c | c | c | c ||} 
    \hline
    $\phi_1$ &  $\phi_\text{0}$  & $r_\text{ps}$ & $r_\text{\text{shadow}}$ & $r_\text{ISCO}$ & $L^2_\text{ISCO}$ \\[1.5ex]
    \hline \rule{0pt}{3ex}  
    0.5 & 0.3 & 1.5013 & 2.6978 & 2.934149 & 3.091750  \\ [1.5ex]
    \hline  \rule{0pt}{4ex}  
    2.0 & 0.7 & 1.4596 & 2.6765 & 3.170782 & 1.601612  \\ [1.5ex]
    \hline  \rule{0pt}{4ex}  
    3.0 & 2.0 & 1.6144& 9.0979  & 7.977254 & 32.485091   \\ [1.5ex]
    \hline  \rule{0pt}{4ex}
    5.0 & 1.2 &1.2784 & 4.5925 & 3.551564 & 5.658222 \\ [1.5ex]
    \hline
    \end{tabular}
    \caption{Values of $r_\text{ps}$, $r_\text{shadow}$, $r_\text{ISCO}$, and $L^2_\text{ISCO}$ for some specific SHBHs solutions. We use units with $r_H = 1.0$.}
    \label{table:ISCO}
    \end{table}
    \end{center}


    \begin{figure*}
    \centering
    \mbox{
     \includegraphics[scale=0.4]{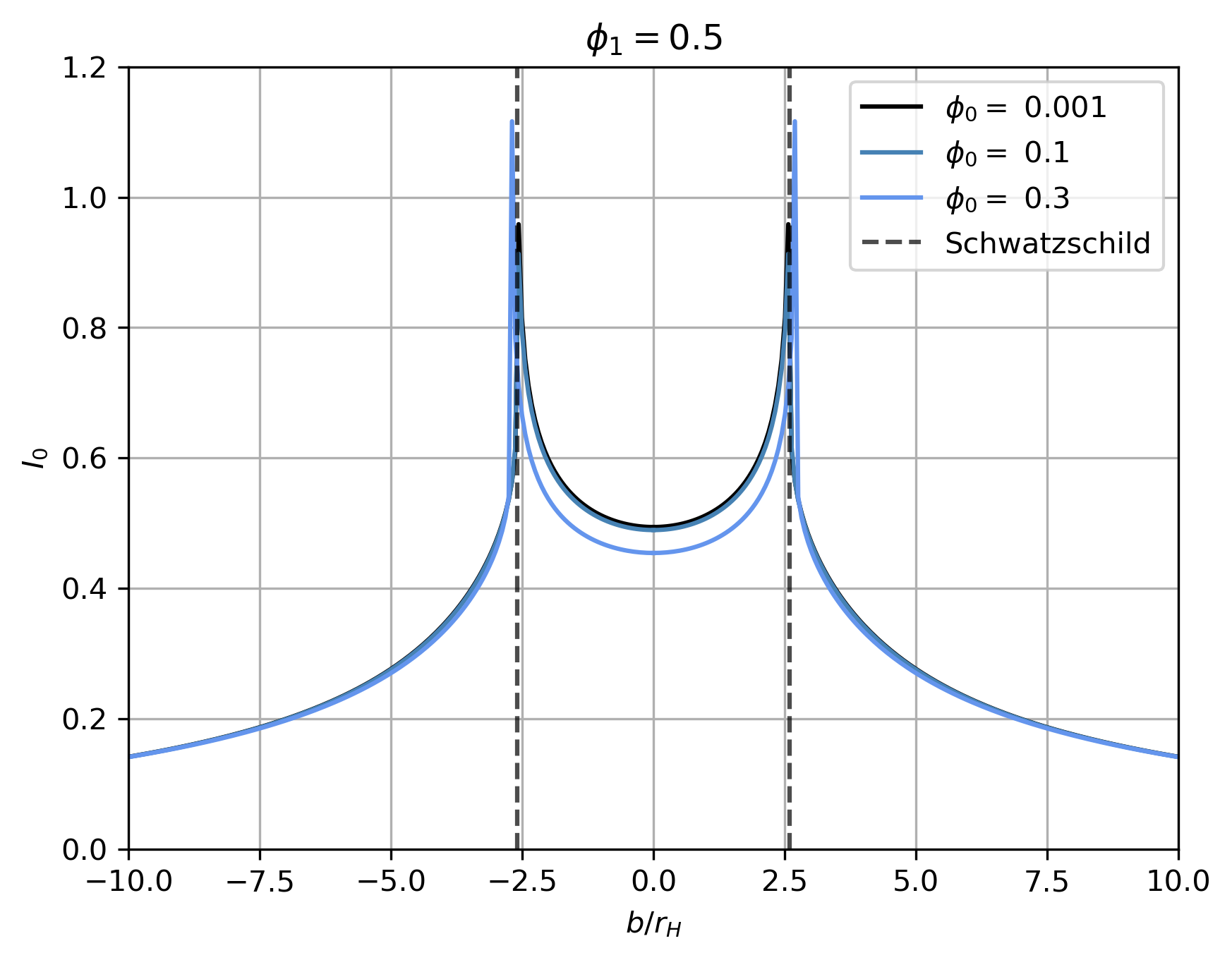}
     \includegraphics[scale=0.4]{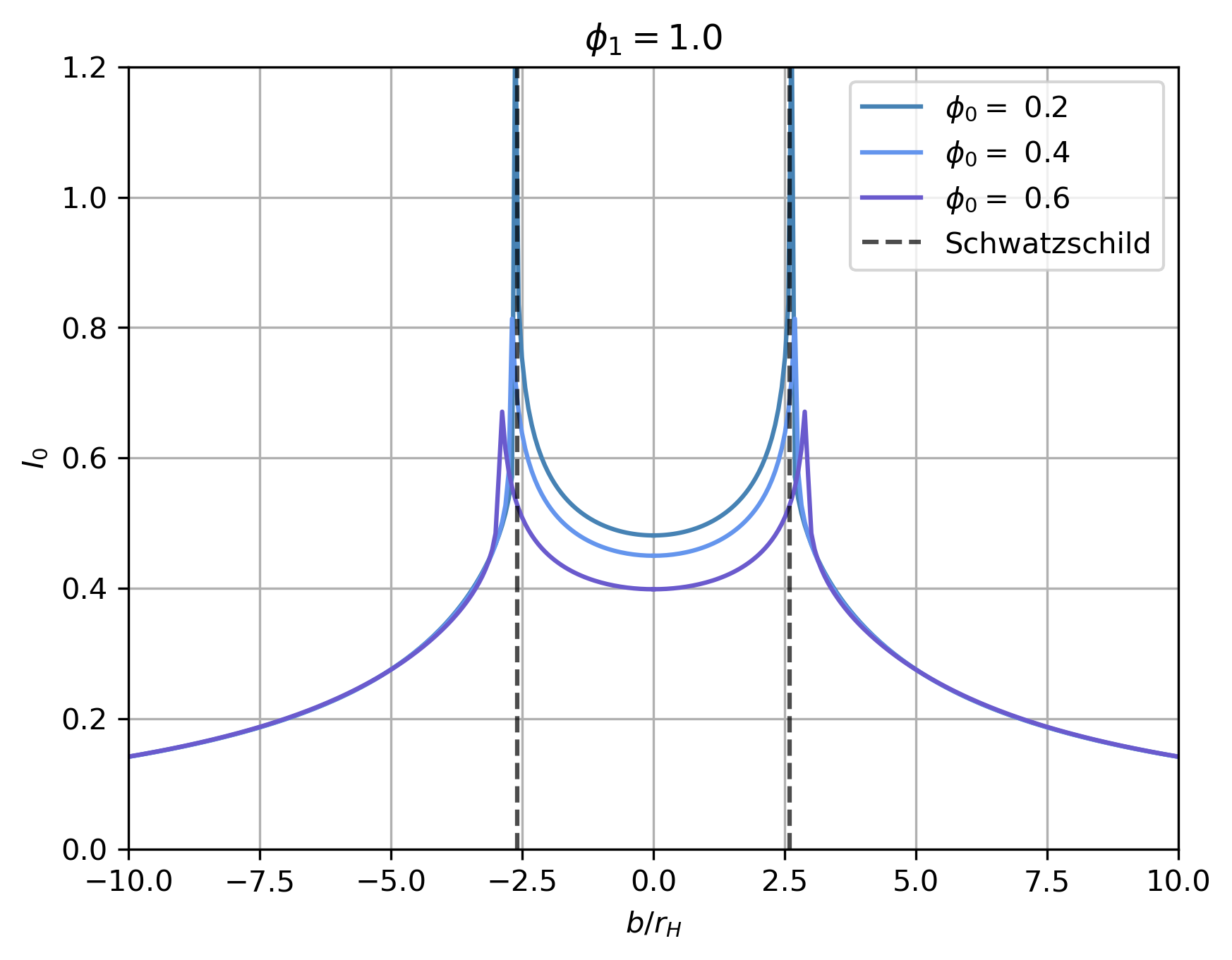}
     \includegraphics[scale=0.4]{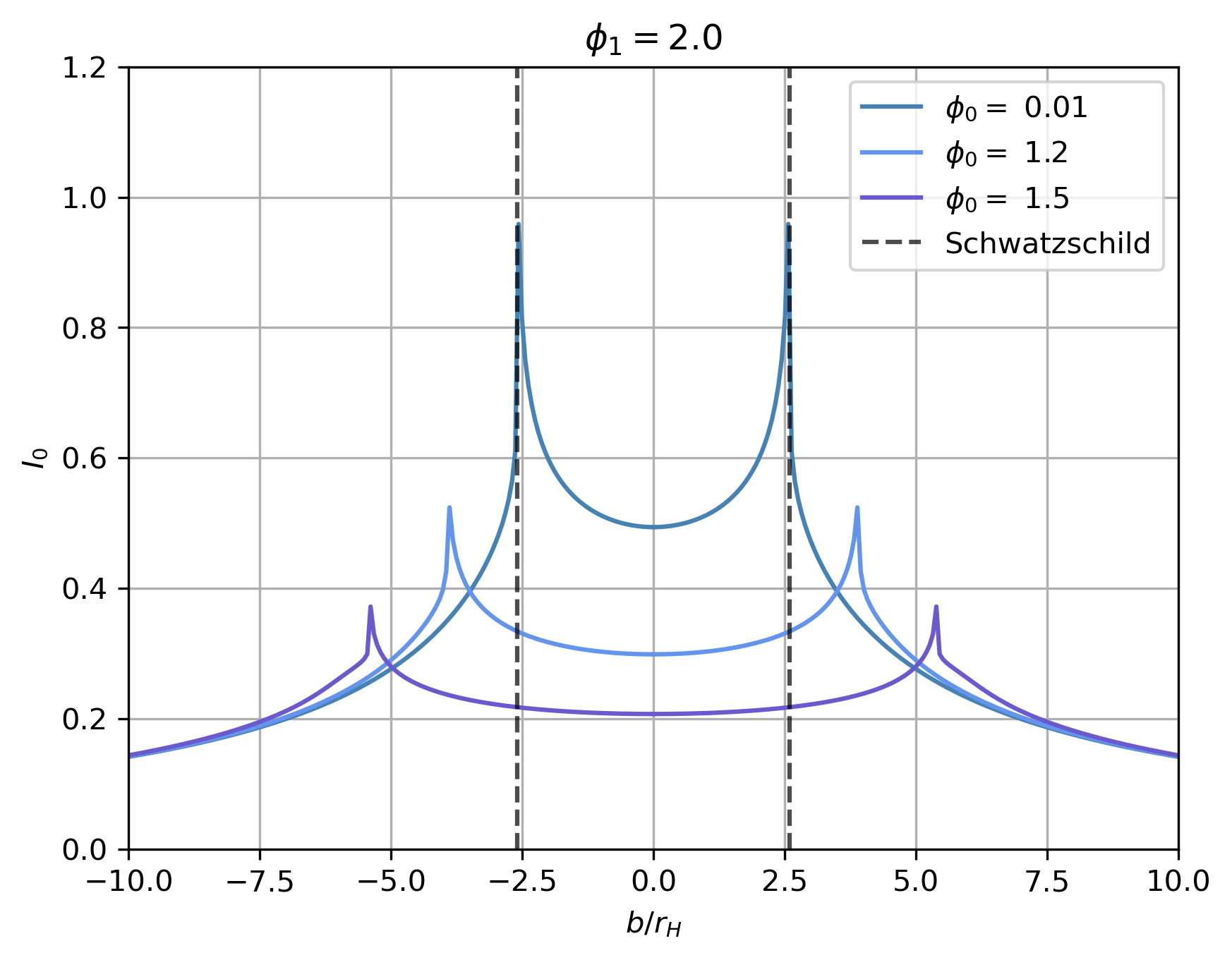}}
     \mbox{
     \includegraphics[scale=0.44]{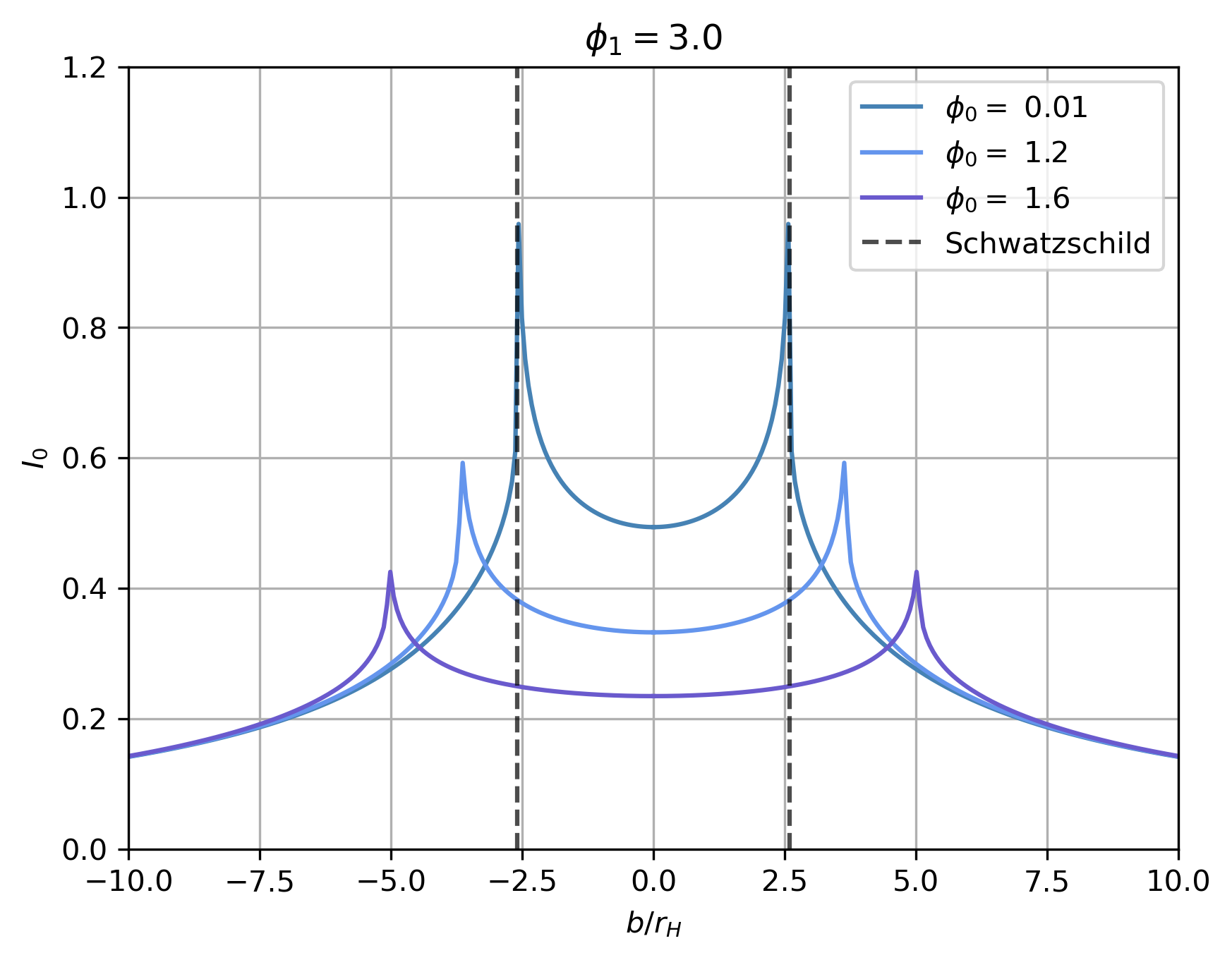}
     \includegraphics[scale=0.44]{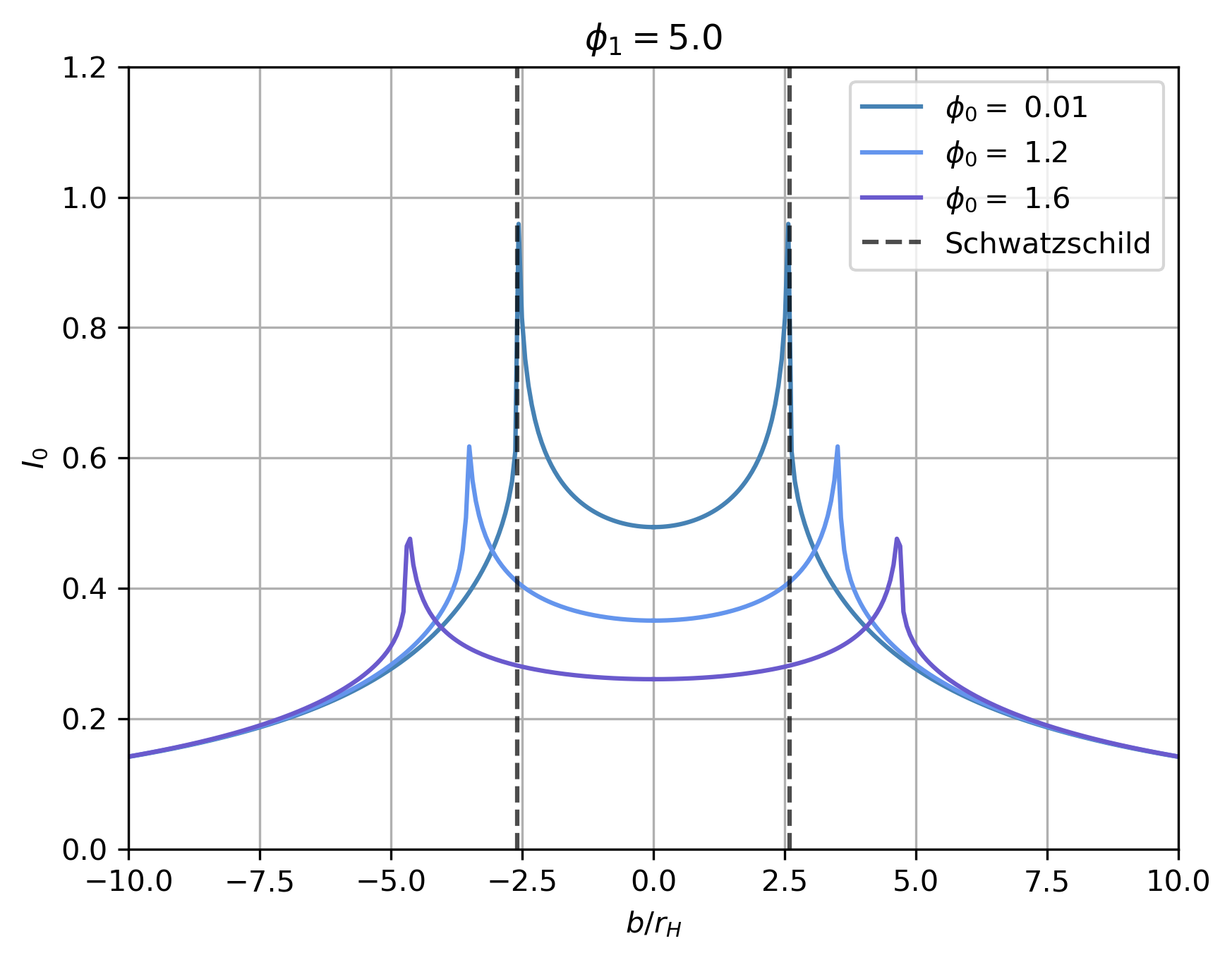}}
    \caption{Observed specific intensity $I_0$ as function of the impact parameter for different values of $\phi_0$ for the SHBHs with $\phi_1 = 0.5, 1.0, 2.0, 3.0$ and $5.0$. The dashed black lines indicate the radius of the shadow for a Schwarzschild black hole.}
    \label{fig:shadow_sizes}
    \end{figure*}

\subsection{Shadow and Photon Rings for Spherical Accretion Flow}

    As mentioned in the introducation, to  investigate the shadow, we will consider a static, spherically symmetric, optically, and geometrically thin accretion flow in the surroundings of the SHBH. The observed specific intensity $I(\nu_0)$ is defined as the integral of the specific emissivity $j(\nu_e)$ along the photon trajectory $\gamma$~\cite{Zeng:2020dco, Jaroszynski:1997bw, Bambi:2013nla}
\begin{equation}
I(\nu_0)  = \int_\gamma g^3 j(\nu_e) d\ell_\text{prop},
 \end{equation}
where $\nu_0$ and $\nu_e$ are the observed and emitted photon frequencies; respectively. The redshift factor, $g$, is the ratio between these frequencies and is written in terms of the metric tensor as
\begin{equation}
g=\frac{\nu_0}{\nu_e} = \sqrt{-g_{tt}}.
\end{equation}
The proper length, $d\ell_\text{prop}$, can be computed from the line element \eqref{eq:metric} taking $dt=d\theta =0$ and $\theta = \frac{\pi}{2}$,
\begin{equation}
d\ell_\text{prop} = \sqrt{g_{rr} + r^2 \left( \frac{d\varphi}{dr} \right)^2} dr,
\end{equation}
where the derivative inside the square root comes from the Lagrangian \eqref{eq:Lagrangian} and equations \eqref{eq:ConservedQuantities} and \eqref{eq:impactParameter} as
\begin{equation}
\left(\frac{d\varphi}{dr}\right)^2 = -\frac{g_{rr}}{r^2} \left( \frac{b^2 g_{tt}}{r^2 + b^2 g_{tt}}\right). 
\end{equation}

	Usually,  the emissivity per unit volume (measured in the static frame) for the static spherically symmetric accretion flow is modeled as $j(\nu_e) \propto \frac{1}{r^2}$ for monochromatic radiation emitted with frequency $\nu_e$. Hence, using these results, the specific intensity takes the form
\begin{equation}
I(\nu_0)  = \int_\gamma \frac{\sqrt{-g_{tt}^3 g_{rr}}}{r^2}\sqrt{1 - \frac{b^2 g_{tt}}{r^2 + b^2 g_{tt}}} dr.
\end{equation}
    \begin{figure*}
    \centering
     \includegraphics[scale=0.5]{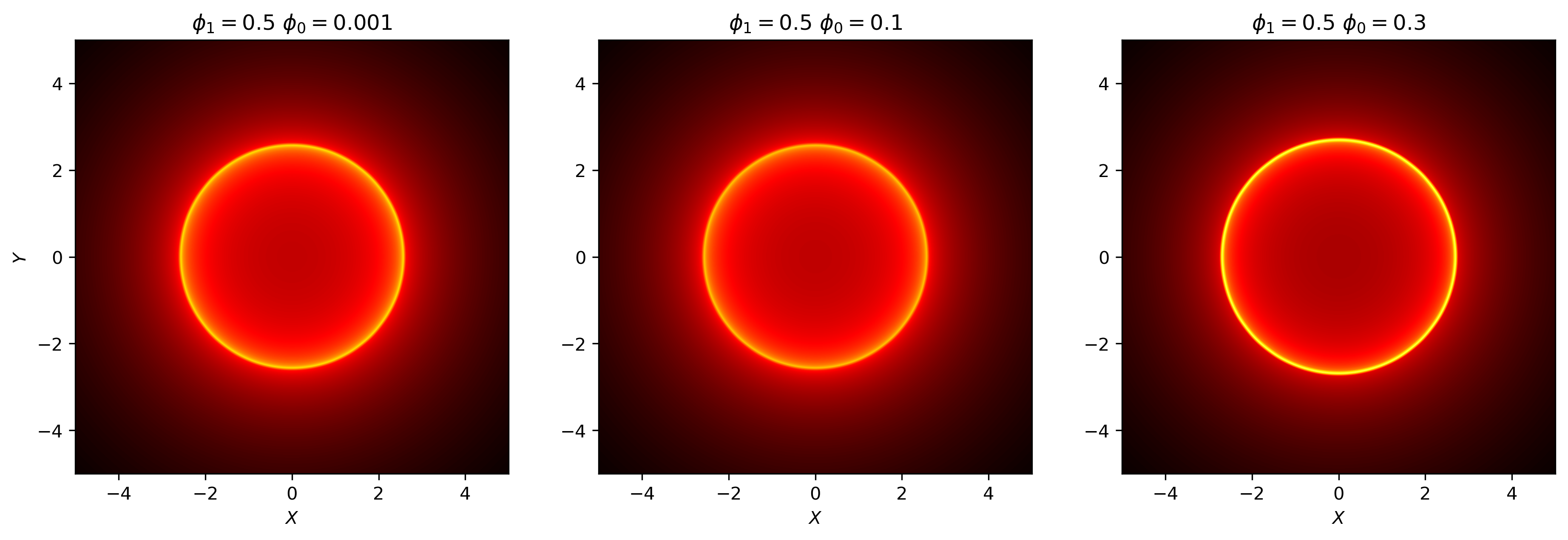}\\
     \includegraphics[scale=0.5]{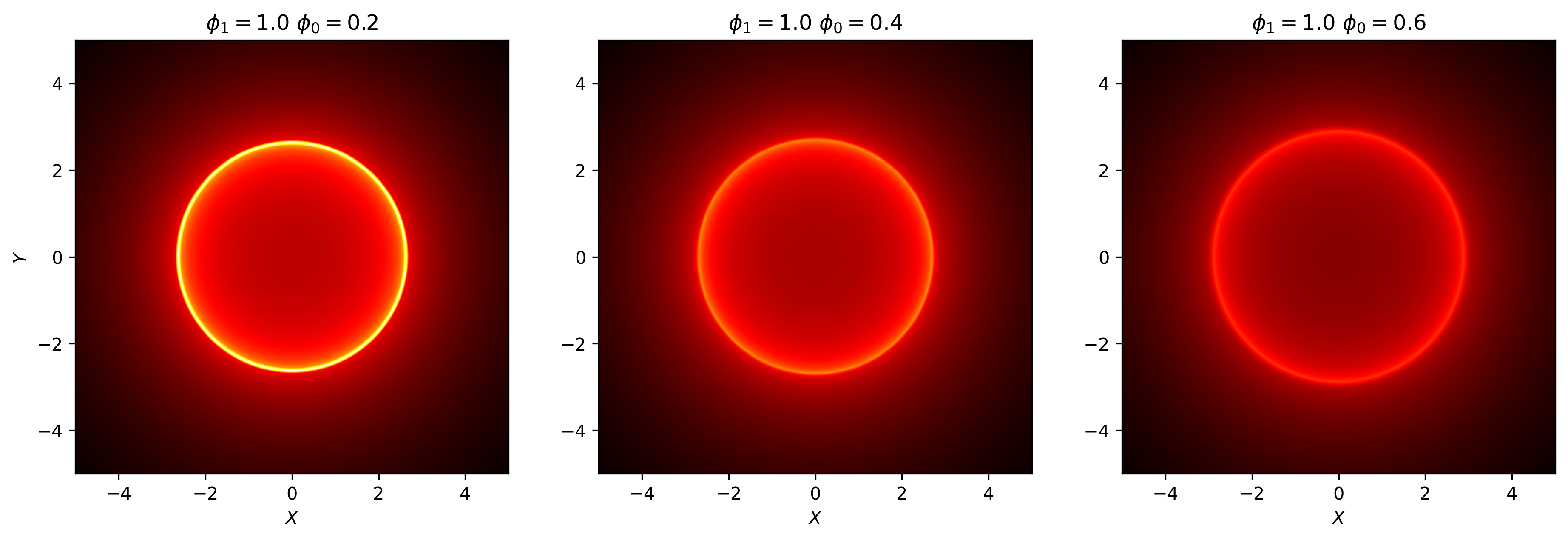}\\
     \includegraphics[scale=0.5]{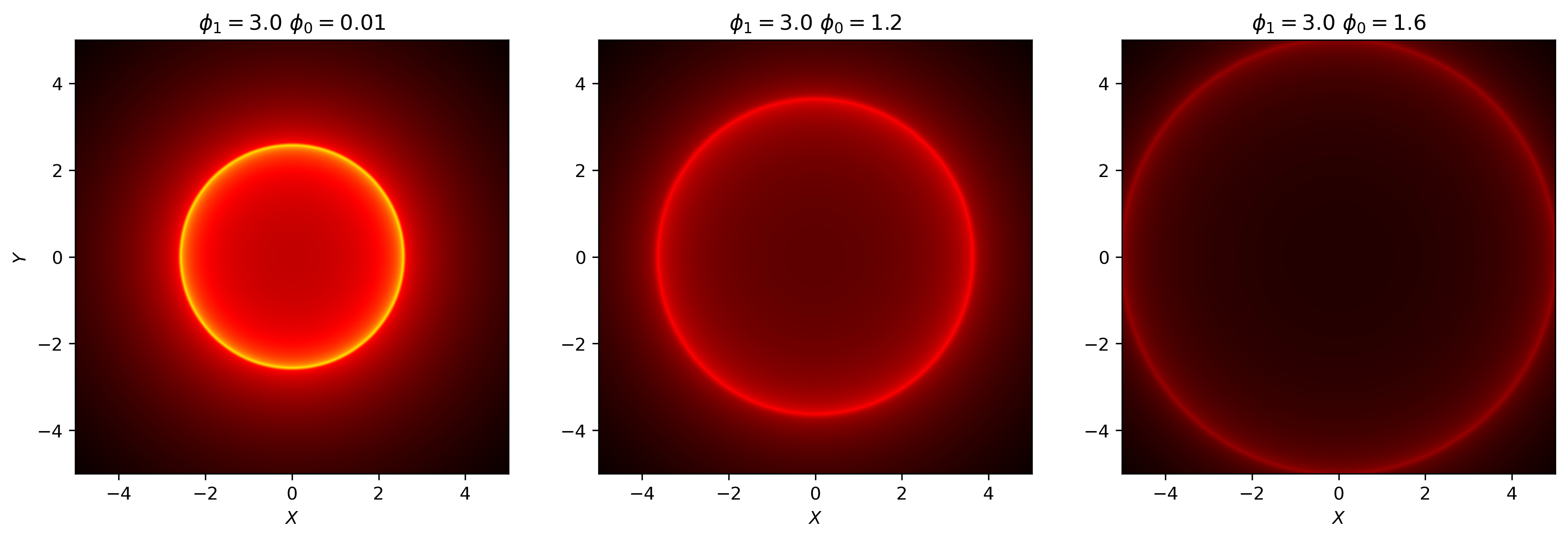}\\
     \includegraphics[scale=0.5]{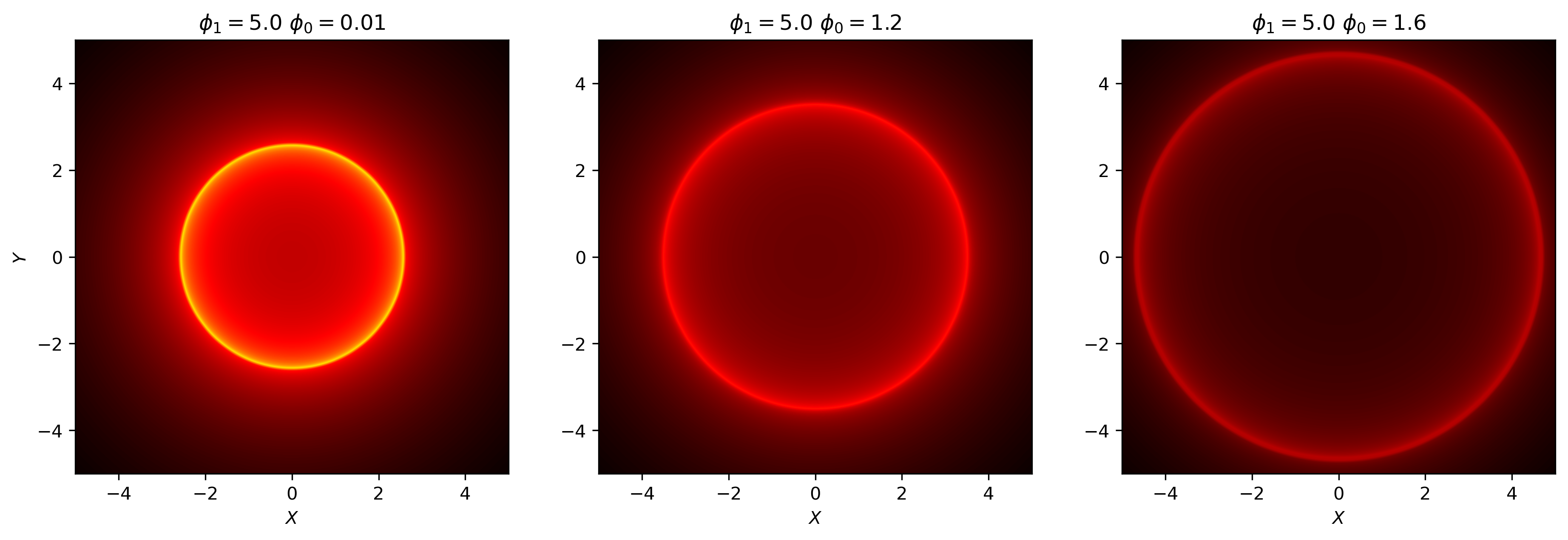}
    \caption{Shadows and photon rings of the SHBH  in the case of static spherical accretion. To illustrate the behavior, we chose the solutions with $\phi_0= 0.001, 0.1$ and $0.3$; $\phi_0= 0.02, 0.4$ and $0.6$ for $\phi_1 = 0.5$ and $\phi_1 = 1.0$ respectively; and $\phi_0= 0.01, 1.2$ and $1.6$ for both $\phi_1 = 3.0$ and $\phi_1 = 5.0$.}
    \label{fig:shadows}
    \end{figure*}
    
Figure~\ref{fig:shadow_sizes} shows the profile of the specific intensity in terms of the impact parameter for different values of $\phi_1 = 0.5$, $1.0$, $2.0$, $3.0$, and $5.0$. Note that, for any of the values of $\phi_0$, the intensity increases from its value at $b/r_H=0$ to reach a peak at the value $b(r_\text{ps})/r_H$, corresponding to the edge of the shadow; then, it drops for greater values of the impact parameters. One can understand this general behavior by noting that, for $b<b(r_\text{ps})$, the radiation produced by the accretion flow is mostly absorbed by the black hole. Meanwhile, radiation at $b(r_\text{ps})/r_H$ is moving in the photon sphere (unstable circular orbit) so that it may orbit the black hole many times before escaping to reach the observer. In fact, from a theoretical point of view, the intensity at this point should be infinity. In the region with $b>b(r_\text{ps})$, the observed intensity is produced only by refracted radiation, and therefore it rapidly decreases when increasing the impact parameter. 

	From Fig.~\ref{fig:shadow_sizes}, it is also clear that increasing the parameter $\phi_0$ produces weaker intensity profiles. We can see this behavior easily by projecting the intensity profile in the 2-dimensional celestial plane of the observer, as seen in Fig.~\ref{fig:shadows}. Note that the bright ring around the shadow (darker region at the center) corresponds to the photon sphere. Note that the radius of these rings is large for greater values of $\phi_0$; see the third and fourth rows of Fig.~\ref{fig:shadows}. 

    \begin{figure}[t]
    \centering
    \mbox{
    \includegraphics[scale=0.2]{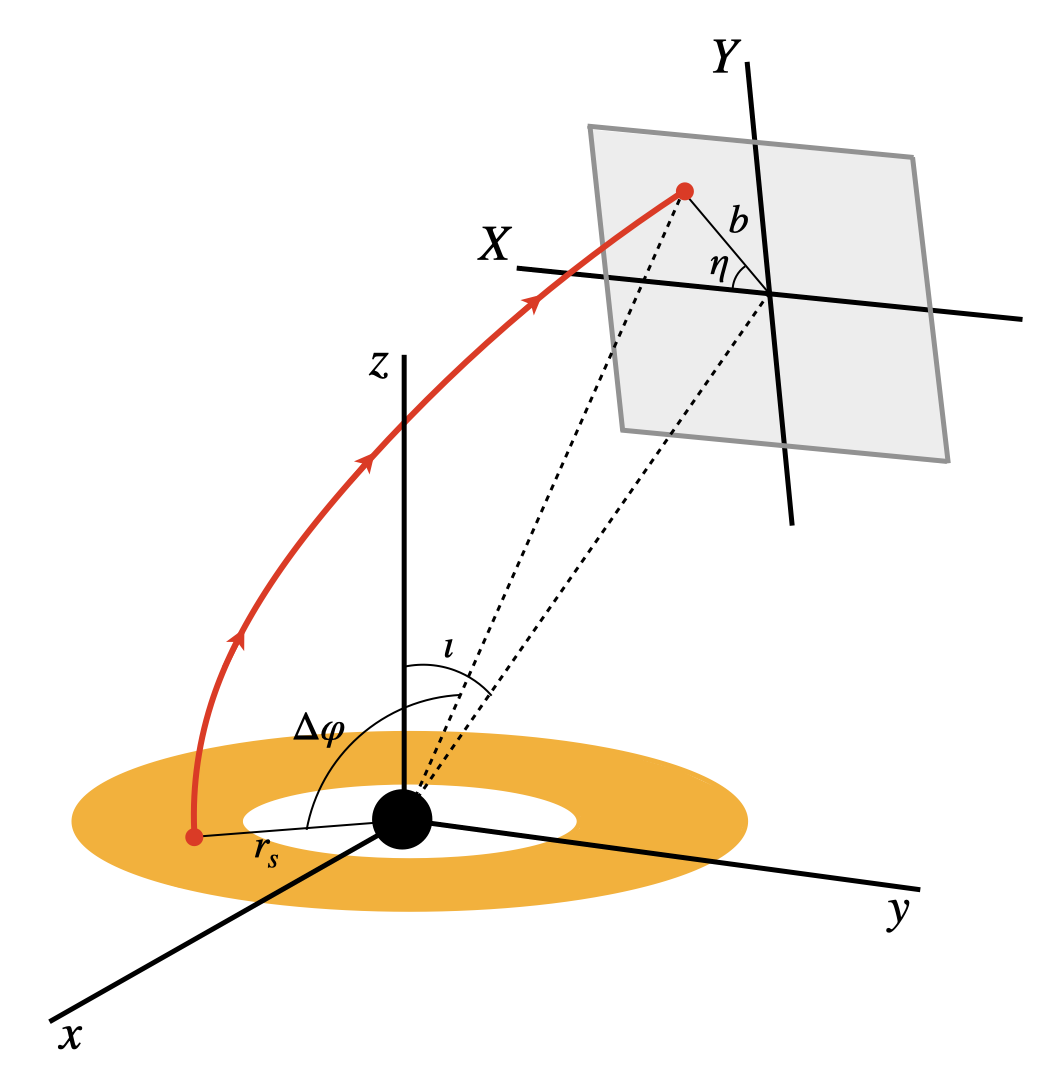}
     }
    \caption{Coordinate system at the image plane of the observer with Cartesian coordinate $(X,Y)$ and polar coordinates $(b,\eta)$.}
    \label{fig:imageplane}
    \end{figure}    

    \begin{figure*}[t]
    \centering
    \includegraphics[scale=0.45]{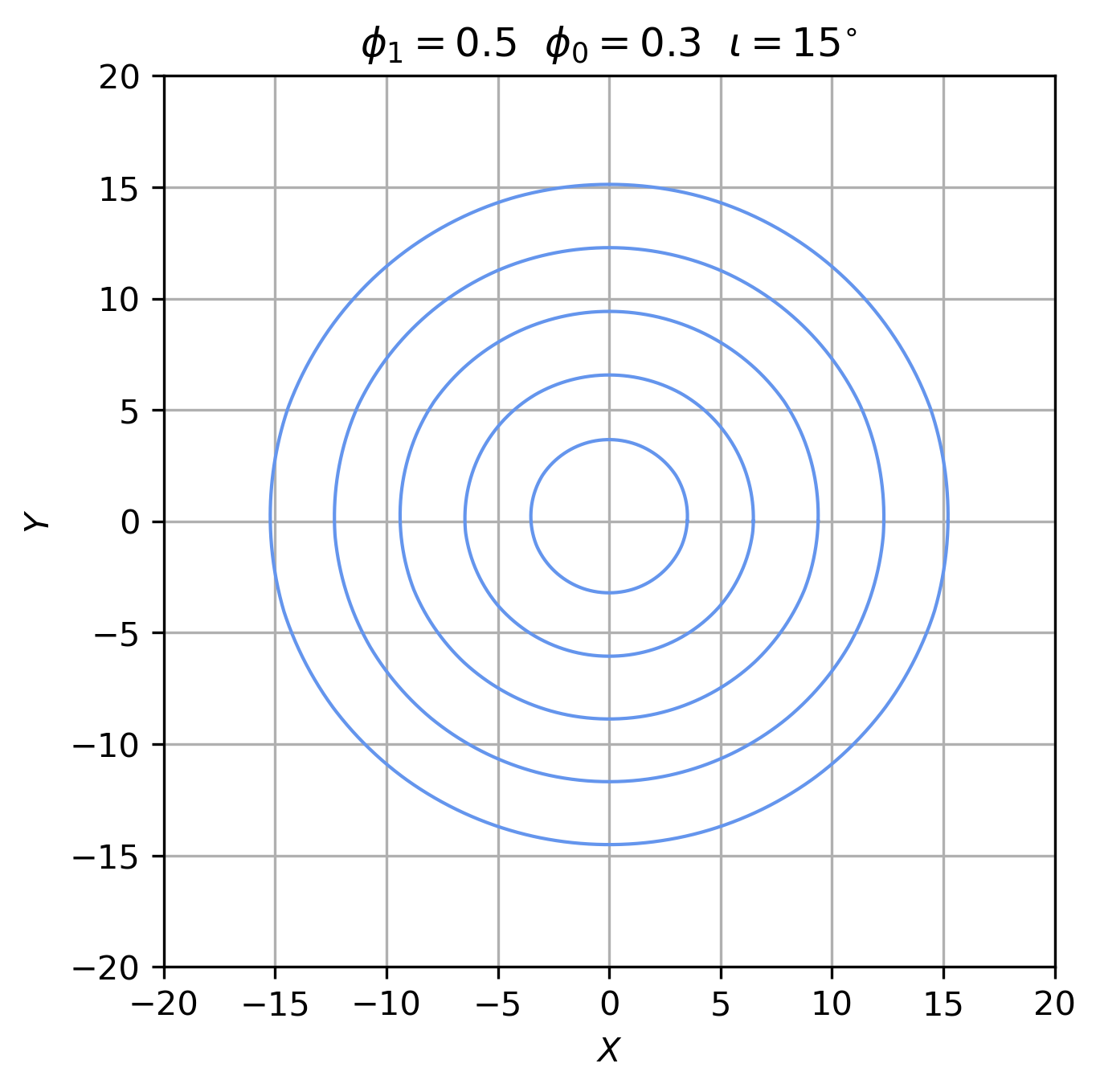}
    \includegraphics[scale=0.45]{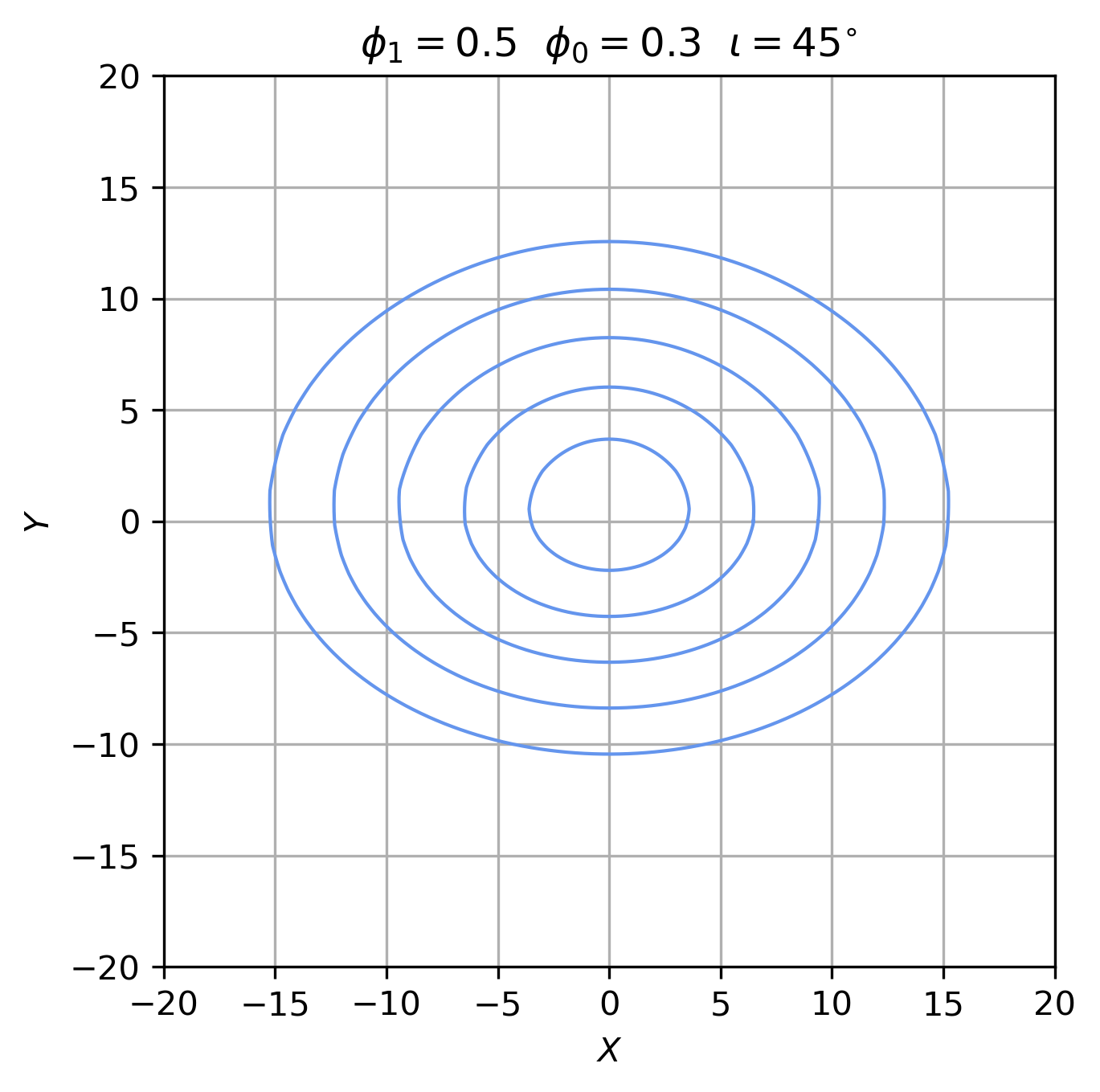}
    \includegraphics[scale=0.45]{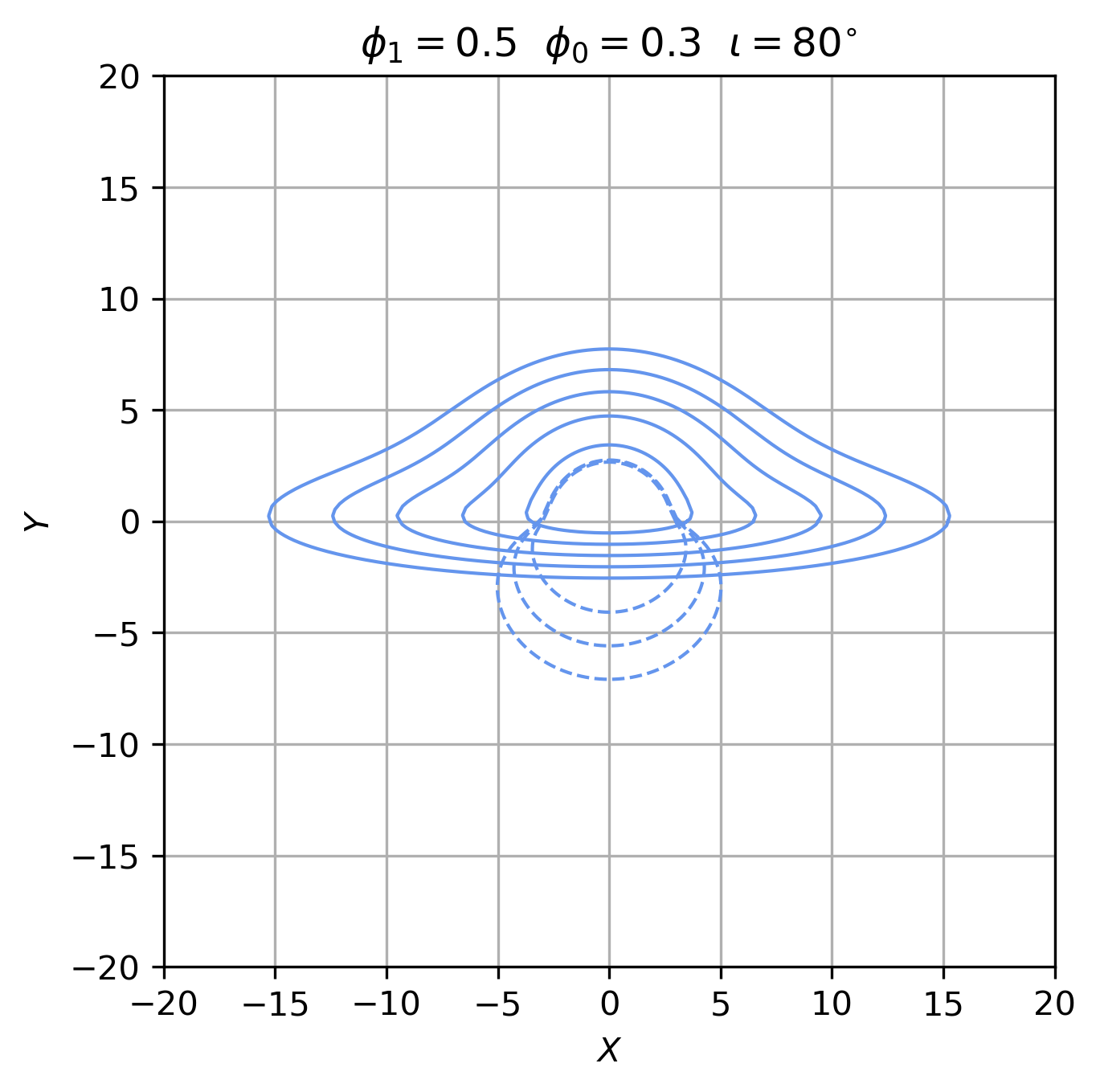}\\
    \includegraphics[scale=0.45]{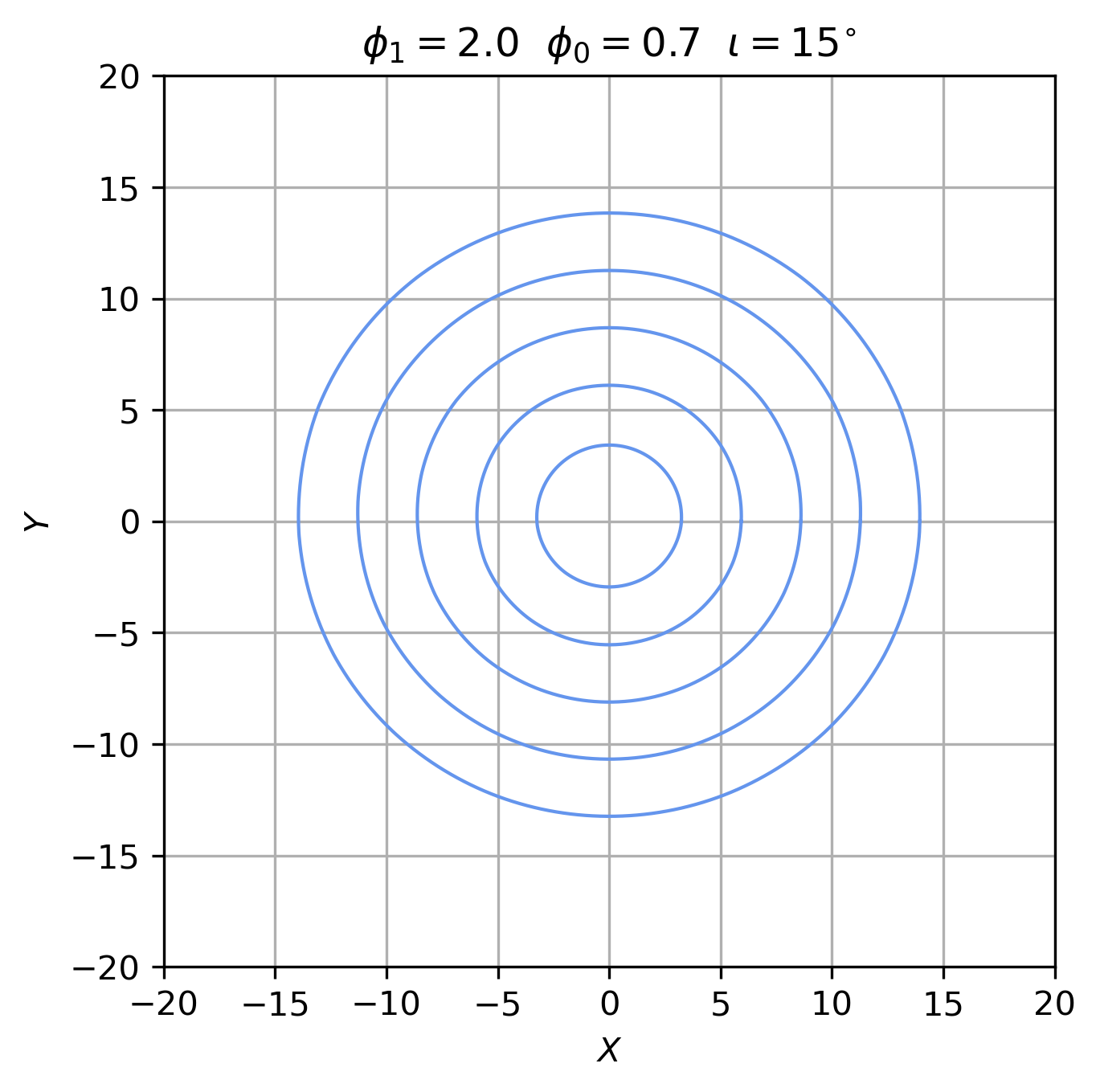}
    \includegraphics[scale=0.45]{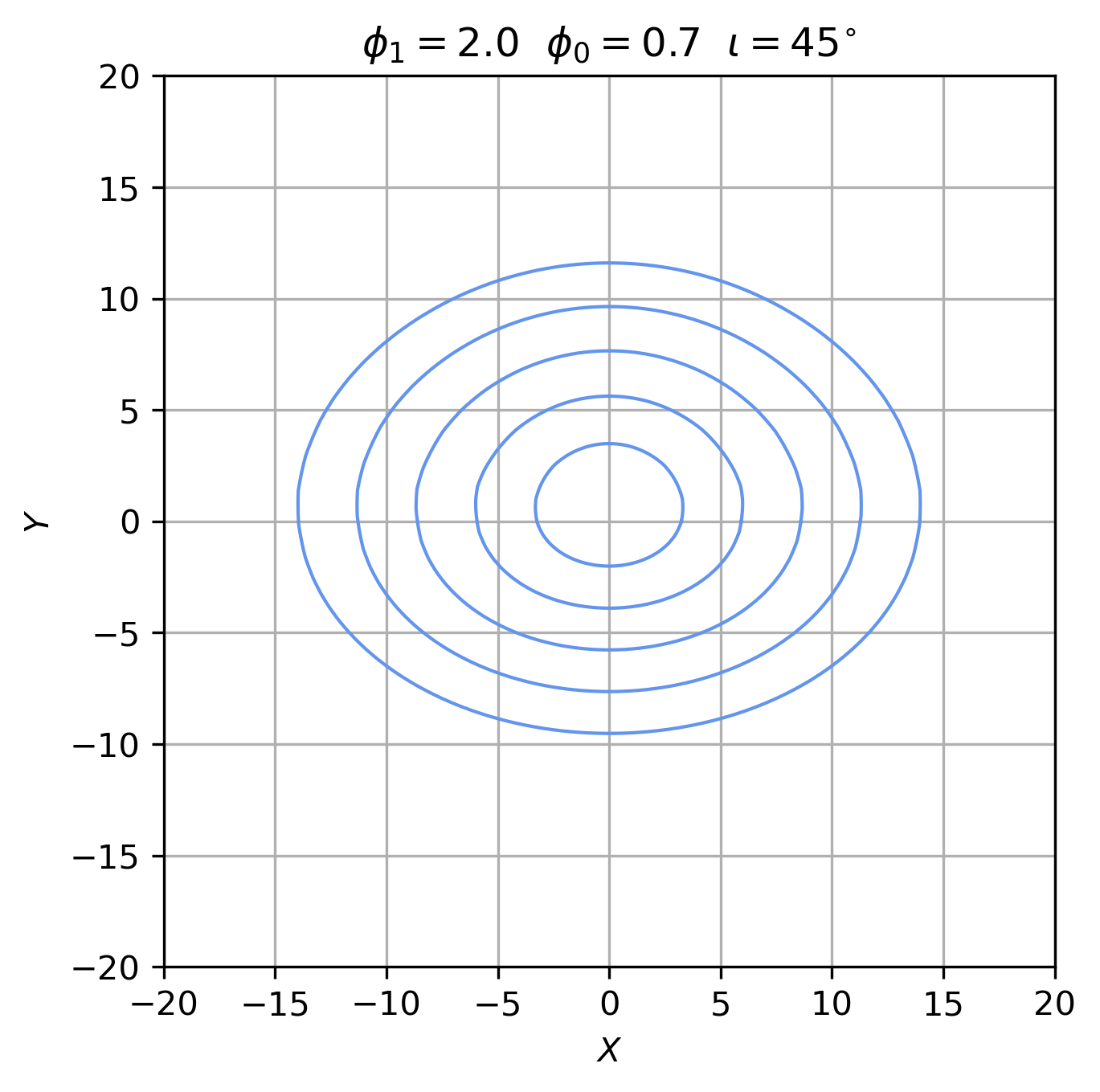}
    \includegraphics[scale=0.45]{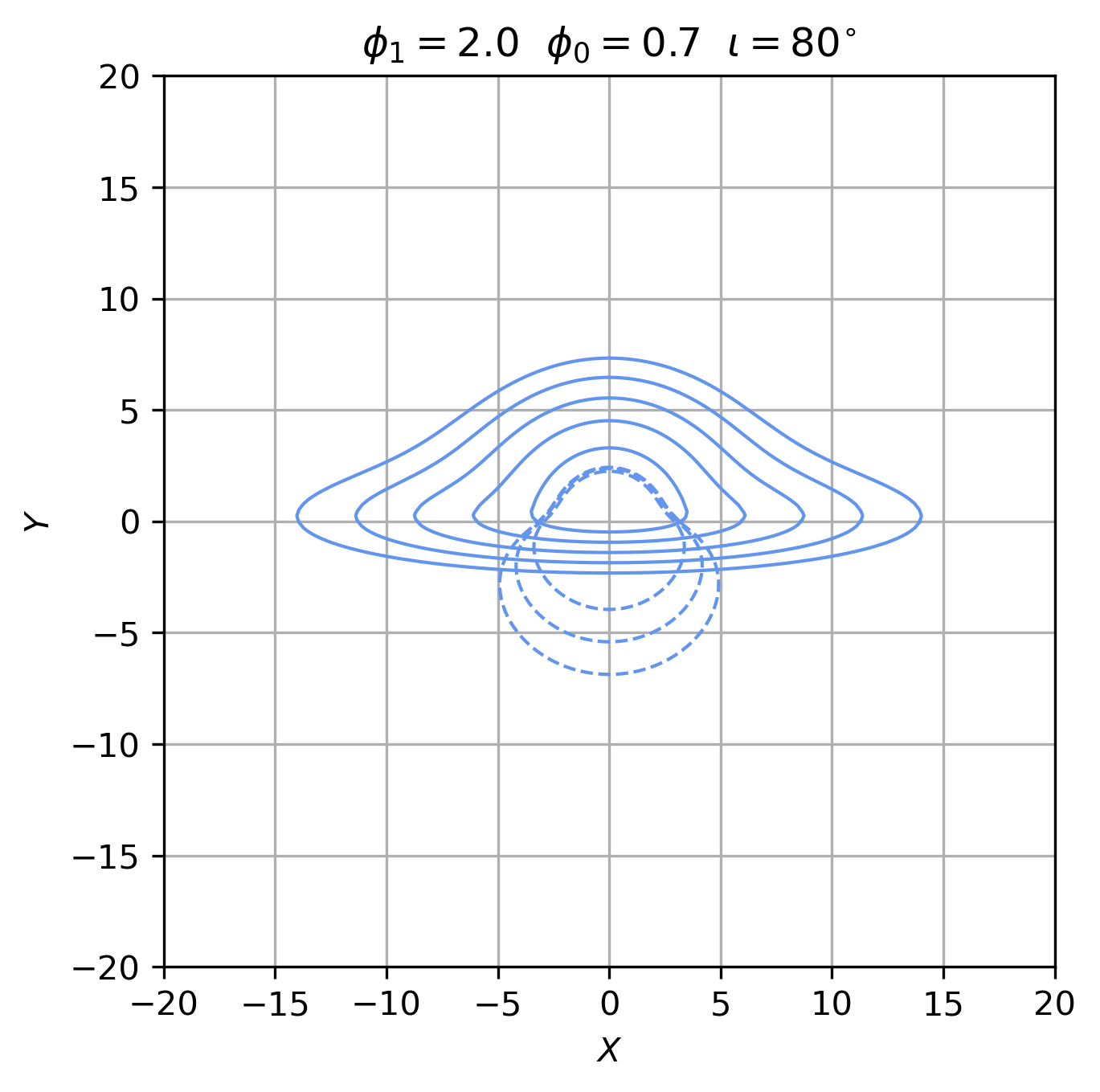}\\
    \includegraphics[scale=0.45]{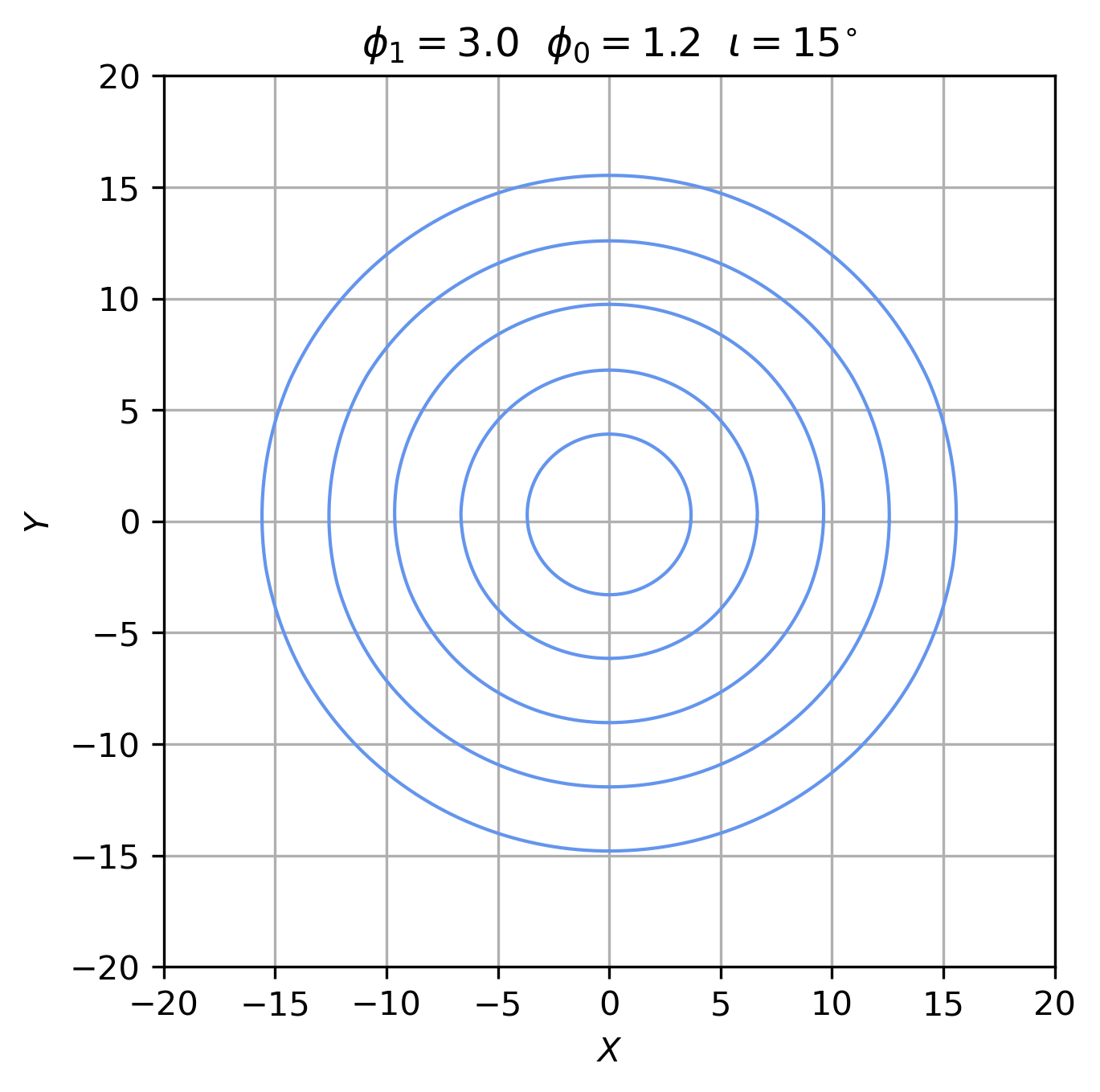}
    \includegraphics[scale=0.45]{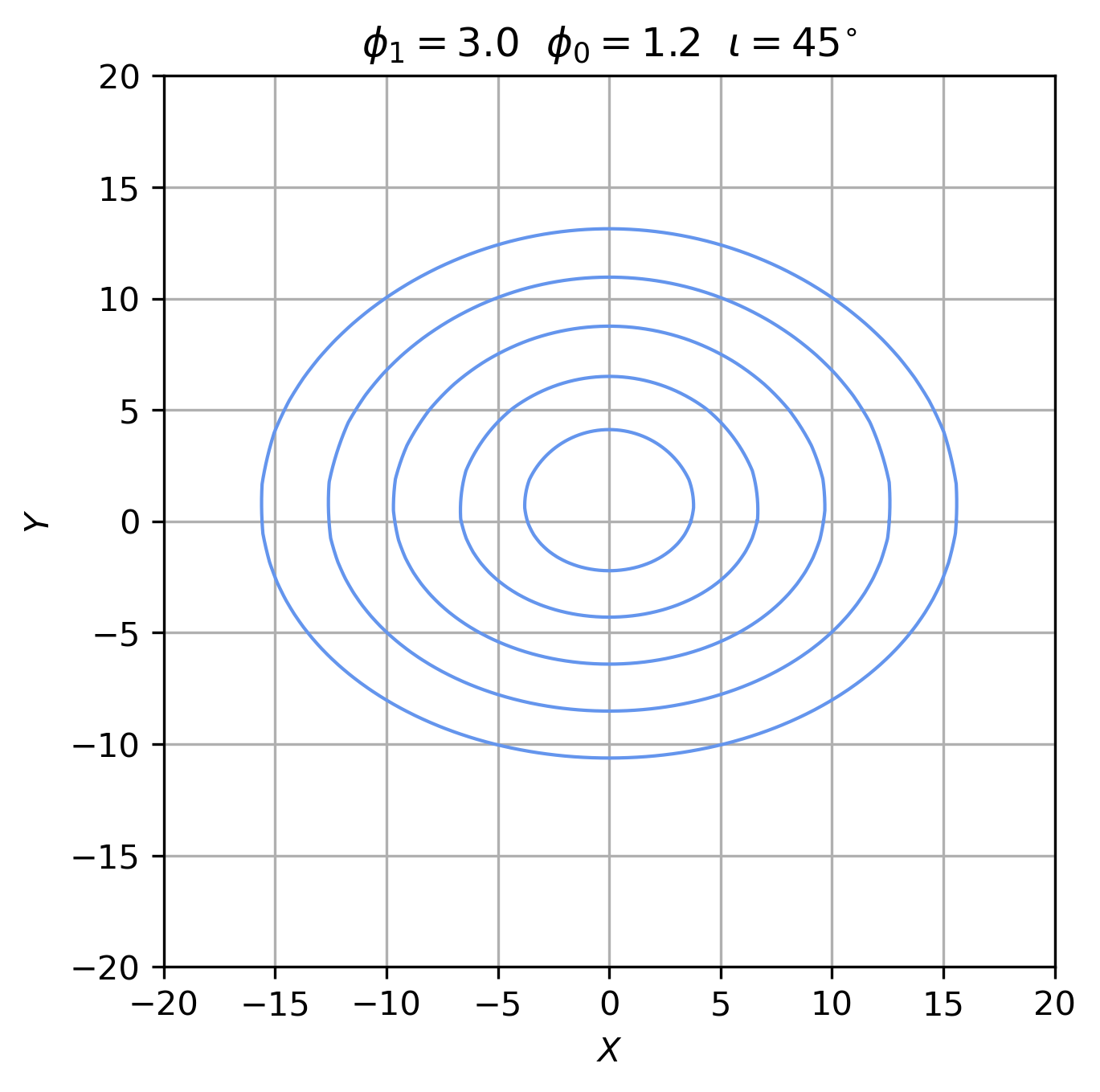}
    \includegraphics[scale=0.45]{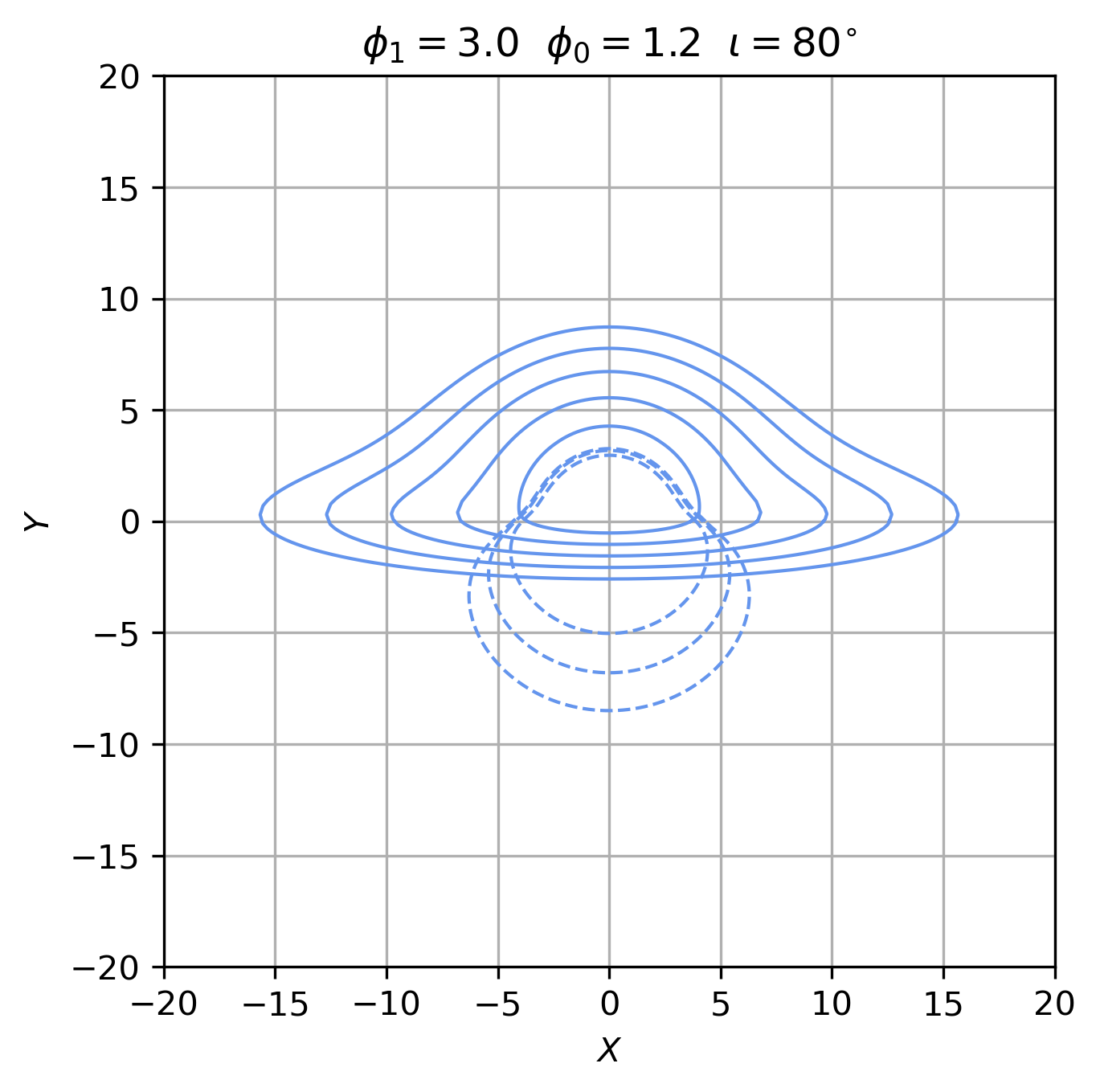}\\
    \includegraphics[scale=0.45]{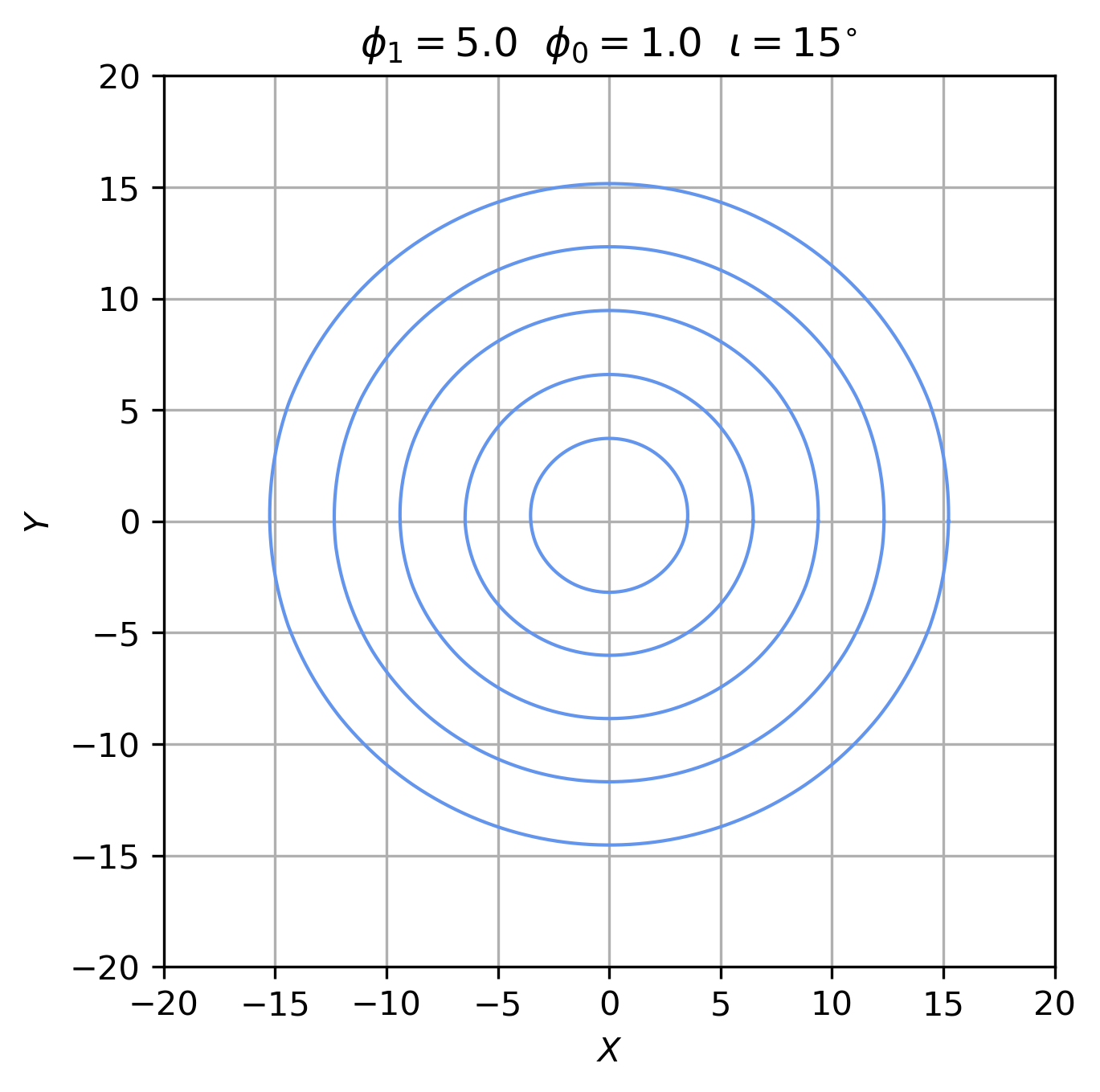}
    \includegraphics[scale=0.45]{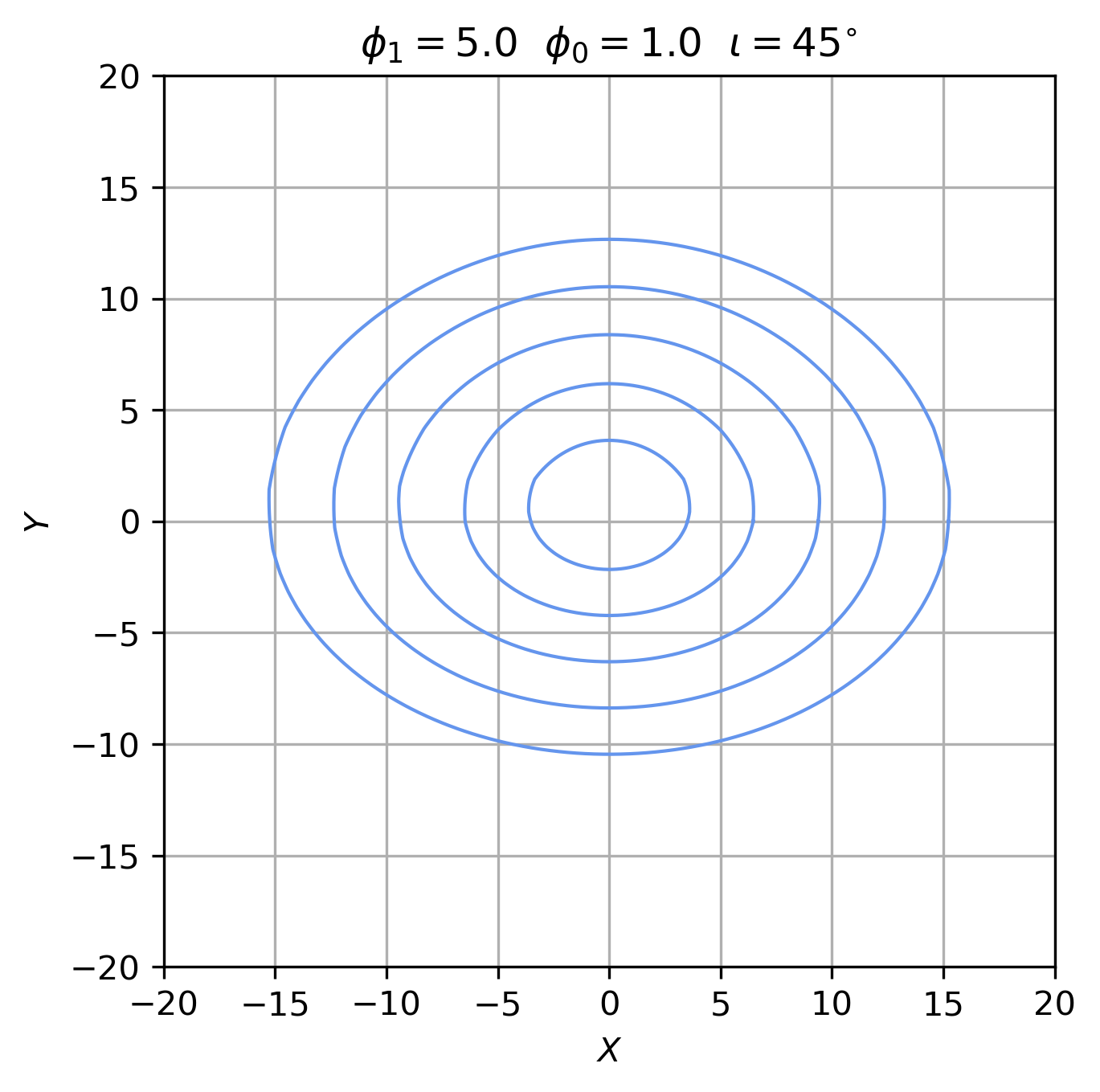}
    \includegraphics[scale=0.45]{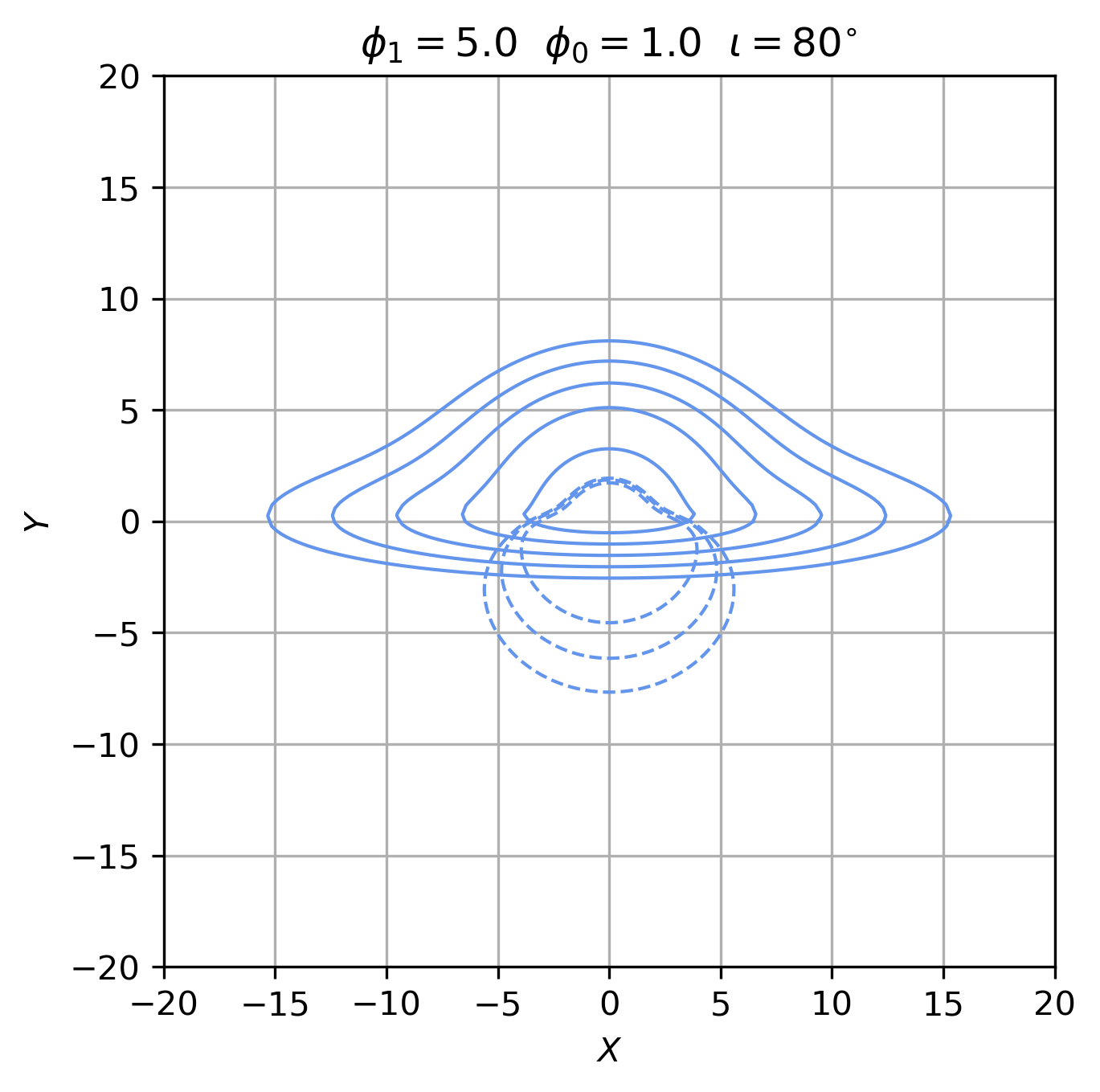}
    \caption{Direct images of a thin accretion disk surrounding the SHBH under different values of the parameters $\phi_1$ and $\phi_0$.  From the center of each panel outward, the profiles correspond to the stable circular orbit with radii $r = r_\text{ISCO}, 2r_\text{ISCO}, 3r_\text{ISCO}, 4r_\text{ISCO}$ and $5r_\text{ISCO}$ (isoradial curves). In the last column we include the secondary image ($n=2$ in Eq.~\eqref{eq:mainIntegral}) for $r = 2r_\text{ISCO}, 5r_\text{ISCO}$, and $10r_\text{ISCO}$. }
    \label{fig:accretionDiskProfiles}
    \end{figure*}

    \begin{figure*}[t]
    \centering
    \includegraphics[scale=0.44]{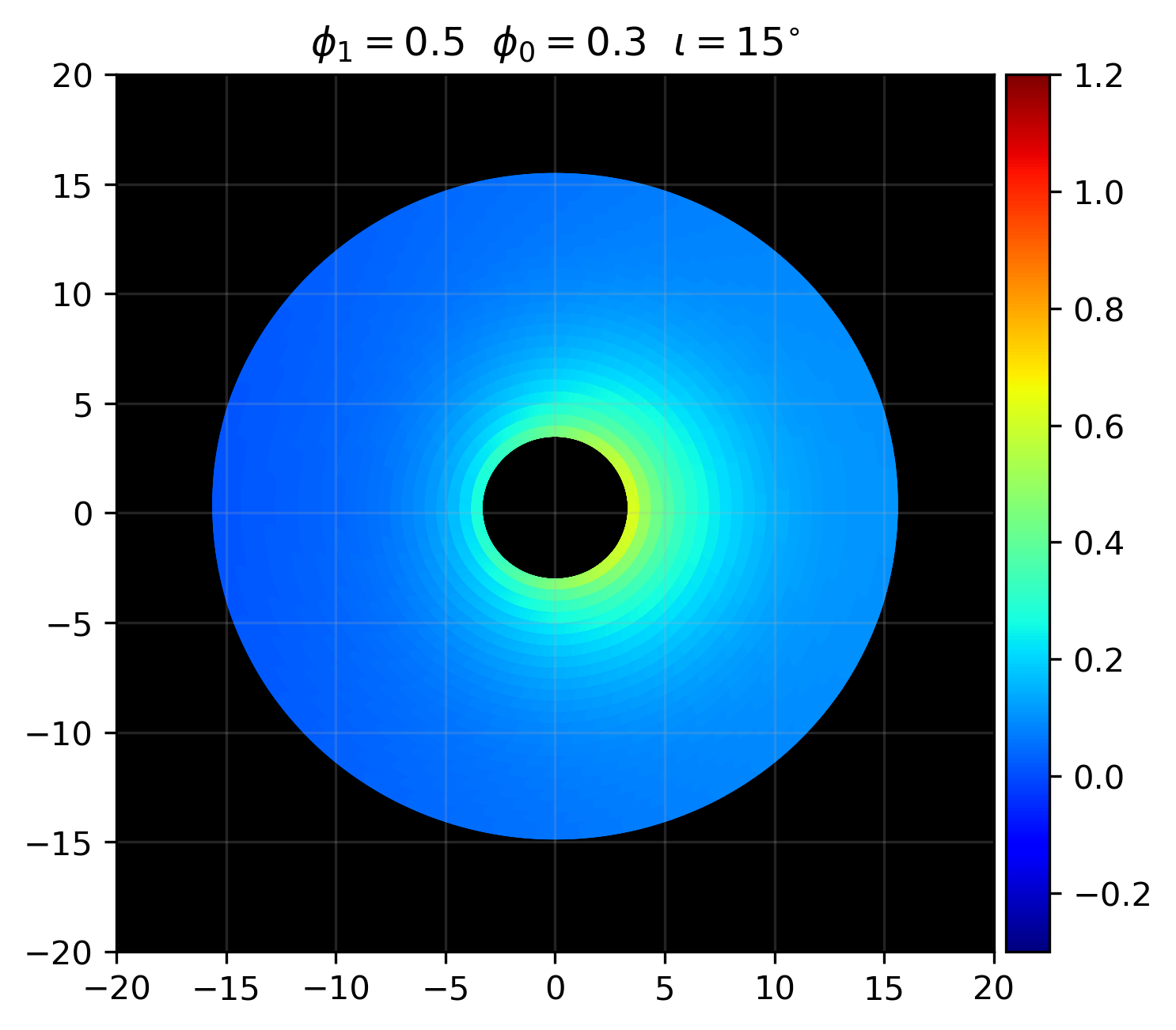}
    \includegraphics[scale=0.44]{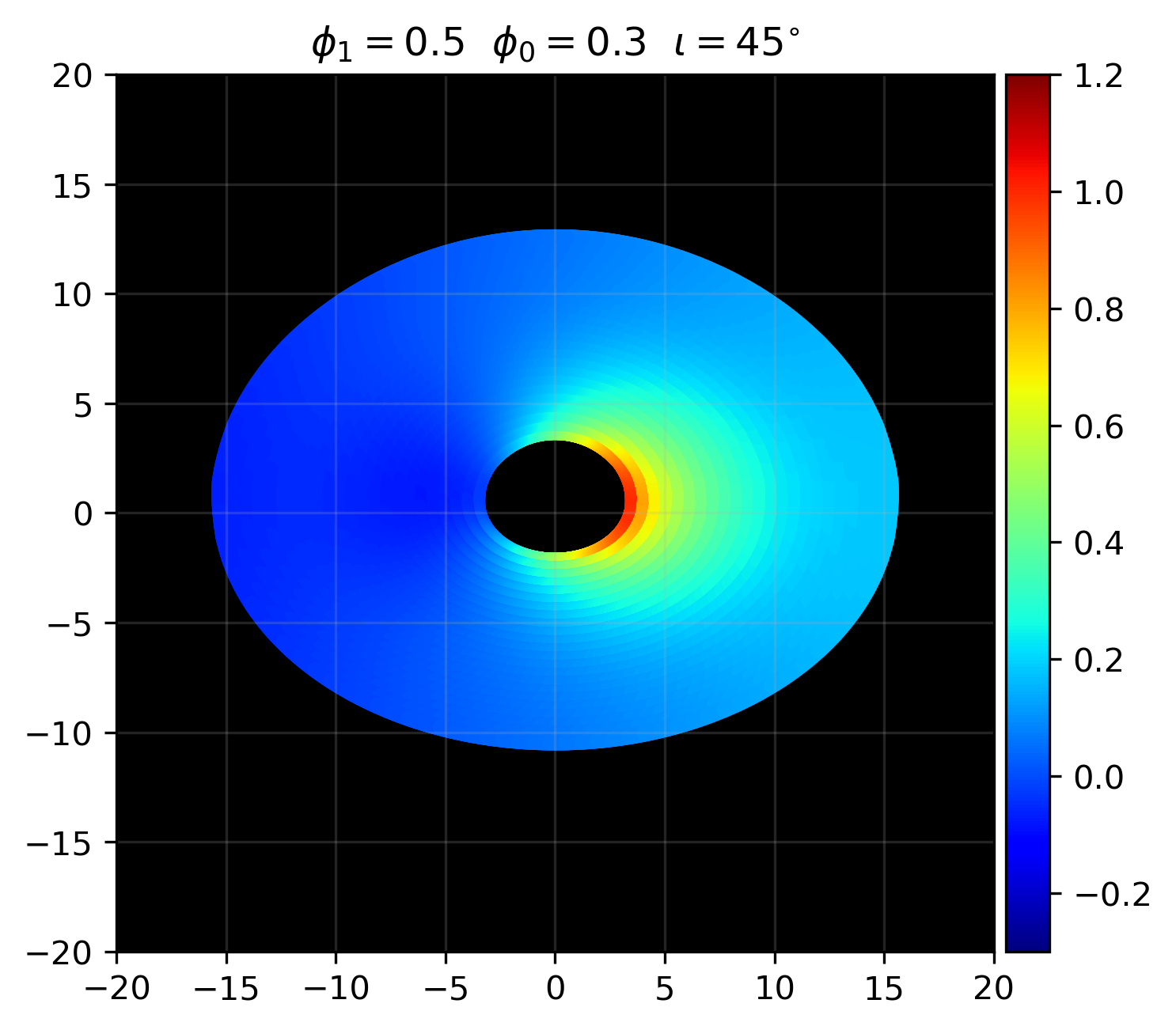}
    \includegraphics[scale=0.44]{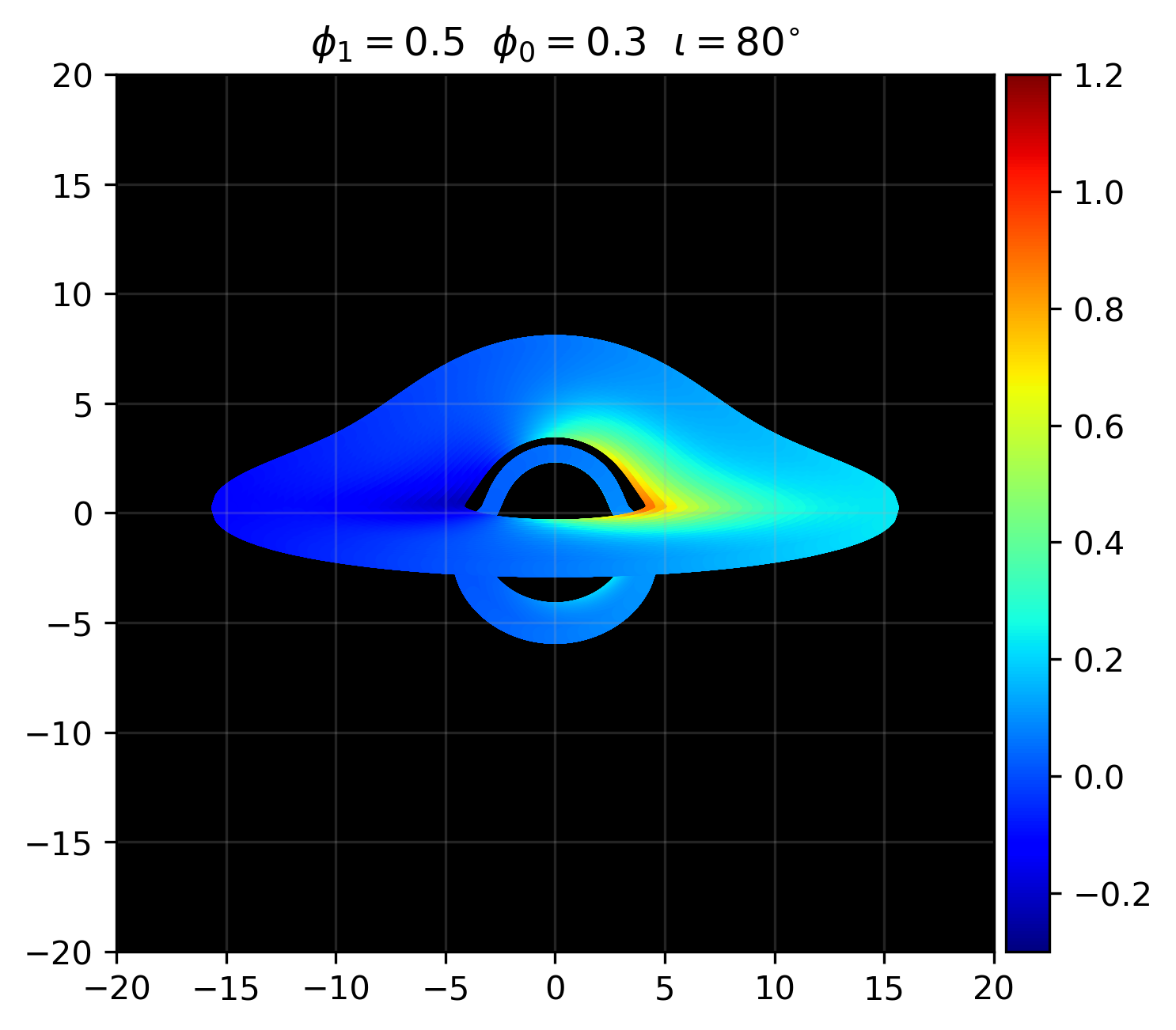}\\
    \includegraphics[scale=0.44]{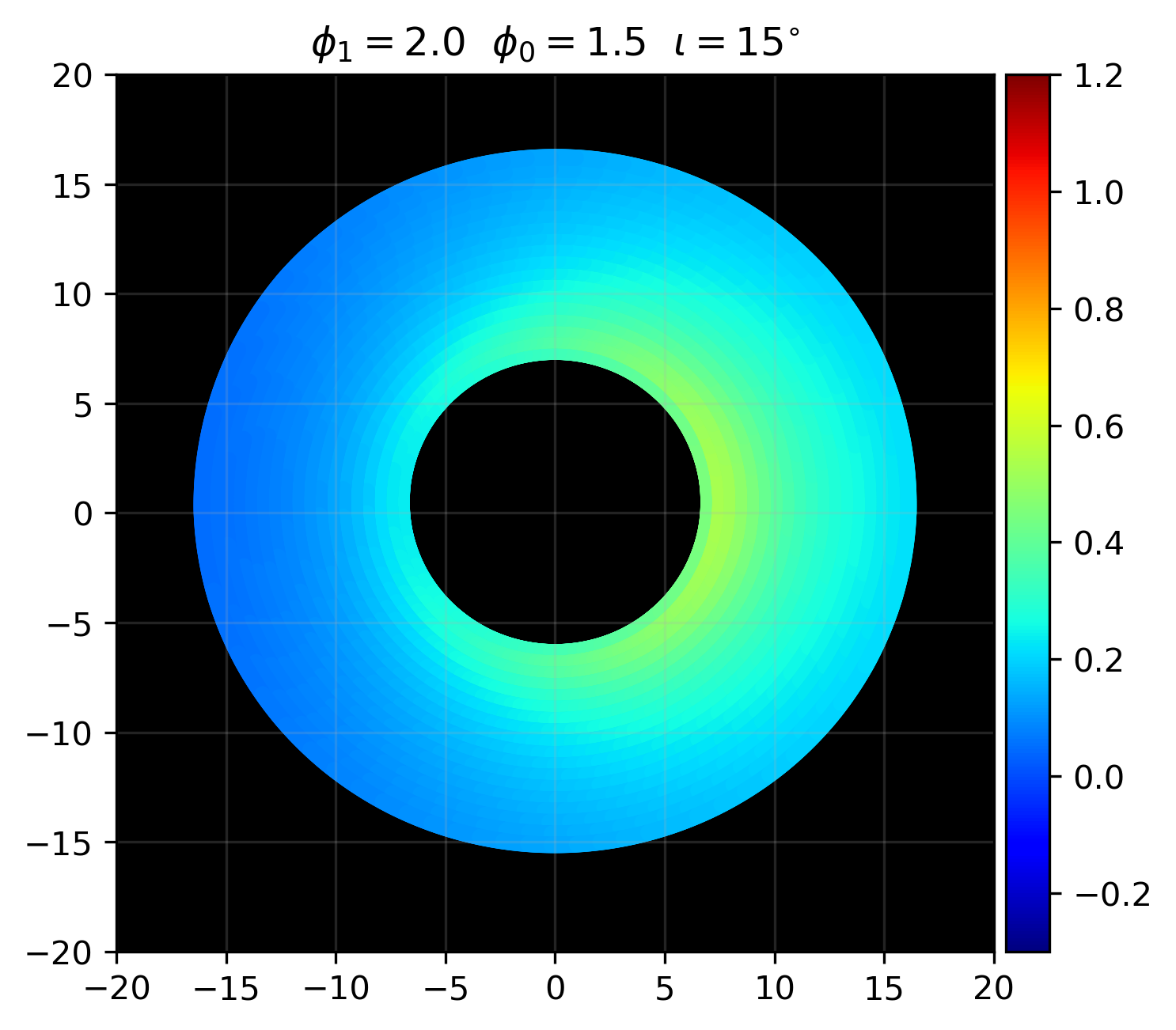}
    \includegraphics[scale=0.44]{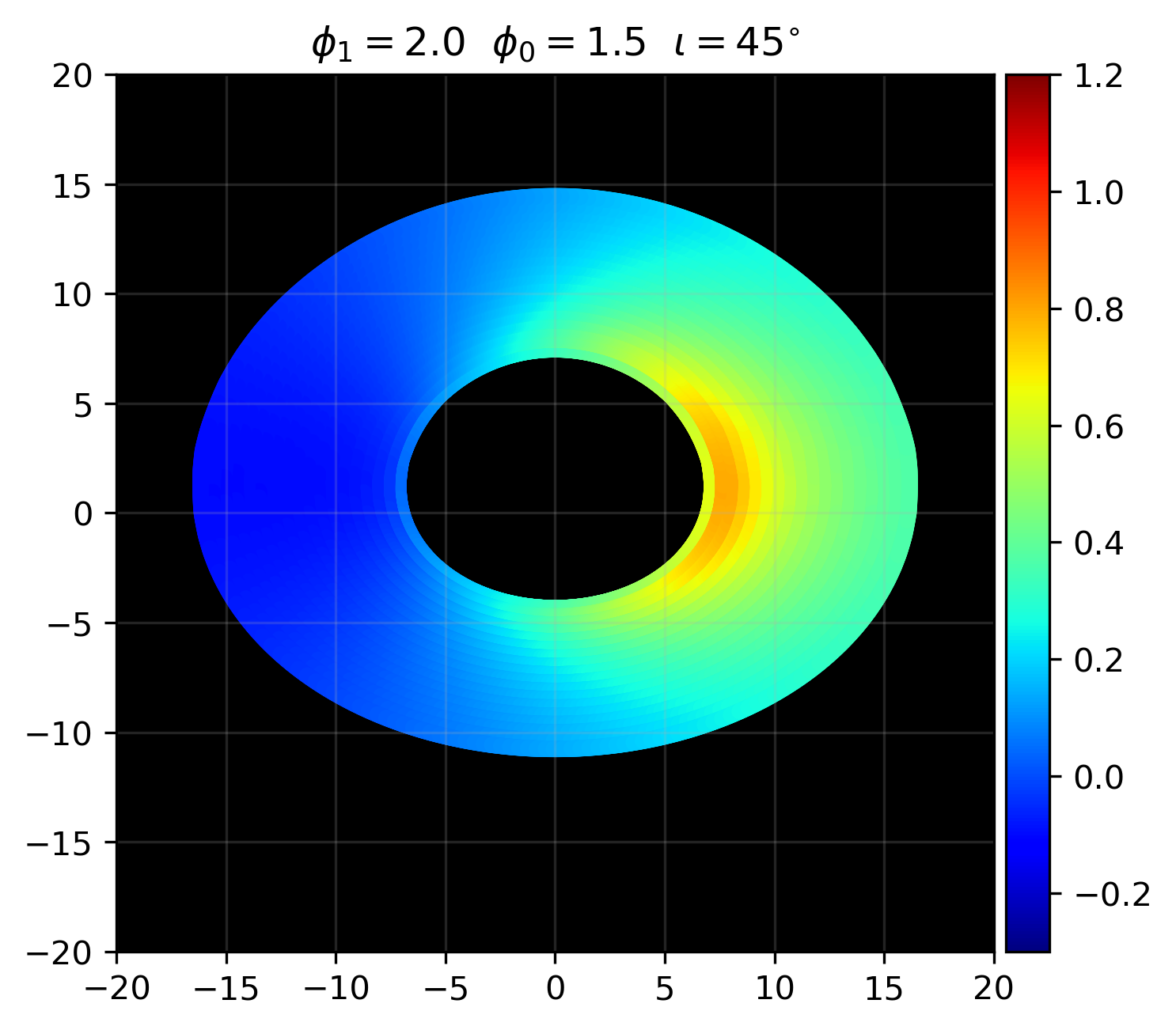}
    \includegraphics[scale=0.44]{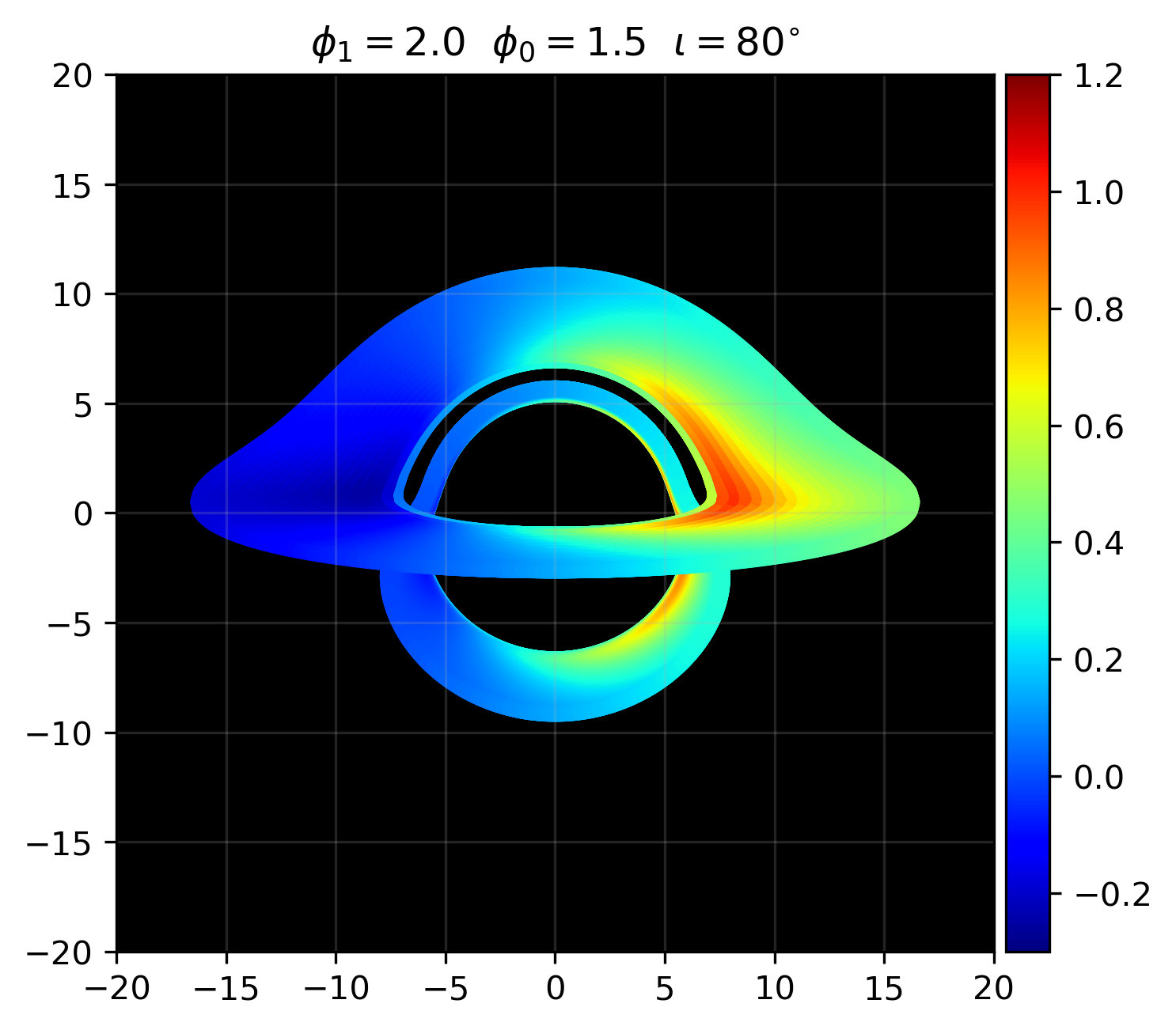}\\
    \includegraphics[scale=0.44]{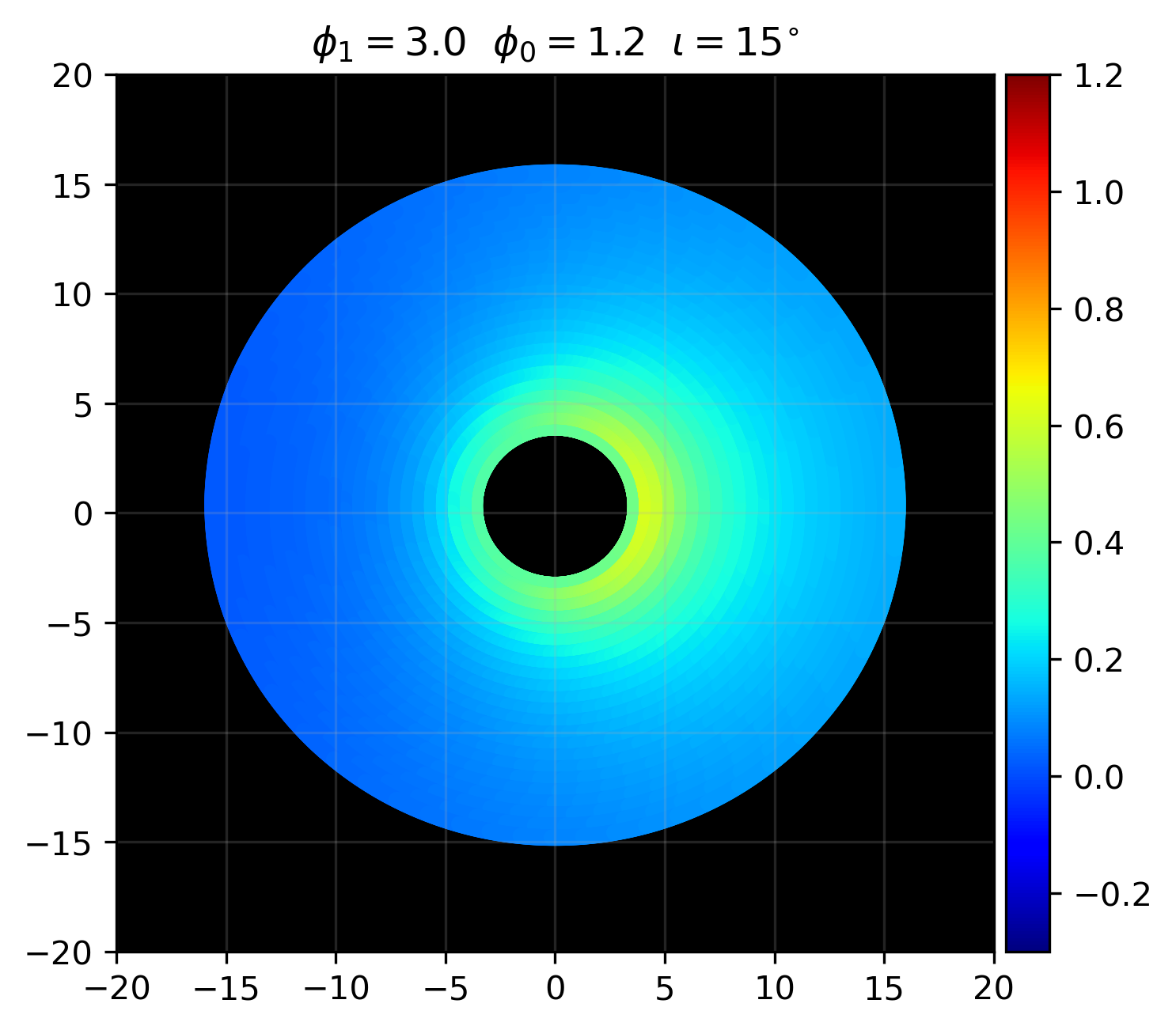}
    \includegraphics[scale=0.44]{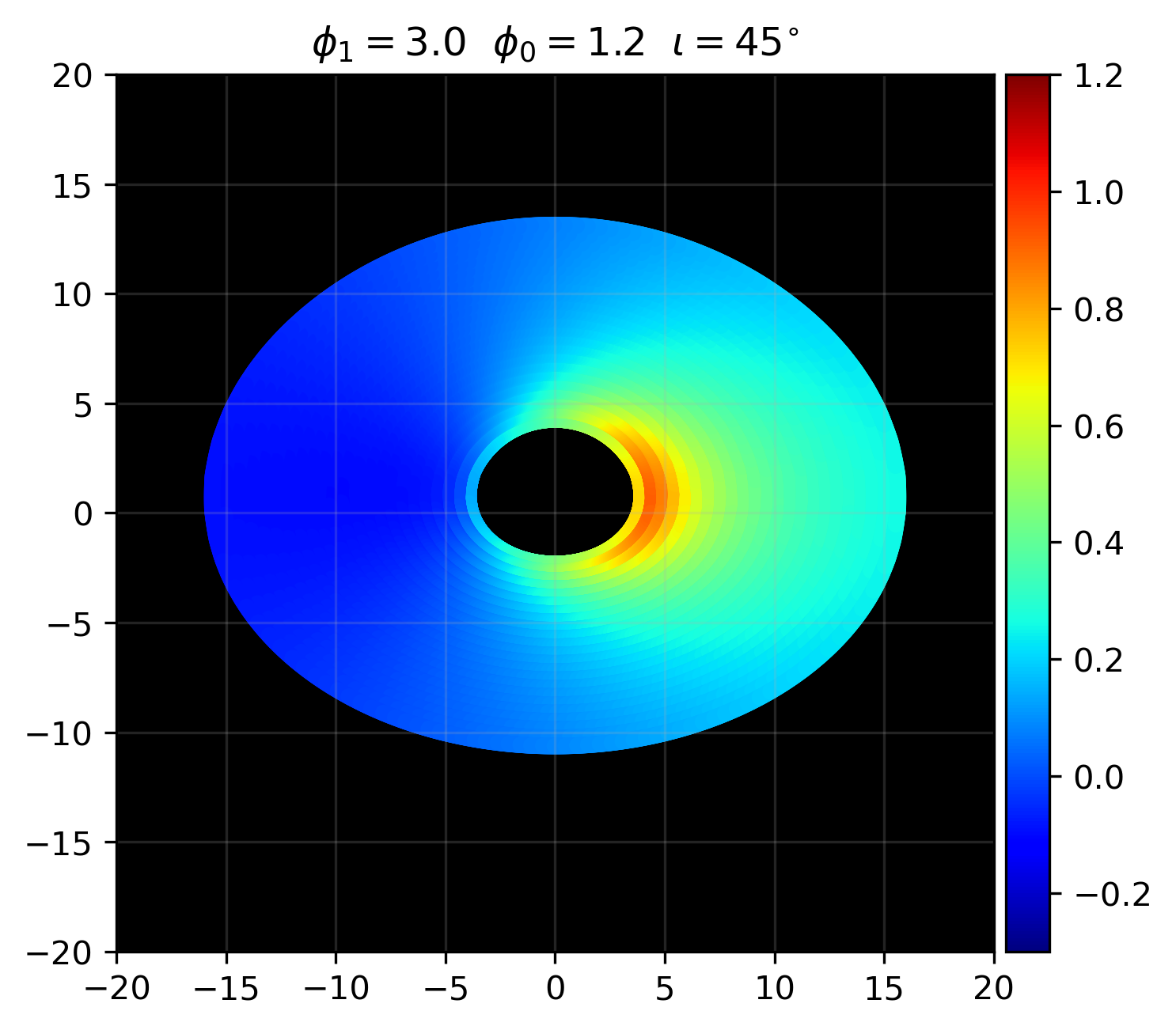}
    \includegraphics[scale=0.44]{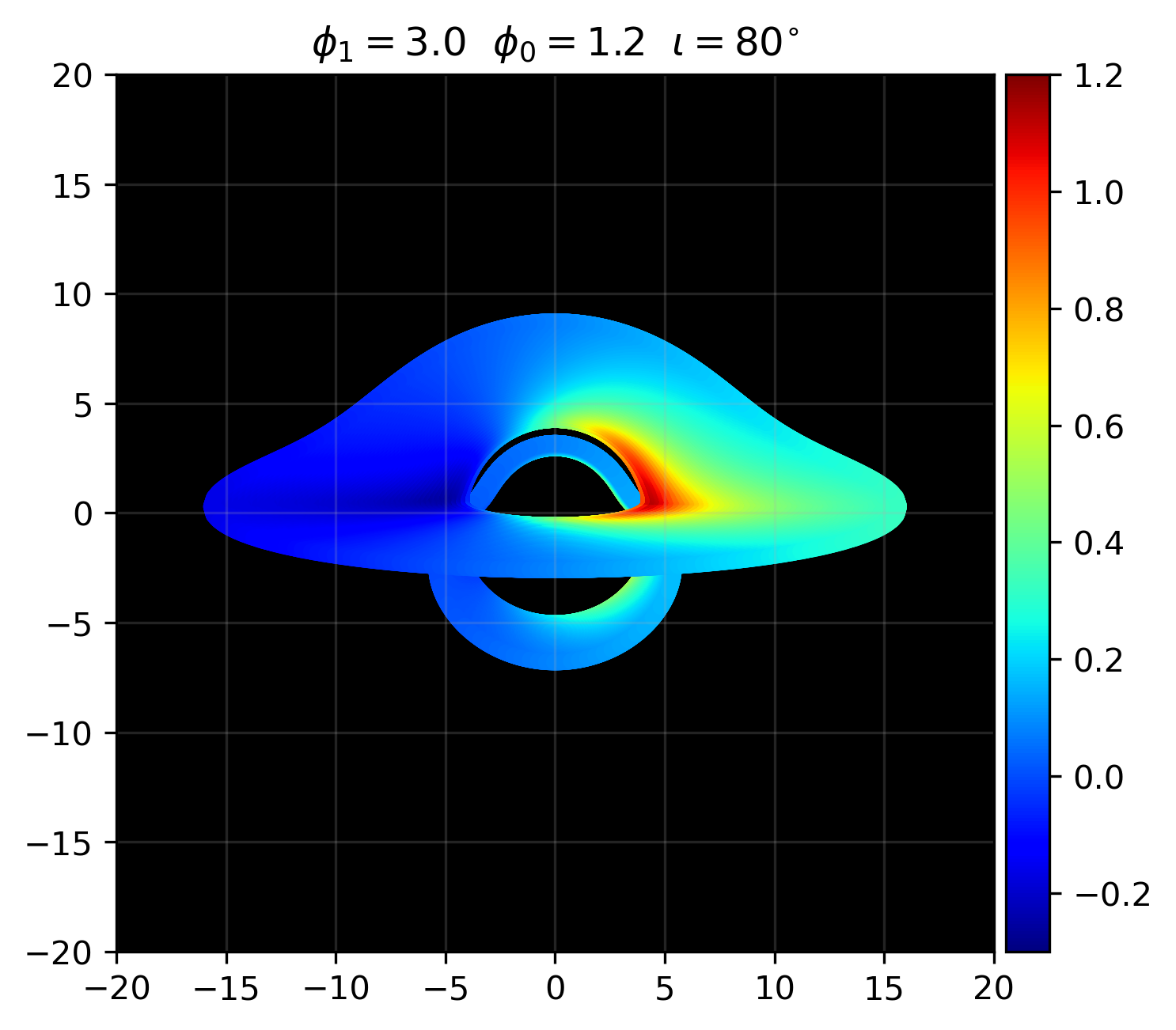}\\
    \includegraphics[scale=0.44]{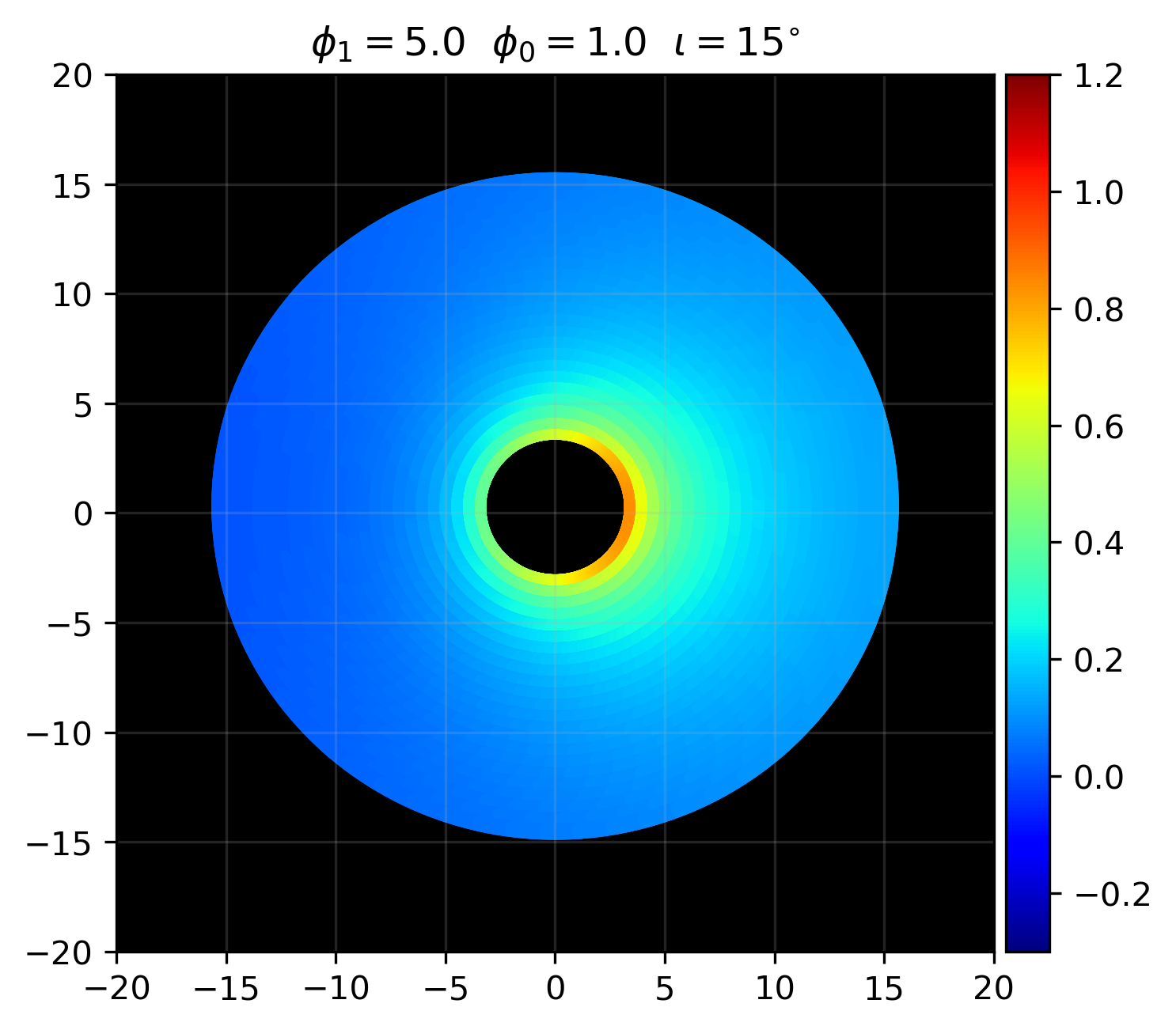}
    \includegraphics[scale=0.44]{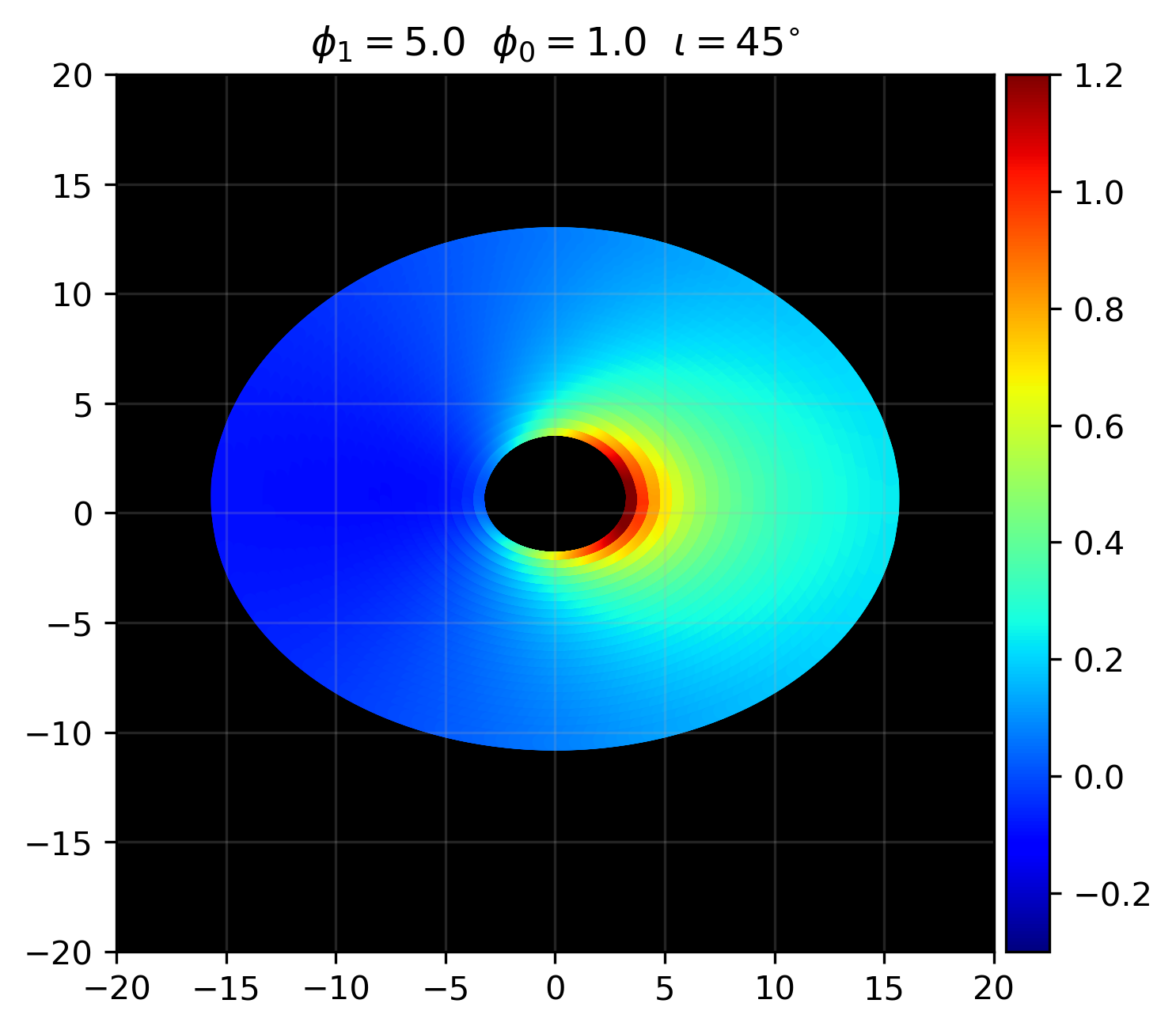}
    \includegraphics[scale=0.44]{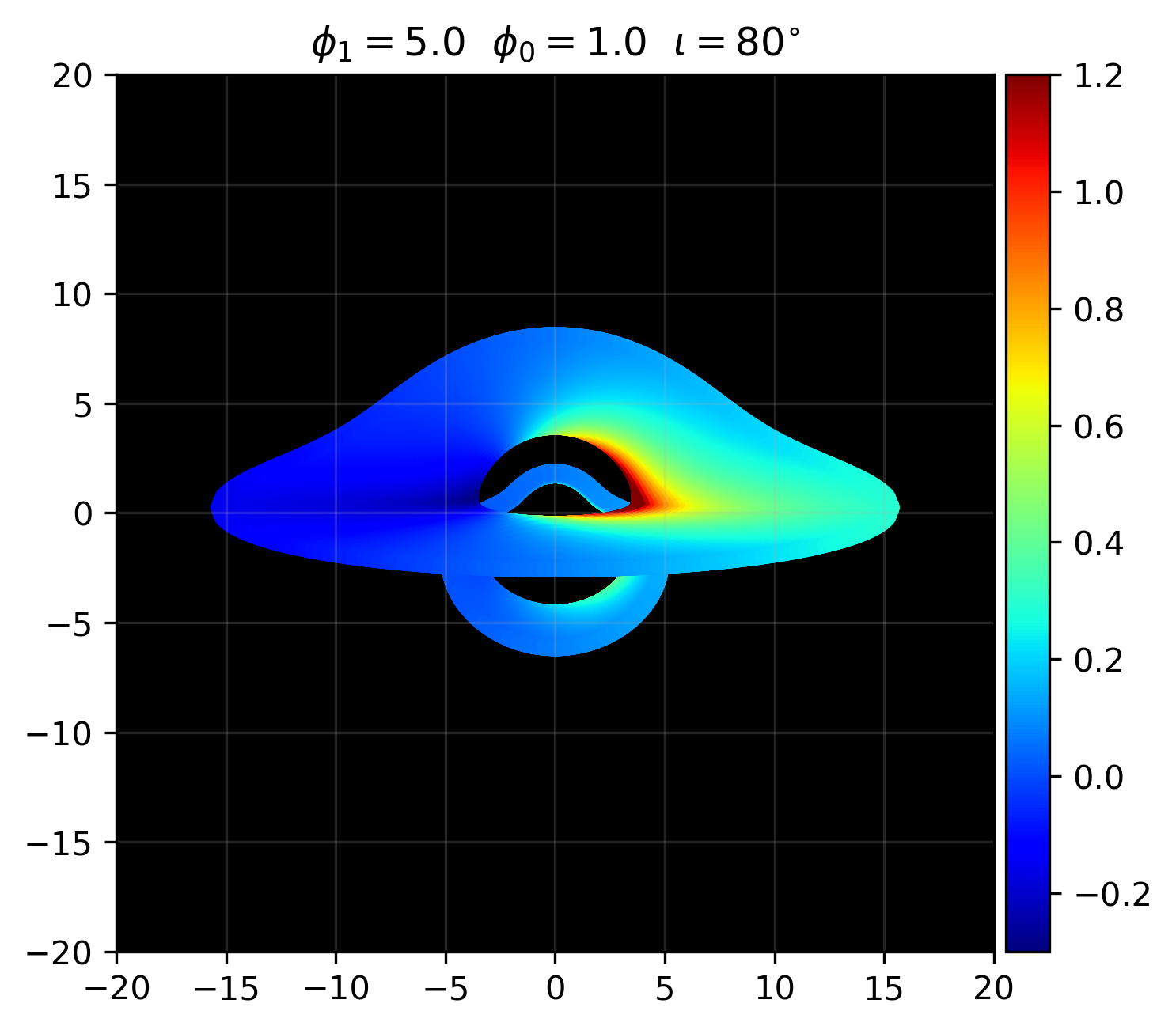}
\caption{Redshift $z$ in the direct images cast by the SHBH for different values of the parameters $\phi_1$ and $\phi_0$ with different inclination angles. The inner edge of the accretion disk corresponds to $r_\text{ISCO}$ while the outer edge is chosen as $5r_\text{ISCO}$. In the figures, we assume the observer is located at infinity and at rest in the gravitational field of the BH.}
\label{fig:accretionDiskRedshift}
 \end{figure*}

    \begin{figure*}[t]
    \centering
    \includegraphics[scale=0.48]{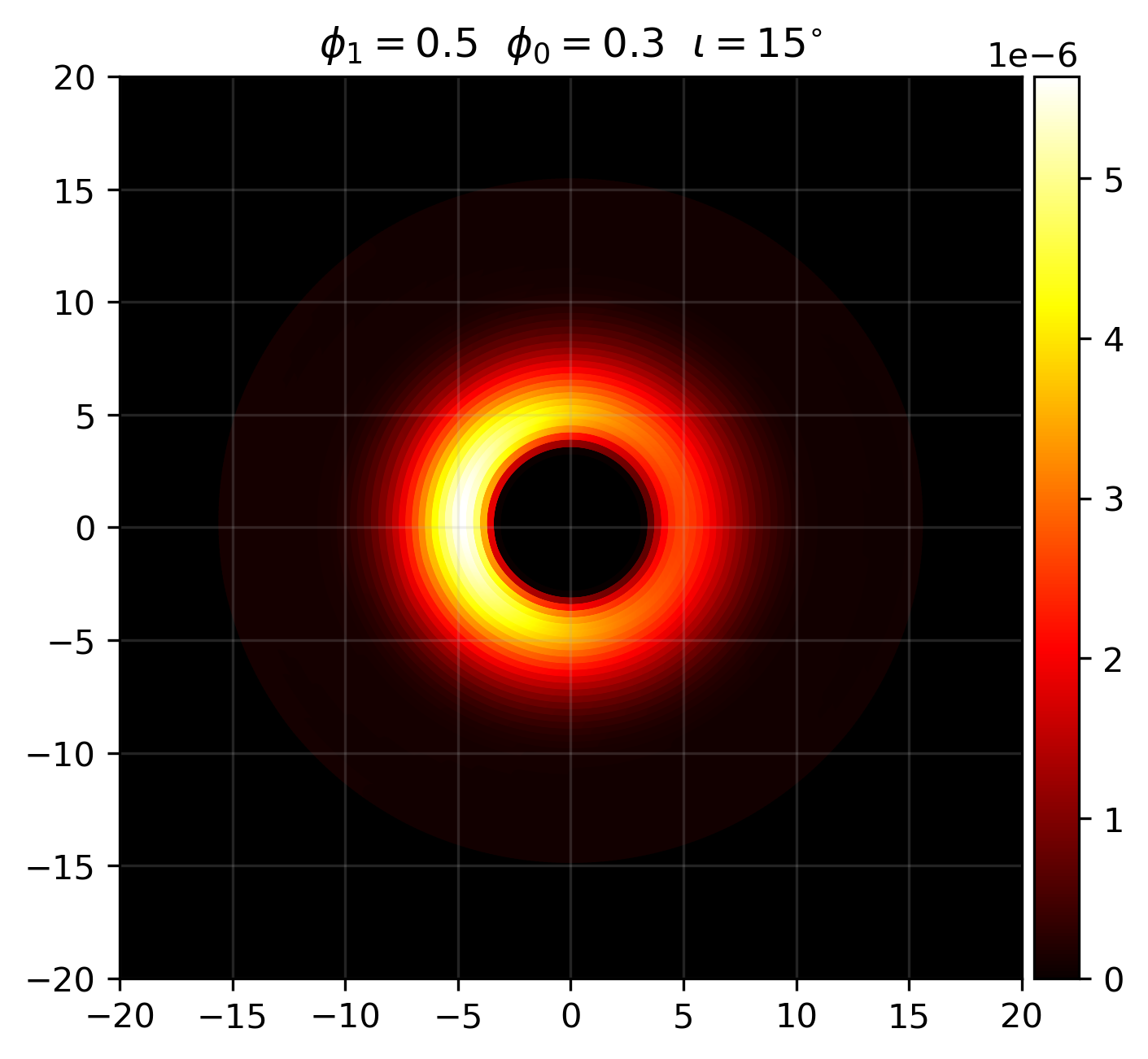}
    \includegraphics[scale=0.48]{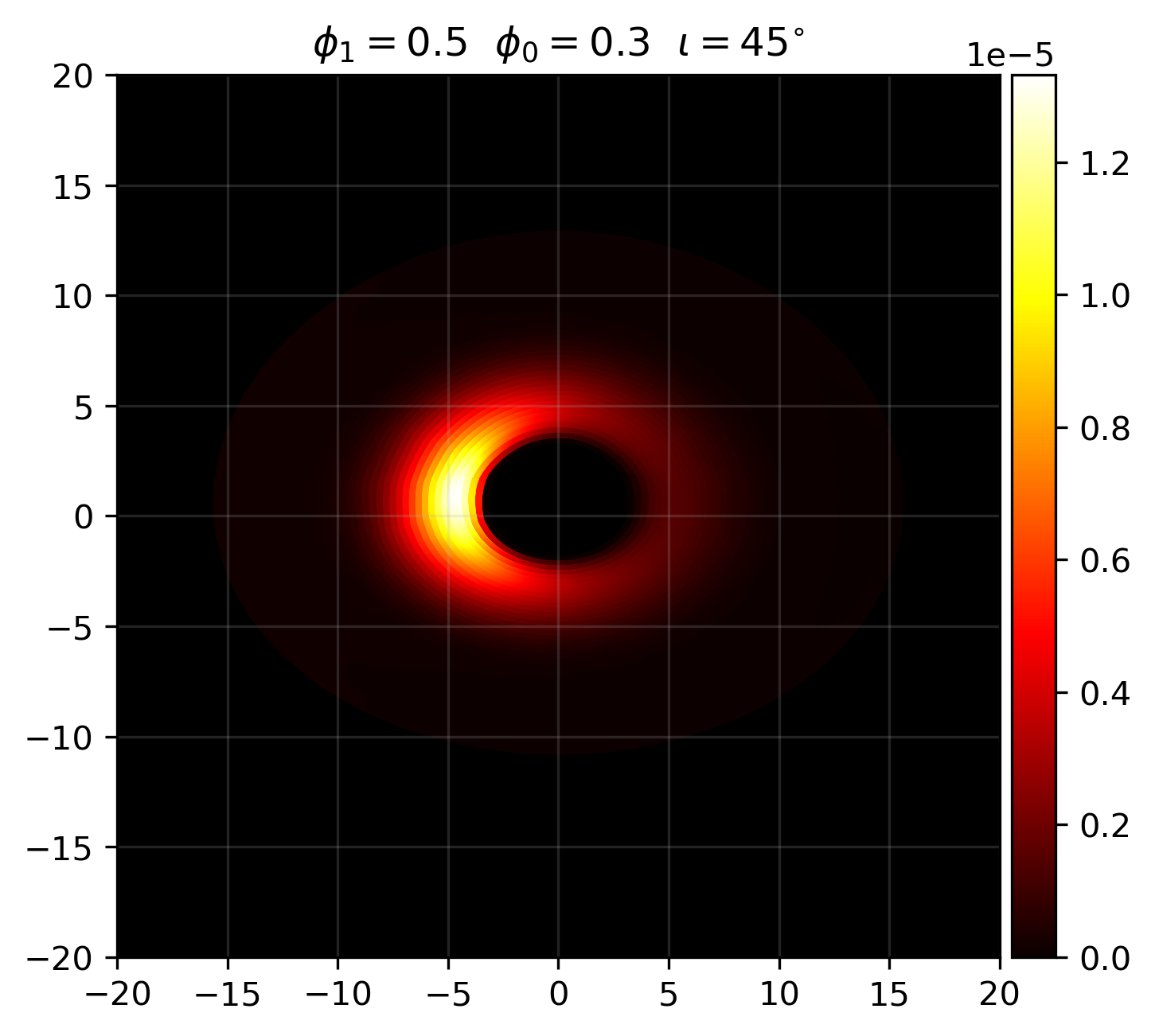}
    \includegraphics[scale=0.48]{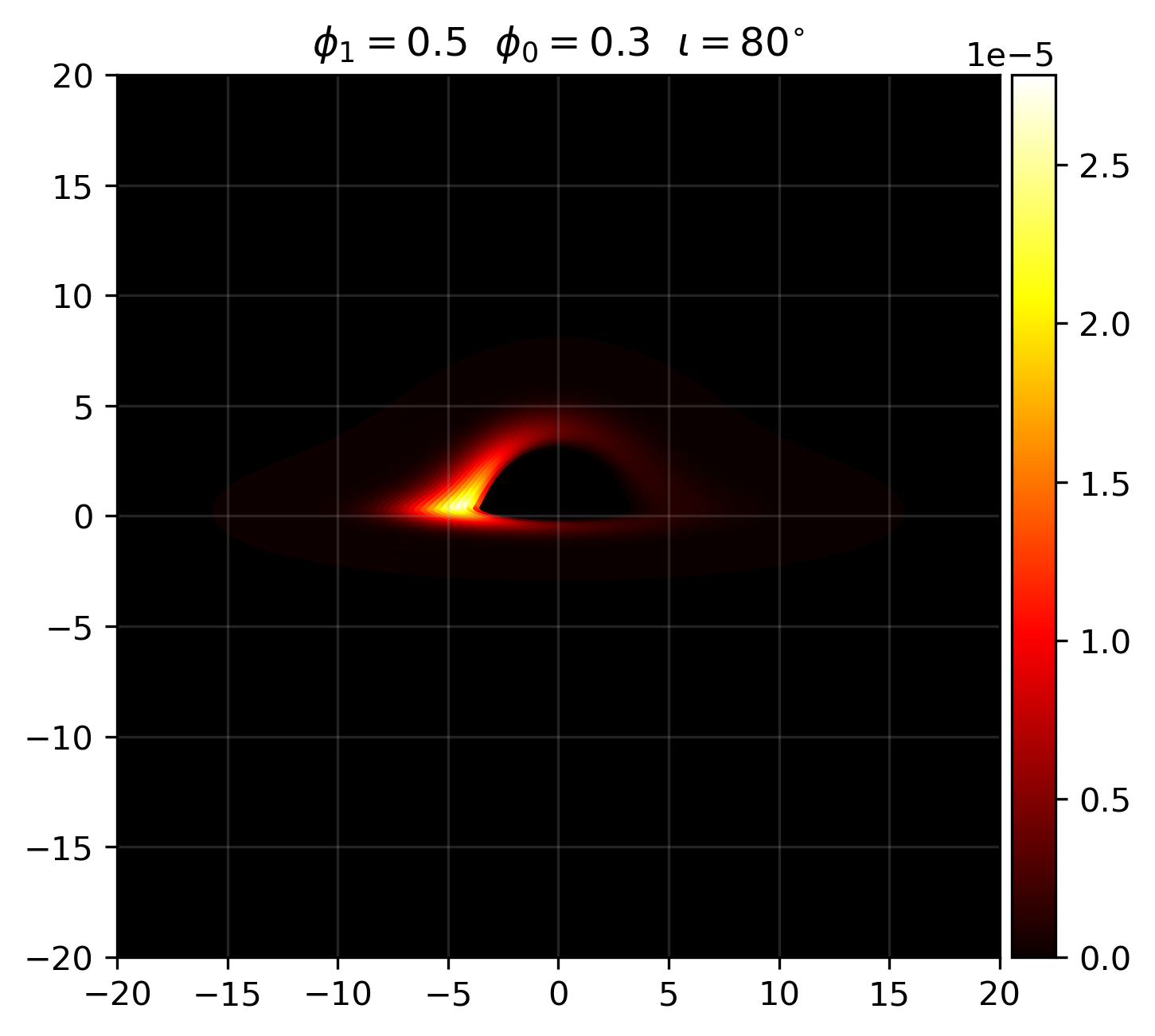}\\
    \includegraphics[scale=0.48]{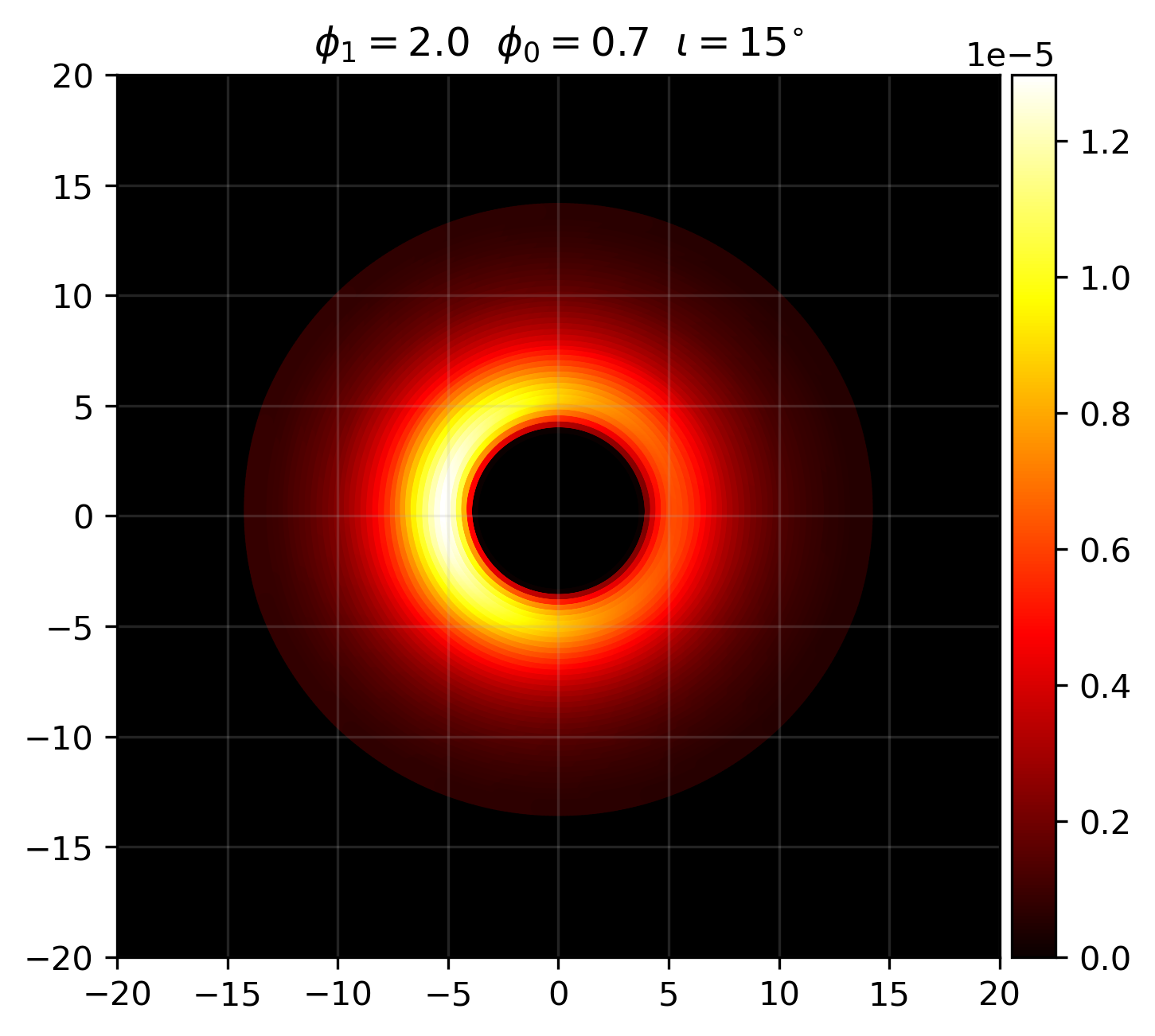}
    \includegraphics[scale=0.48]{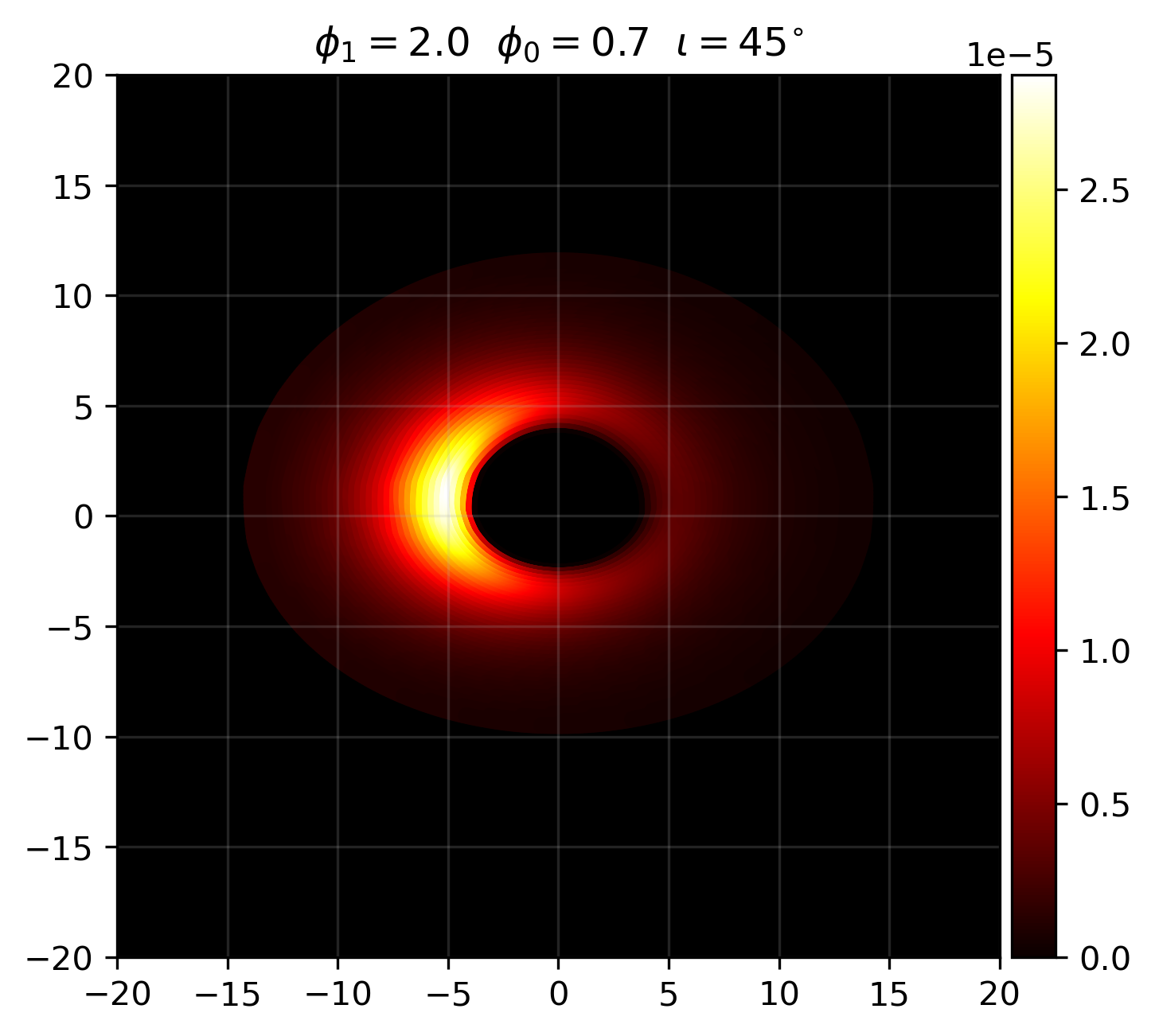}
    \includegraphics[scale=0.48]{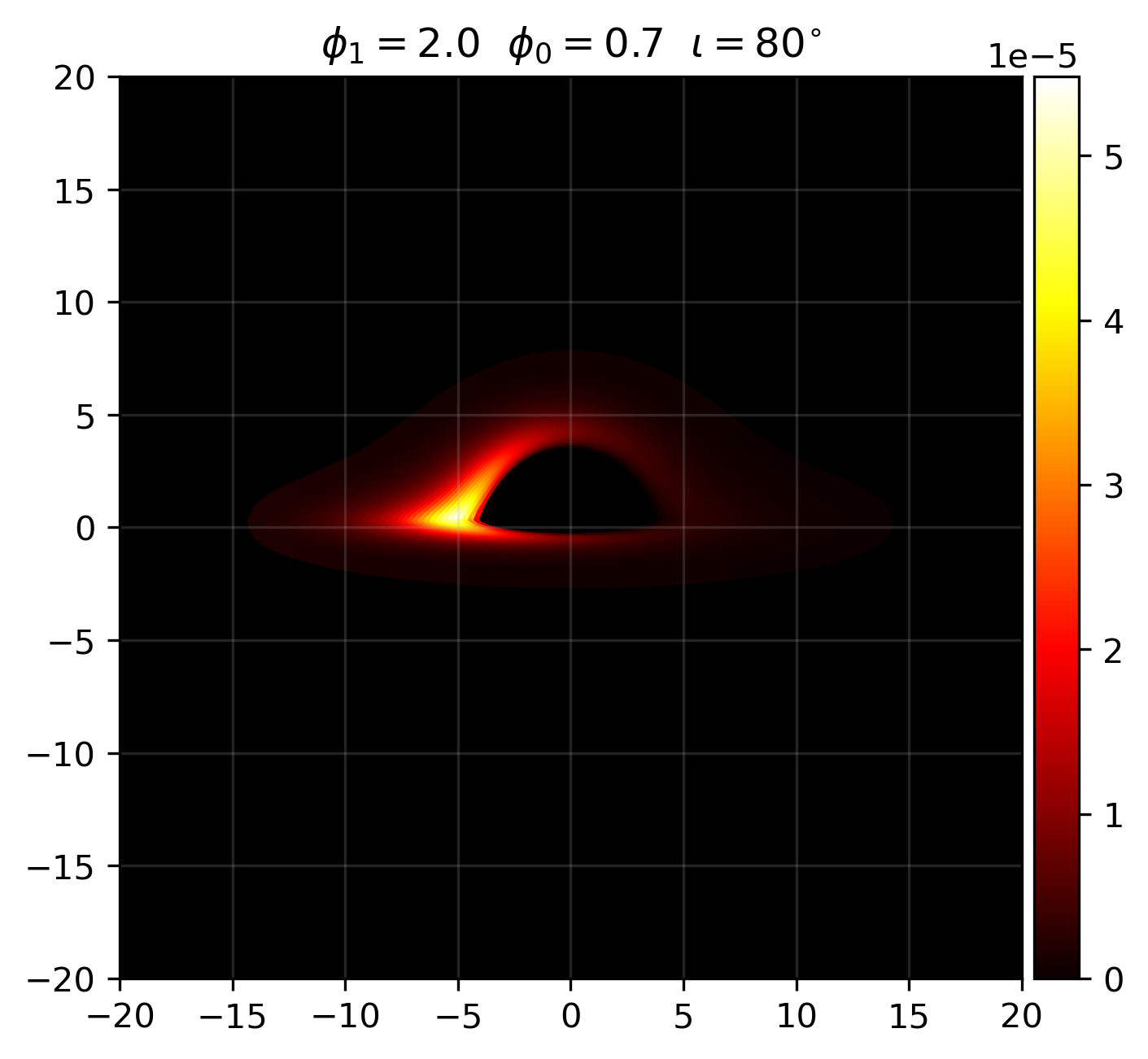}\\
    \includegraphics[scale=0.48]{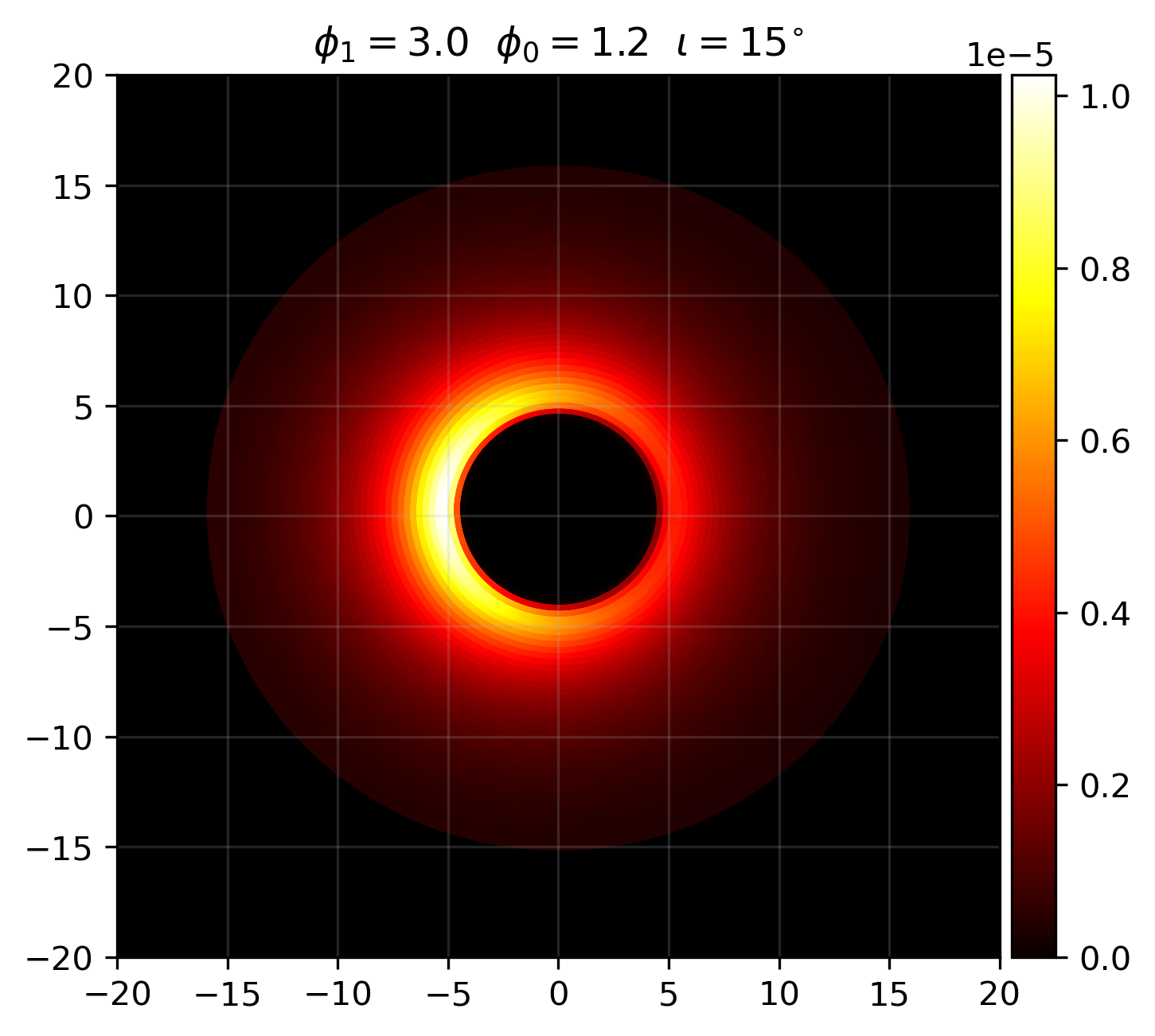}
    \includegraphics[scale=0.48]{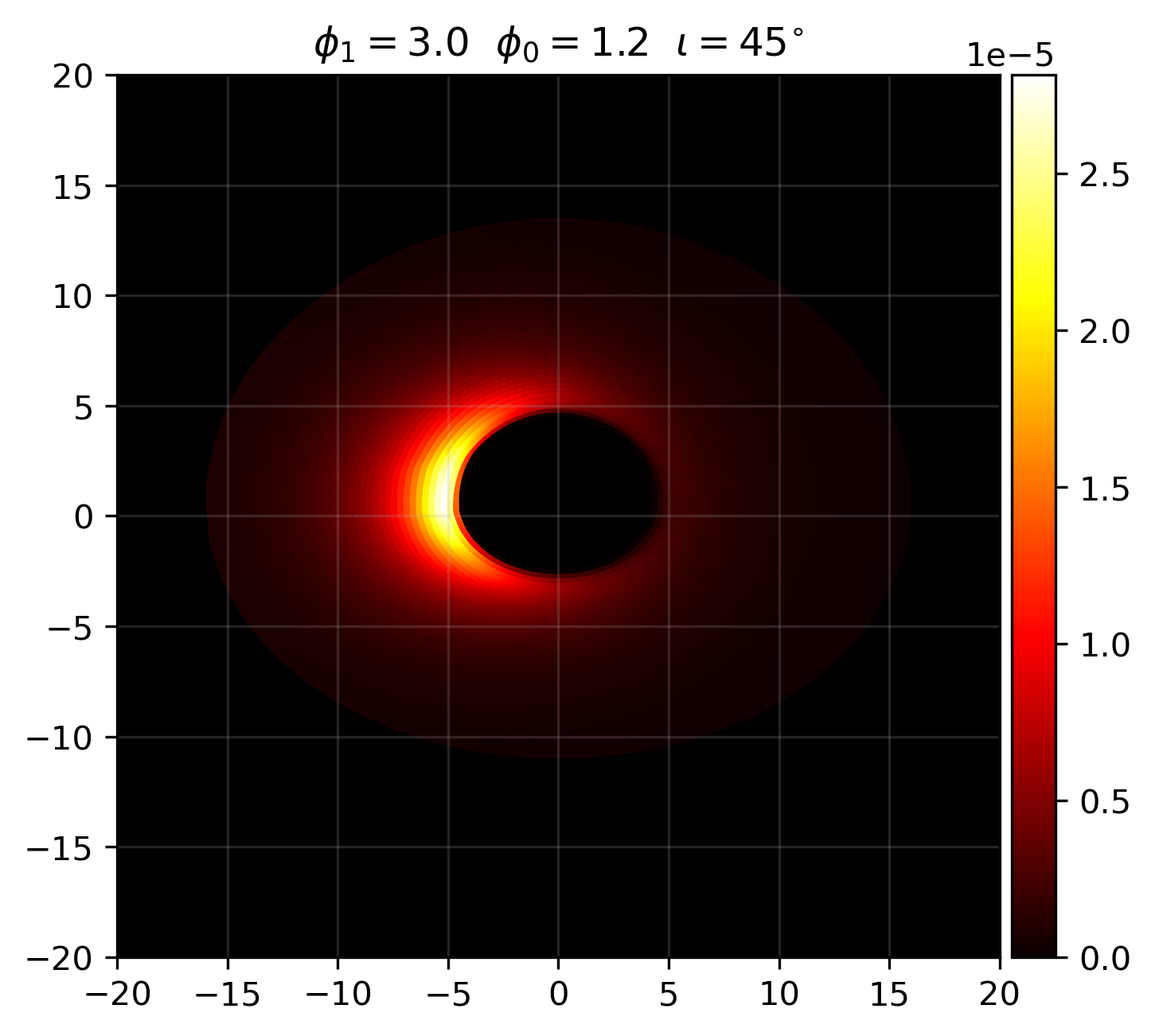}
    \includegraphics[scale=0.48]{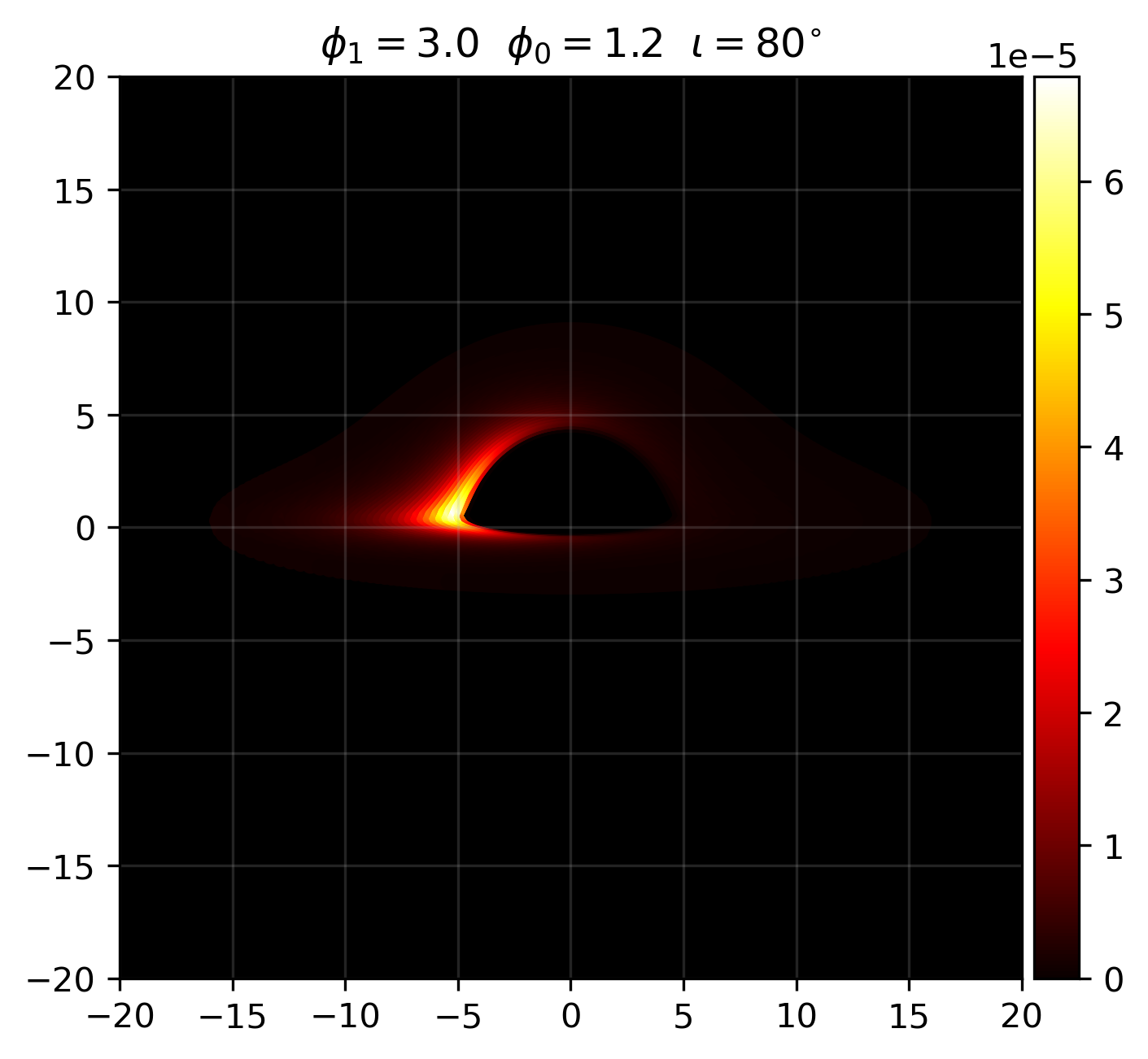}\\
    \includegraphics[scale=0.48]{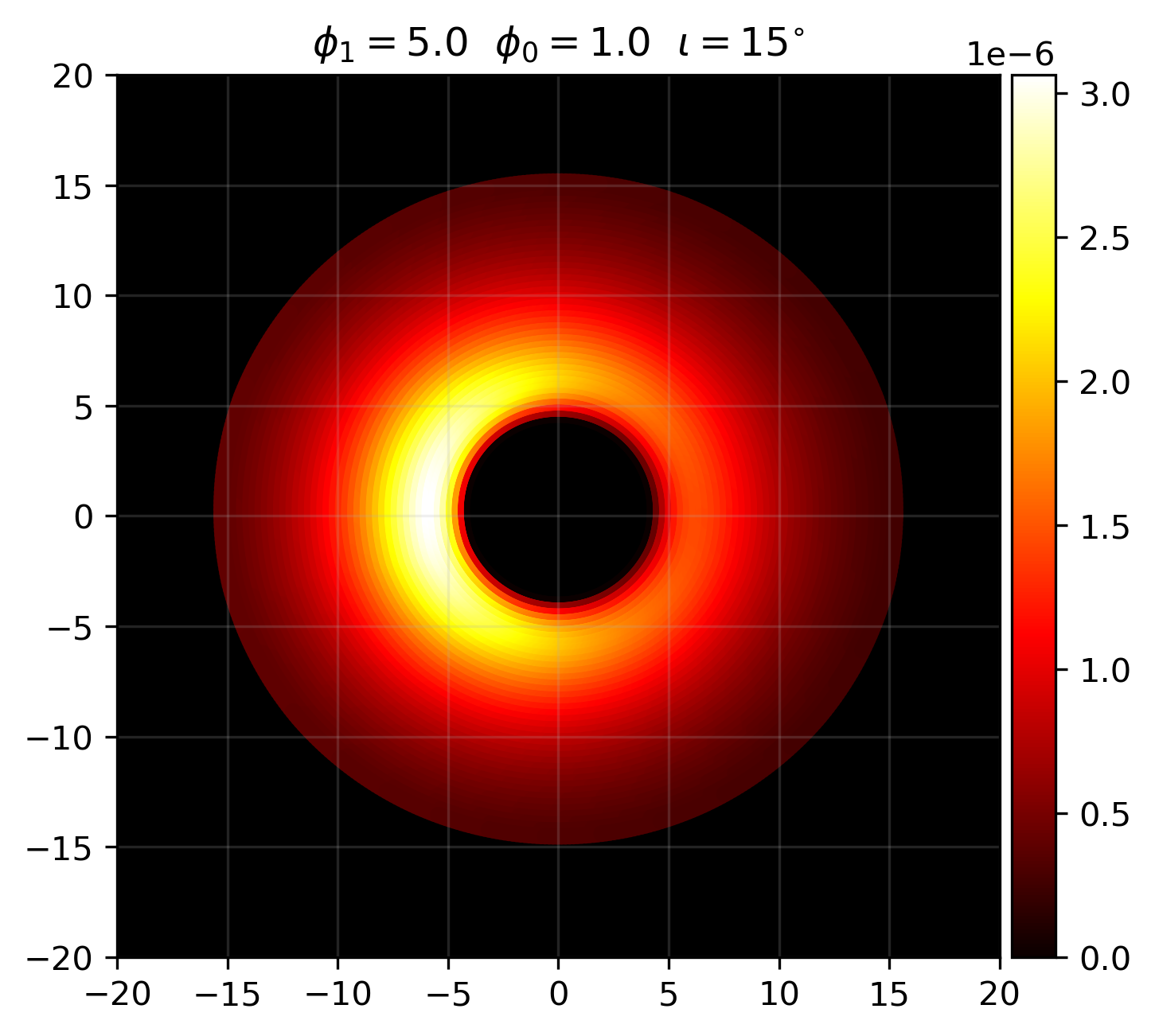}
    \includegraphics[scale=0.48]{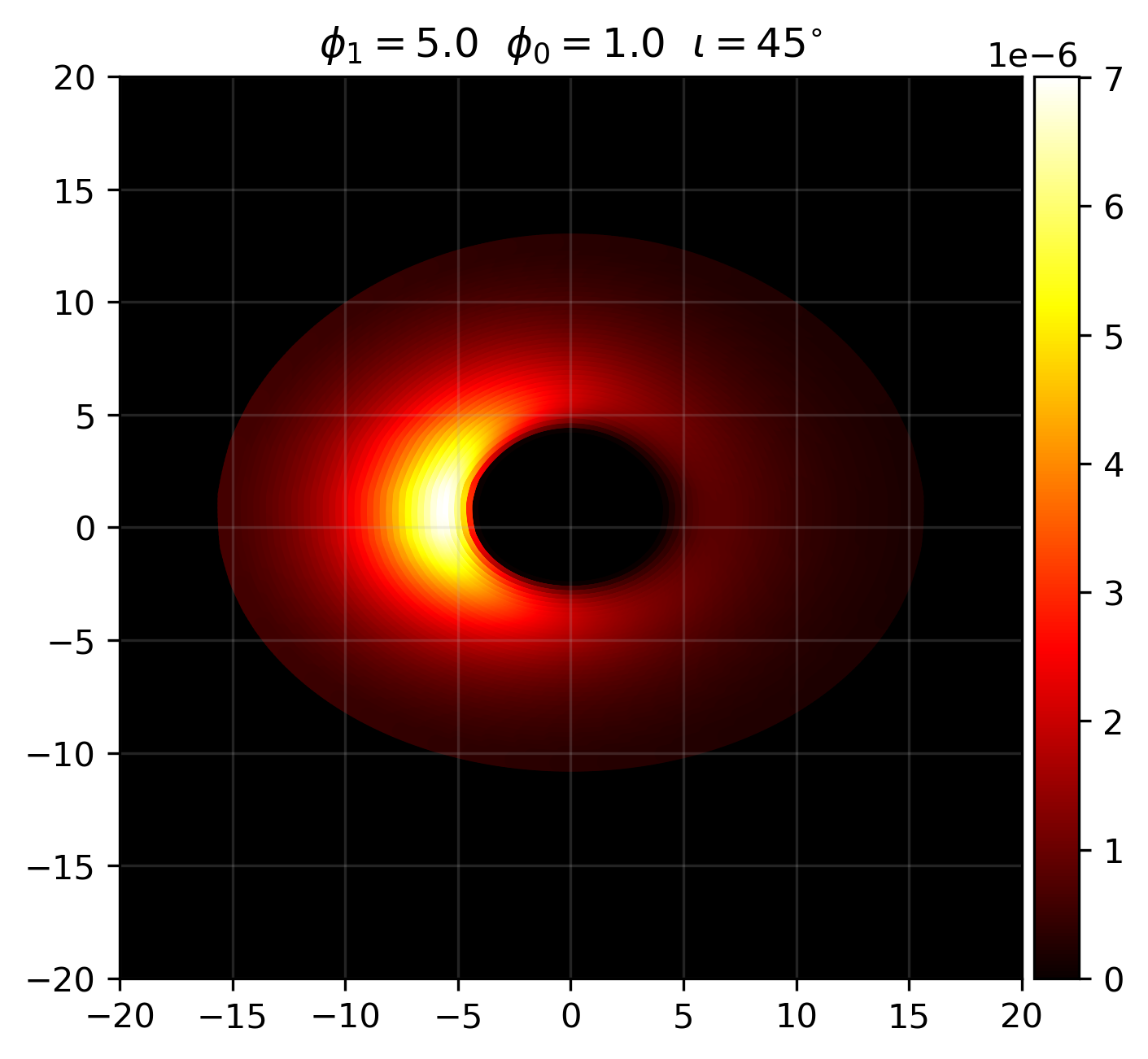}
    \includegraphics[scale=0.48]{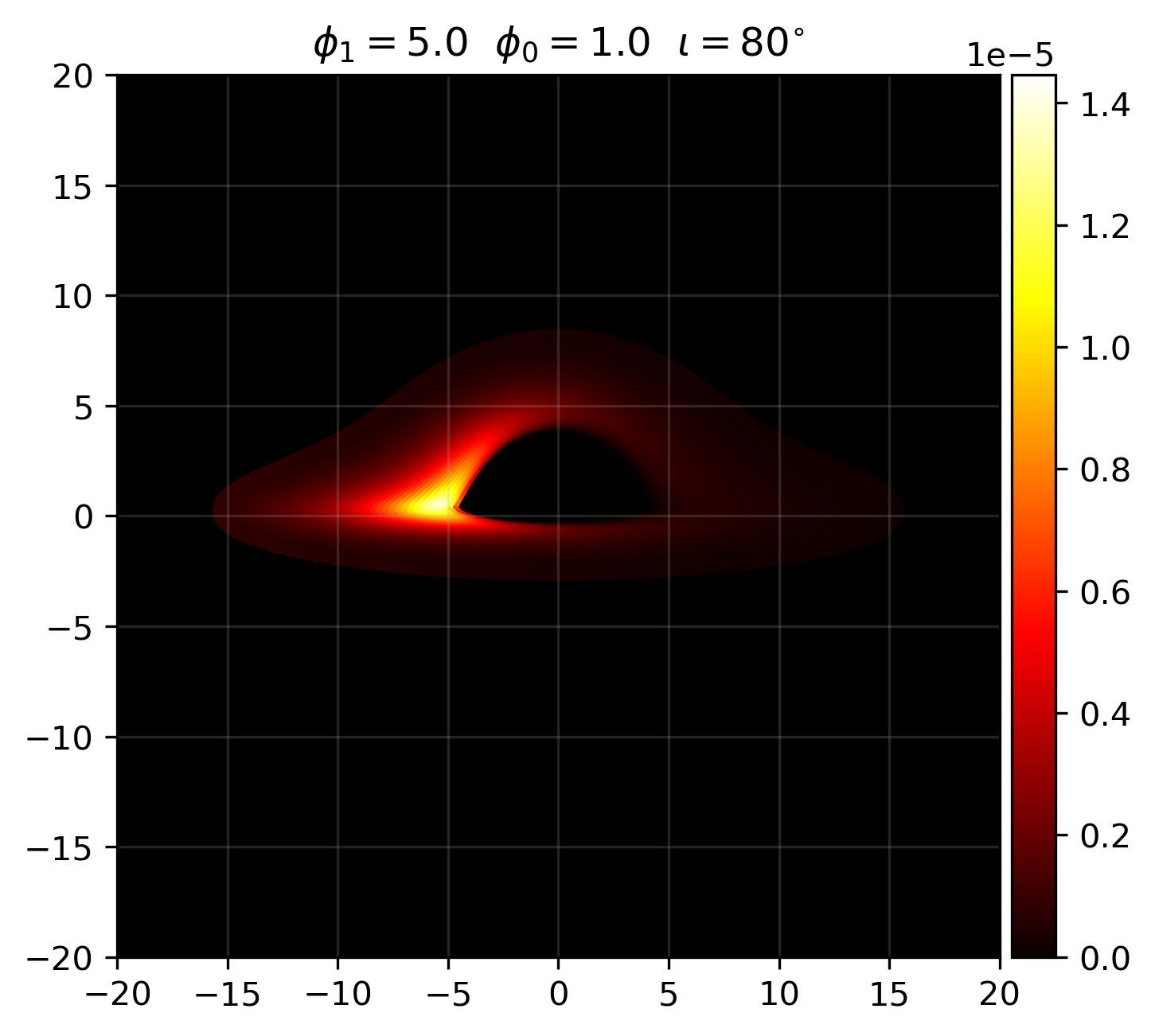}
\caption{Observed energy flux of the direct images cast by the SHBH for different values of the parameters $\phi_1$ and $\phi_0$ with different inclination angles. In the figures, we assume the observer is located at infinity and at rest in the gravitational field of the BH.}
\label{fig:accretionDiskFlux}
 \end{figure*}

\subsubsection{Accretion Disk}

	From an astrophysical standpoint, we consider models of BH with accreting material in the context of supermassive BH, which are thought to be located at the center of some active galactic nuclei. The accreting material probably has high angular momentum, resulting in non-spherical, but rather axisymmetric accretion, with the material forming a thin, flattened disk~\cite{Luminet1979}.
	
	According to Luminet, to construct a realistic image, it is necessary to build a precise model for the proper luminosity of the disk, taking into account the rotation of the disk, which induces a Doppler effect shift; this requires the following assuptions~\cite{Luminet1979}:
\begin{enumerate}
\item The BH has an external space-time geometry in which the disk lies with negligible self-gravity; this geometry is static, spherically symmetric, uncharged, and asymptotically flat. 
\item The disk is geometrically thin and opaque.
\item On average, the gas elements of the disk move very close to equatorial circular geodesic orbits around the BH.
\item Effects such as possible absorption by distant diffuse clouds surrounding the BH and those from the secondary heating of the disk by reabsorption of some of its light are neglected.
\end{enumerate}

   We begind by obtaining the isoradial curves seen by an observer located at infintiy. Hece, from Eqs.\eqref{eq:radialEoM} and \eqref{s3e10}, it is possible to write
\begin{equation}
\frac{dr}{d\varphi} = \frac{\dot{r}}{\dot{\varphi}} = \pm r^2 \sqrt{\frac{e^{2\sigma (r)}}{b^2} - \frac{N(r) }{r^2}},
\end{equation}
where the $\pm $ sign determines the direction of the motion for the photon. Usually, to integrate this equation, one introduces the variable $u =1/r$ (see for example~\cite{Chandrasekhar:1998}), and therefore, solving for $\varphi$ gives:
\begin{equation}
\label{eq:varphiIntegral}
\Delta \varphi = \pm \int_{u_o}^{u_s} \frac{du}{\sqrt{\frac{e^{2\sigma (u)}}{b^2} - u^2 N(u)}},
\end{equation}
where $\Delta \varphi$ is the deflection angle of the light ray traveling from the source position, $u_s$, to the distant observer at $u_o$. Light rays with $b>b(r_{ps})$ may have a radial turn point $u_t$ (the perihelion~\cite{Luminet1979}) defined by the condition  
\begin{equation}
\frac{e^{2\sigma (u_t)}}{b^2} - u_t^2 N(u_t) = 0.
\end{equation}
    In this case, the integral in (\ref{eq:varphiIntegral}) can be calculated as
\begin{align}
\Delta \varphi = \pm \biggl[ & \int_{u_o}^{u_t} \frac{du}{\sqrt{\frac{e^{2\sigma (u)}}{b^2} - u^2 N(u)}}\notag  \\ 
&+ \int_{u_t}^{u_s} \frac{du}{\sqrt{\frac{e^{2\sigma (u)}}{b^2} - u^2 N(u)}} \biggr].\label{eq:varphiIntegral2}
\end{align}

	To project the image of the accretion disk into the screen of an asymptotic observer, we define a cartesian coordinate system $(X, Y)$ as shown in Fig.~\ref{fig:imageplane}. Since we define the impact parameter on the observer's screen, it is possible to introduce the polar coordinates $(b, \eta)$. Therefore, the relation between the deflection angle $\Delta \varphi$, $\eta$ and the the inclination angle $\iota$ is given by (see references \cite{Muller2009} and \cite{Gyulchev2021} for details)
\begin{equation}
\cos \Delta \varphi = - \frac{\sin \eta \tan \iota}{\sqrt{\sin^2 \eta \tan^2 \iota + 1}},
\end{equation}
where the angular coordinate in the observer's screen takes the values $\eta \in [0,2\pi)$.

	From Fig.~\ref{fig:imageplane}, it is clear that the possible values for the deflection angle are $\frac{\pi}{2} - \iota \leq \Delta \varphi \leq \frac{\pi}{2} + \iota$. However, some light rays may turn around many times around the black hole before reaching the asymptotic observer, creating high-order relativistic images. Hence, we can establish that 
\begin{equation}
\Delta \varphi = n \pi - \arccos\left[ \frac{\sin \eta \tan \iota}{\sqrt{\sin^2 \eta \tan^2 \iota + 1}} \right],
\end{equation}
where the index $n = 1,2,3,...$ labels the order of the image. Identifying the deflection angle with the result in Eq.~\eqref{eq:varphiIntegral2} gives the equation that we will solve numerically to obtain the image of the accretion disk in the observer's screen,
\begin{widetext}
\begin{equation}
\label{eq:mainIntegral}
\pm \biggl[  \int_{u_o}^{u_t} \frac{du}{\sqrt{\frac{e^{2\sigma (u)}}{b^2} - u^2 N(u)}} + \int_{u_t}^{u_s} \frac{du}{\sqrt{\frac{e^{2\sigma (u)}}{b^2} - u^2 N(u)}} \biggr] = n\pi - \arccos\left[ \frac{\sin \eta \tan \iota}{\sqrt{\sin^2 \eta \tan^2 \iota + 1}} \right]. 
\end{equation}
 \end{widetext}
	
	Using this equation, in Fig.~\ref{fig:accretionDiskProfiles}, we plot some stable circular orbits on the equatorial plane. The procedure to obtain the image implies taking a specific value for the radius of the circular orbit (we consider the radii $r=r_\text{ISCO}, 2r_\text{ISCO}, 3r_{ISCO}, 4r_\text{ISCO}$ and $5r_\text{ISCO}$) and, for each value of the polar angle in the observer's plane, $0\leq \eta \leq 2\pi$, solve Eq.~\eqref{eq:mainIntegral} to obtain the corresponding polar coordinate $b$. Hence, Fig.~\ref{fig:accretionDiskProfiles} shows the primary images, $n=1$, of these circular orbits around some of the SHBHs solutions with various inclination angles ($\iota = 15^o, 45^o$ and $75^o$). 

\subsubsection{The Redshift Factor}
	
	The redshift factor,
\begin{equation}
g = 1 + z = \frac{E_e}{E_o},
\end{equation}
is obtained from the comparison frequency (energy) of the emitted and the observed photons, and these are obtained by projecting the momentum of the corresponding photon along the adequate velocity vector. For example, the observer is characterized by the velocity vector $p_o^\mu = \frac{\partial}{\partial t}$, indicating that he is static in the asymptotic region. Therefore, we have the energy of the observed photon as $E_o = k_\mu p_o ^\mu = k_t$. 
    
    On the other hand, for the emitted photon, we consider the rest frame of the emitting  region and write
\begin{equation}
E_e = k_\mu p_s^\mu = k_t p_s^t \left( 1 + \Omega \frac{k_\varphi}{k_t} \right).
\end{equation}
	
	As shown in \cite{Gyulchev2021, Tian2019, Liu2022, Hul2023}, the ratio $\frac{k_\varphi}{k_t} = b \sin \iota \cos \eta$ is an invariant along the trajectory of the photon. Hence, we may obtain the redshift factor as
\begin{equation}
g = \frac{1 + \Omega b \sin \iota \cos \eta}{-g_{tt}(r_s)}E, \label{eq:g_factor}
\end{equation}
where the energy and the angular frequency of the photon at the circular orbit are obtained from Eqs.~\eqref{eq:EnergyatCirc} and \eqref{eq:OmegaatCirc}, respectively.

	It is clear that the redshift factor depends on the radius $r_s$ of the source, and it is related to the coordinates $(b,\eta)$ on the observer's plane through Eq.~\eqref{eq:mainIntegral}. In Fig.~\ref{fig:accretionDiskRedshift}, we show the redshift distribution for the same SHBHs shown in Fig.~\ref{fig:accretionDiskProfiles}, considering an inner radius $r_\text{in} = r_\text{ISCO}$ and an outer radius $r_\text{out} = 5r_\text{ISCO}$ for various inclination angles.


\subsubsection{Energy Flux and the Image of the Accretion Disk}

Using the above results, we can obtain the image of a steady-state thin accretion disk on the equatorial plane of the SHBH. We will consider that the gas is in hydrodynamics equilibrium and described by the Page-Thorne model \cite{Page1974, Thorne1974}. Hence, the energy flux emitted by the disk is given by
\begin{equation}
F_e = - \frac{\dot{M}}{4\pi \sqrt{-\gamma}} \frac{\frac{d\Omega}{dr}}{(E - \Omega L )^2} \int_{r_\text{ISCO}}^{r_s} (E-\Omega L) \frac{dL}{dr} dr,
\end{equation}
where $\dot{M}$ represents the accretion rate, $\gamma = - e^{2\sigma(r)} r^4$ is the determinant of the induced metric in the equatorial plane,  $L$, $E$ and $\Omega$ given by Eqs.~\eqref{eq:AngMomentumCircular}, \eqref{eq:EnergyatCirc} and \eqref{eq:OmegaatCirc}, respectively.

This emitted energy flux will be affected by the redshift factor $g$ in Eq.~\eqref{eq:g_factor} to give the observed energy flux,
\begin{equation}
F_o = \frac{F_e}{g^4}.
\end{equation}

In Fig.~\ref{fig:accretionDiskFlux}, we show the observed energy flux in the direct images of a thin accretion disk around SHBHs with different values of $\phi_1$ and $\phi_0$. We consider the inclinations angles $\iota = 15^{\circ}, 45^{\circ}$ and $80^{\circ}$. The intensity of the energy flux is represented with a continuous color map, showing the existence of a brighter region on the left side of the disk caused by the Doppler effect produced by the gas rotation. The images also show that the intensity of the observed energy flux changes depending on the value of the parameters defining the black hole; i.e., $\phi_1$ and $\phi_0$.


\section{Conclusion\label{secV}}       

In this paper, we investigated the optical appearance of scalar hairy black holes with asymmetric potential, a numerical solution obtained in Ref.~\cite{Corichi:2005pa} and discussed in Ref.~\cite{Chew:2022enh}. To do so, we used the Hamiltonian formalism to obtain the equations of motion and define the potential for massive and massless particles (photons), from which we computed the photon sphere and the ISCO for different values of $\phi_1$ and $\phi_0$, parameters related to the asymmetric potential, see Eq.~\eqref{Sec_II_eq2}. In the case of photons, our results show that the smaller the value of $\phi_0$, the closer to the Schwarzschild case. Moreover, the location of the photon sphere changes as $\phi_1$ and $\phi_0$ increase; the change is evident for values of $\phi_1>1.0$. In the case of $\phi_1=0.5$, the photon sphere maintains closer to that of Schwarzschild, i.e., $r_\text{ps}/r_H\approx 1.5$. On the other hand, for SHBH solutions with $\phi_1>1.0$, we found that the photon sphere decreases, reaches a minimum value, and then increases again as $\phi_0$ increases.

When considering the ISCO, our results show a similar behavior: the smaller the values of $\phi_0$, the closer to the Schawarzchild black hole solution, being $\phi_0=0.5$ the closest, keeping the values of $r_\text{ISCO}/r_H\approx 3.0$. When $\phi_1$ and $\phi_0$ increase, our results show differences from Schwarzschild; however, we found that solutions with $\phi_1=2.0$ also keep the values of the ISCO close to that of Schwarzschild, see Fig.~\ref{fig:rISCO}. Similarly to the photon sphere, we also found that the ISCO reduced its value, reaching a minimum and increasing again as $\phi_0$ increases for values of $\phi_1>2.0$.

In the case of the shadow, our analysis shows it increases as the value of $\phi_0$ increases. We found this behavior for all values of $\phi_1$. As a consequence, the angular diameter of the shadow increases for all values of the asymmetric potential parameters $\phi_1$ and $\phi_0$. In particular, the increment is prominent for SHBH solutions with $\phi_1=1.0$, see Fig.~\ref{fig:EHTshadows2}. Using the values of the angular diameter as a function of $\phi_0$ for different values of $\phi_1$, we were able to constrain the asymmetric potential parameters. However, in the case of $\phi_1=0.5$, it was impossible to constrain the SHBH solutions since the numerical values did not reach the constraint set by the EHT observations for both M87* and SgrA*.  Therefore, the case $\phi_1=0.5$ is excluded by EHT observations.

In this work, we also investigated the profile of the specific intensity. In this case, we found an interesting behavior. Although the intensity $I_0$ has the usual profile (especially for small values of $\phi_0$), it becomes weaker when the value of $\phi_0$ increases. Moreover, our analysis shows that the photon sphere is sharper for small values of the $\phi_0$, while it blurs as $\phi_0$ increases and its size becomes large. In this sense, we can conclude that asymmetric parameter $\phi_0$ does affect the motion of photons, preventing them from circling the SHBH several times in contrast to the Schwarzschild case; this is shown clearly in Fig.~\ref{fig:shadows}.

The iso-radial curves for the SHBH solutions have similar behavior as the Schwarzschild case; i.e., they all have the famous ``mushroom'' shape described by J.~P.~Luminet in Ref.~\cite{Luminet1979}. Since the ISCO changes for different values of the asymmetric parameters $\phi_0$ and $\phi_1$, the size of the images is affected, growing for larger values of the ISCO. In the case of the primary image (continuous blue line in Fig.~\ref{fig:accretionDiskProfiles}), we see the usual behavior for different values of the inclination angle $\iota$. Nevertheless, when analyzing the secondary image, we found differences. For example, for small values of $\phi_0$ and $\phi_1$, the secondary image (blue dashed line in Fig.~ref{fig:accretionDiskProfiles}) resembles that of Schwarzschild. However, as the values of $\phi_0$ and $\phi_1$ increase, the upper part of the secondary image becomes flattened. This effect can be seen clearly in the last column, the fourth row of Fig.~ref{fig:accretionDiskProfiles}), where we show the SHBH solution with $\phi_0=1.0$ and $\phi_1=5.0$.  

In the case of the redshift, similar to the Schwarzschild case, we see that the Doppler blueshift contribution due to the rotation of the disk in the left half-part of the observer's plate can exceed the overall gravitational redshift part due to the presence of the SHBH. Therefore, the corresponding photons have a resulting blueshift. In the right half-part, on the other hand, we see a resulting redshifting effect.  The order of magnitude for the blueshift/redshift is the same for all the SHBH solutions. Hence, the asymmetric parameters, $\phi_0$ and $\phi_1$, do not affect the behavior of $z$.

Using the values of $z$ obtained at each point in the observer's photographic plate, we computed the bolometric flux relative to the total factor $F_e$. Our analysis shows, as in the Schwarzchild case, that the flux is maximum in the regions where the spectral shift is a blueshift. On the other hand, we found the flux is affected by the parameters $\phi_0$ and $\phi_1$, and the inclination angle $\iota$. For example, when the inclination angle changes from $80^\circ$ to $15^\circ$, the flux decreases its value. We also found that the smaller values for the flux occur for the SHBH solution with $\phi_0=1.0$, $\phi_1=5.0$ and $\iota=15^\circ$, where the order of magnitude for the flux is $10^{-6}$. In contrast to the SHBH solution with $\phi_0=1.2$, $\phi_1=3.0$ and $\iota=80^\circ$, where the order of magnitude for the flux is $10^{-5}$.

Finally, it is interesting to extend this work and consider axially symmetric black holes, which, from the astrophysical point of view, are more realistic and could help to test some black hole solutions in modified theories of gravity~\cite{Volkov:1998cc, Herdeiro:2014goa, Sullivan:2019vyi, Sullivan:2020zpf,Gao:2021ubl}. In particular, it would be interesting to consider the numerical solutions obtained by A.~Sullivan et al. in Refs.~\cite{Sullivan:2019vyi, Sullivan:2020zpf}, where the authors developed a numerical code to obtain stationary and axisymmetric solutions that describe rotating black hole spacetimes in a wide class of modified theories of gravity, such as scalar–Gauss-Bonnet gravity with a linear (linear scalar–Gauss-Bonnet) and an exponential (Einstein-dilaton–Gauss-Bonnet) coupling.

\section*{Acknowledgements} \label{sec:acknowledgements}
The authors thank Xiao Yan Chew for kindly sharing the numerical data. This work was supported by the Universidad Nacional de Colombia, Hermes Grant Code 57057, and by the Research Incubator No.64 on Computational Astrophysics of the Observatorio Astronómico Nacional. C.A.B.G. acknowledges the support of the Ministry of Science and Technology of China (grant No.~2020SKA0110201) and the National Science Foundation of China (Grants No.~11835009).

\textbf{Data Availability Statement} This manuscript has no associated data. (There is no observational data related to this article.)



	\end{document}